\documentclass[12pt]{article}
\usepackage{amsmath}
\usepackage{amssymb}
\usepackage{enumerate}
\usepackage{latexsym}
\usepackage{amsfonts}
\usepackage{ulem} 
\usepackage{graphicx}
\usepackage{amsfonts}
\usepackage{amssymb}
\usepackage{amsmath,amsthm}
\usepackage{stmaryrd}
\usepackage[linkcolor=blue,colorlinks,hypertexnames=false,plainpages=false]{hyperref}

\usepackage{picture}
\usepackage{a4wide}
\usepackage{color}

\numberwithin{equation}{section}
\theoremstyle{plain}
\newtheorem{thm}{Theorem}
\newtheorem{lemma}[thm]{Lemma}
\newtheorem{corollary}[thm]{Corollary}

\theoremstyle{definition}

\newcommand{\mmmintone}[1]{{\dis{\int\kern -.36cm
-}}_{\kern-.21cm\substack{#1}}\;\;}
\newcommand{\mmmintwo}[2]{{\dis{\int\kern -.43cm
-}}_{\kern-.21cm\substack{#1}}^{\substack{#2}}\;\;}
\newcommand{\submint}{{\scriptstyle{\int\kern -.66em -}}}
\newcommand{\submintone}[1]{{\scriptstyle{\int\kern -.66em
-}}_{\scriptscriptstyle{\kern-.21em\substack{#1}}}}
\newcommand{\fracmint}{{\textstyle{\int\kern -.88em -}}}
\newcommand{\fracmintone}[1]{{\textstyle{\int\kern -.88em
-}}_{\scriptscriptstyle{\kern-.21em\substack{#1}}}\;}

\newcommand{\Ii}{\text{\bf 1}}

\newcommand{\dis}{\displaystyle}

\newcommand{\nn}{\nonumber}
\newcommand{\und}{\underline}
\newcommand{\la}{\lambda}
\newcommand{\La}{\Lambda}

\newcommand{\eps}{\epsilon}
\newcommand{\ga}{\gamma}
\newcommand{\Ga}{\Gamma}
\newcommand{\Om}{\Omega}
\newcommand{\om}{\omega}
\newcommand{\si}{\sigma}

\newcommand{\nada}[1]{}

\newcommand{\cL}{\mathcal{L}}

\newcommand{\cR}{\mathcal{R}}

\def\square{\ifmmode\sqr\else{$\sqr$}\fi}
\def\sqr{\vcenter{
         \hrule height.1mm
         \hbox{\vrule width.1mm height2.2mm\kern2.18mm\vrule width.1mm}
         \hrule height.1mm}}                  
	\definecolor{deepcerise}{rgb}{0.85, 0.2, 0.53}
	\definecolor{darkspringgreen}{rgb}{0.09, 0.45, 0.27}
\definecolor{darkpastelgreen}{rgb}{0.01, 0.75, 0.24}
\definecolor{carmine}{rgb}{0.59, 0.0, 0.09}
\definecolor{caraibbengreen}{rgb}{0.0, 0.8, 0.6}
\definecolor{greyd}{cmyk}{0,0,0,0.4}
\newcommand{\red}{\textcolor{black}}

\usepackage{authblk}

\author[1]{Marzio Cassandro}
\author[2]{Immacolata Merola}
\author[1]{Errico Presutti}
\affil[1]{\small GSSI, viale Francesco Crispi, 7 - 67100 L'Aquila (AQ)}
\affil[2]{University of L'Aquila, DISIM, Via Vetoio, Coppito, (AQ)}
\begin{document}

%
%
%
%
%
%
%
%
%
%
%
%
%
\title{Renewal properties of the $d=1$ Ising model}
\makeatletter
\newcommand{\subjclass}[2][201]{%
  \let\@oldtitle\@title%
  \gdef\@title{\@oldtitle\footnotetext{#1 \emph{Mathematics subject classification.} #2}}%
}
\newcommand{\keywords}[1]{%
  \let\@@oldtitle\@title%
  \gdef\@title{\@@oldtitle\footnotetext{\emph{Key words and phrases.} #1.}}%
}
\makeatother
\newcommand*\MSC[1][2010]{ #1 Mathematical subject classification: }

\maketitle
 \begin{abstract}
We consider the $d=1$ Ising model with Kac potentials at inverse temperature $\beta>1$ where
mean field predicts a phase transition with two possible equilibrium magnetization  $\pm m_\beta$,
$m_\beta>0$.  We show that when the Kac scaling parameter $\ga$ is sufficiently small  typical spin configurations are  described (via a coarse graining) by an infinite sequence of successive plus and minus intervals where the empirical magnetization is ``close'' to $m_\beta$ and respectively $-m_\beta$. We prove that the corresponding marginal
of the unique DLR measure is
a renewal process.

\MSC{82B05, 82B20, 60K05}
 \end{abstract}
%
%
%
%
%
%

\vskip2cm

\section {Introduction}
\label{sec:intro}

In this paper we consider the  Ising model 
with a
ferromagnetic Kac potential.
The formal  hamiltonian is
      \begin{equation}
        \label{intro.1}
H_\ga(\si) = -\frac 12 \sum_{x\ne y} J_\ga(x,y) \si(x)\si(y)
     \end{equation}
$\si(x)\in \{-1,1\}$ is the spin at site $x\in \mathbb Z^d$ and
      \begin{equation}
        \label{intro.2}
J_\ga(x,y)= \ga c_\ga J(\ga|y-x|)
     \end{equation}
where $J(r)\ge 0$ is a smooth probability density  supported by $r\le 1$;
$c_\ga$ a normalization constant such that $\sum_{y\ne x}J_\ga(x,y)=1$, $c_\ga\to 1$ as $\ga\to 0$.

The mean field version of the model has a free energy
     \begin{equation}
        \label{intro.3}
f_{\beta}(m)=  -\frac{m^2}2 - \frac{S(m)}{\beta}
     \end{equation}
where $\dis{S(m)= -\frac{1-m}{2}\log \frac{1-m}{2}- \frac{1+m}{2}\log \frac{1+m}{2}}$
is the entropy when the magnetization is $m$. When $\beta>1$, $f_{\beta}(m)$ is a symmetric two wells function with minima at $\pm m_\beta$, where
 $ m_\beta$ is the positive solution of the mean field equation
      \begin{equation}
        \label{intro.4}
m_\beta = \tanh\{\beta m_\beta\},\quad \beta>1
      \end{equation}
This suggests that when $\ga$ is small the typical spin
configurations should be close to $m_\beta$ or $-m_\beta$.  Indeed when
$d\ge 2$ it is proved (\cite{CP}, \cite{BZ}, \red{\cite{BMPZ}}) that for any $\ga$ small enough the plus DLR measure (obained as limit of Gibbs measures with plus boundary conditions) has typical configurations described by a ``sea'' where the ``local magnetization'' is close to $+m_\beta$ with small and rare islands where the local magnetization is close to $-m_\beta$. The spin flip of the above picture describes the minus DLR measure.

In this paper we study the $d=1$ case. In one dimension with finite range interaction and  any inverse temperature $\beta$   there is a unique
DLR measure, thus typical configurations cannot be as in $d\ge 2$ predominantly close to $m_\beta$ or to $-m_\beta$ and therefore they must be close to alternating $+m_\beta$ and $-m_\beta$ intervals.  The problem has been first studied in \cite{COP}.  The original idea  in \cite{COP} was to see this in the context of  metastability, see for instance \cite{OV05}, \cite{BDeH},  namely  to relate
 the typical spin configurations to the trajectories of a random walk in the two well potentials $\ga^{-1}f_\beta(m)$.  For $\ga$ small the two wells are separated by a very high barrier and typically the random walk stays close to the bottom of a well with small fluctuations and it will very rarely jump to the other well; it will then keep doing that for ever, namely alternating from one well to the other.

Since the barrier height scales as $\ga^{-1}$ the waiting time for jumps from one well to the other   scales as  $e^{c \ga^{-1}}$ while the time it takes for the actual jump is much smaller as it scales as $\ga^{-1}$.

\red{In a first  version of our paper we have considered the problem in the presence of a magnetic field $h$. The magnetic field modifies the two wells potential so that the intervals with magnetization $m_\beta$ and $-m_\beta$ get different lengths. The paper however was getting too long and we decided to restrict to $h=0$. We did not examine the case when  the magnetic field is random: the length of the intervals then scales as $\ga^{-2}$ (see \cite{COP1} and \cite{COP2}) in contrast with the exponential behaviour at $h=0$. It is an open interesting question whether  the renewal properties that we prove in this paper extend to the case with the random magnetic field. }


 The relation with metastability however is not straightforward because we are studying a Gibbs process while metastability is usually framed in the context of Markov processes.  It is true that in $d=1$ Gibbs processes (with finite range interaction as in our case) are Markov but the transition probability of the latter is not simply related to the Hamiltonian of the Gibbs measure. Indeed to get to the transition probability one needs to know spectral properties of the transfer matrix, in particular the eigenvectors
of the maximal eigenvalue.  This is the approach used by Kac et al. \red{(\cite{KUH1}, \cite{KUH2}, \cite{KUH3})}  to derive the van der Waals theory from systems with Kac potentials, however to carry out the whole program along these lines looks maybe possible but not easy at all.

\red{An alternative approach, used in
  \cite{COP}, and which  goes back to Lebowitz and Penrose, \cite{LP},
shifts the mathematical context  from the spectral analysis of the transfer matrix to a variational problem wth a non local free energy functional.
The tunnelling problem for the corresponding non local ``penalty functional" has been studied in \cite{BDDP}, \cite{BDP} in the $d\ge 2$ case.
Going back to  Lebowitz and Penrose, the reduction to a variational problem comes from a coarse graining which gives rise to the free energy functional }
     \begin{equation}
        \label{intro.5}
\mathcal F(m)= \int dr\{f_{\beta}(m(r))-f_{\beta}(m_\beta)\} + \int\int drdr'
J(|r-r'|) [m(r)-m(r')]^2
     \end{equation}

\red{Lebowitz and Penrose used  } coarse graining to describe spin configurations in terms of a sequence of successive intervals in $\mathbb Z$ where a plus interval is followed by an interface interval then by a minus interval, then by another interface interval and this structure is repeated endlessly.  In the plus and minus intervals the empirical magnetization of the spins is close to $m_\beta$, respectively $-m_\beta$, the interface intervals separate the pluses from the minuses.  A precise definition is given in the next section. Calling $\mu_\ga$ the DLR measure (at inverse temperature $\beta>1$) in \cite{COP} it was proved that:

\begin{itemize}

\item  The probability  $\mu_\ga[B]$ of the event $B$ that the origin belongs to an interface interval vanishes as $\ga\to 0$.

\item  In the set $B^c$ call  $\ell_\ga$ the length of the plus or minus interval which contains the origin, then for any $\delta>0$
    \begin{equation}
        \label{intro.6}
\lim_{\ga\to 0}\mu_\ga\Big[e^{\ga^{-1}(\bar f -\delta)}\le \ell_\ga \le  e^{\ga^{-1}(\bar f +\delta)}  \Big| B^c  \Big] =1
     \end{equation}
where
    \begin{equation}
        \label{intro.7}
\bar f  = \inf_{m(r) \to \pm m_\beta \;{as}\: r\to \pm \infty} \mathcal F(m)
     \end{equation}

\item  Properly normalized the distribution of $\ell_\ga$ converges to an exponential distribution of mean 1 as $\ga\to 0$ and the lengths of successive intervals become independent.

\end{itemize}

Purpose of this paper is to investigate the structure of the plus, minus and interface intervals  when $\ga$ is small but without taking the limit $\ga\to 0$.  We call $\Om$ the space of sequences of such intervals and $P_\ga$ the measure on $\Om$ induced by $\mu_\ga$.  Namely $P_\ga$ is obtained from $\mu_\ga$ by integrating over all spin configurations which give rise to the same sequence and since we are just interested in the configurations in $\Om$
$P_\ga$  retains exactly the information we are interested in.  The disadvantage  when going from $\mu_\ga$ to $P_\ga$ is that we loose the nice property of $\mu_\ga$ that its conditional probabilities have finite range dependence on the conditioning.

$P_\ga$ however has a very nice structure, in fact (and this is
the main result in this paper)  $(\Om,P_\ga)$ is a renewal process.  More precisely we show that it is possible to add to the configurations in $\Om$   sequences of ``renewal marks'' so that the new space $(\Om^*,P^*_\ga)$ has for all $\ga$ small enough
the following properties:

\begin{itemize}

\item   $(\Om^*,P^*_\ga)$ is a renewal process where the renewal property occurs when a
renewal mark
appears.

\item The marginal of $P^*_\ga$ on $\Om$, namely disregarding the marks, is $P_\ga$.

\end{itemize}

Thus if we want to compute the probability of a given finite sequence of intervals we go to the space $(\Om^*,P^*_\ga)$ and look for the first mark appearing before our sequence, what happens earlier is not relevant in computing the probability.

The paper is organized as follows.
In Section \ref {sec:12} we define the parameters relevant to our analysis and in Section \ref {sec:13} we present our main results and discuss their physical interpretation.
Section \ref{sec:outline} presents an outline of how the proof of the statements of Section \ref {sec:13} is organized. The subsequent sections and  appendices are devoted to detail the full proofs.

\vskip2cm

\setcounter{equation}{0}

\section{Plus, minus and interface intervals}
\label{sec:12}

In this section we make precise the definition of plus, minus and interface intervals.
We will use throughout the paper four main lengths: $\ga^{-1}$, which is the interaction length, $\ell^-_\ga=\delta \ga^{-1}$, $\delta\in (0,1)$, and $\ell^+_\ga=  \ga^{-(1+\alpha)}$, $\alpha\in (0,1/2)$, which are the lengths used in the definition of plus and minus phases, finally $\ga^{-1/2}$ which is the coarse graining length.  We will also use a parameter
$\zeta>0$ in the definition of the plus and minus phases.  The relation between
$\delta$ and $\zeta$ is as follows: $\zeta$ can be any positive number $\le \zeta^*$, $\zeta^*>0$   suitably
small; then for any such $\zeta$ there is $\delta^*=\delta^*(\zeta)$ positive and we can take any
$\delta \le \delta^*$, (all that independently of $\ga$).  For the definition of
$\delta^*$ and $\zeta^*$ we refer to  Chapter 6 of \cite{presutti}.

We want the four lengths commensurable, namely $\ell^+_\ga$ an integer multiple of $\ga^{-1}$ which in turns should be an integer multiple of $\ell^-_\ga$ which should be an integer multiple of $\ga^{-1/2}$.  This could be achieved by taking integer parts but to have simpler notation
we suppose $\alpha = 1/4$, $\ga \in \{2^{-4n}, n\in \mathbb N\}$ and $\delta\in \{ 2^{-k}, k\in \mathbb N\}$.  For $\ga$ small enough the above commensurability requests are fulfilled and we will tacitly suppose that $\ga$ is as small as needed.

We use the standard notation in lattice systems, namely if $\Delta \subset \mathbb Z$ then $\si_\Delta = \{\si(x), x\in \Delta \}$, if $\Delta \cap \La= \emptyset$, $\si_\Delta,\si_\La
= \{\si(x), x\in \Delta \cup \La \}$ and with $\Delta$ and $\La$ as above, $\Delta$ finite,
   \begin{equation}
        \label{12.1}
H_\ga(\si_\Delta | \si_\La) = -\sum_{\{x,y\} \in \Delta  \cup \La: \{x,y\}\cap \Delta \ne \emptyset} J_\ga(x,y) \si(x) \si(y)
    \end{equation}
which is the energy of the spins in $\Delta$ in interaction with those in $\La$.

We  call
   \begin{equation}
        \label{12.2}
C_i^{\pm} = \{x \in  [i\ell^{\pm}_\ga,(i+1)\ell^{\pm}_\ga) \}
    \end{equation}
and shorthand
   \begin{equation}
        \label{12.3}
s_i = \si_{C_i^{+}},\quad \und s = \{s_i, i \in \mathbb Z\}
    \end{equation}
$s_i$ will be called the $i$-th block spin.
Given a spin configuration $\si$ we first define
     \begin{eqnarray}
      \label{12.4}
&&\eta_i =
  \pm 1 \;\;\text{if}\;\;   \big| \frac{1}{\ell^-_\ga} \sum_{y\in C_i^-}(\si(y)\mp m_\beta)| \le \zeta,\quad
  \text{$\eta_i=0$ otherwise}
    \end{eqnarray}
and then
      \begin{eqnarray}
      \label{12.5}
&&\theta_i   =
  \pm 1 \;\;\text{if}\;\; \eta_j \equiv \pm 1 \;\;\text{for all}\;\; C_j^- \subset C_i^+,\quad \text{$\theta_i =0$ otherwise}
  \\&&
  \Theta_i =
  \pm 1 \;\;\text{if}\;\; \theta_j  \equiv \pm 1,\; j=i-1,i, i+1,\quad \text{$\Theta_i (\und s)=0$ otherwise}
     \label{12.6}
  \end{eqnarray}
The plus [minus] phase  in a configuration $\si$ is the set of points where $\Theta = 1$ [$\Theta=-1$].  As we are going to see plus and minus intervals are defined by allowing fluctuations in the plus and minus phases.  

\vskip.5cm

\noindent
{\bf Definition.} [Plus, minus and interface intervals]

\begin{itemize}

\item
An interval
$[m,n]$ is a $+-$ interface if $\Theta_{m-1} = 1$, $\Theta_{n+1}=-1$ and $\Theta_{i}=0$
for all $i \in [m,n]$.

\item
An interval
$[m,n]$ is a $-+$ interface if $\Theta_{m-1} = -1$, $\Theta_{n+1}=1$ and $\Theta_{i}=0$
for all $i \in [m,n]$.

\item
$[m,n]$ is a plus interval if there is  a $+-$ interface which starts at $n+1$ and a
$-+$ interface which ends at $m-1$.

\item
$[m,n]$ is a minus interval if there is  a $-+$ interface which starts at $n+1$ and a
$+-$ interface which ends at $m-1$.

\end{itemize}

\medskip

Thus a plus interval $[m,n]$ starts at $m$ with $\Theta_m=1$ and ends at $n$ with
$\Theta_n=1$ while $\Theta_i\ge 0$ at all $i \in (m,n)$, moreover $[m,n]$ is maximal with such properties.  There could be plus intervals
made of a singleton, i.e.\ with $m=n$.  Minus intervals are defined symmetrically.

A spin configuration determines a partition of $\mathbb Z$ whose elements are the plus, minus and interface intervals.  The partition is denoted by $\om$, the atoms of $\om$ are denoted by $\om_{\ell,m}$, $\ell \in \mathbb Z$, $m\in \{1,..,4\}$, using the following convention.  $\{\om_{\ell,1},\ell \in \mathbb Z\}$, is the collection of all the plus intervals,
$\{\om_{\ell,2},\ell \in \mathbb Z\}$, of the $+-$ interfaces, $\{\om_{\ell,3},\ell \in \mathbb Z\}$, of the minus intervals and finally
$\{\om_{\ell,4},\ell \in \mathbb Z\}$, of the $-+$ interfaces.  The atoms are ordered from left to right in the sense that $\om_{\ell,m}<\om_{\ell',m'}$ if $\ell < \ell'$ or $\ell = \ell', m<m'$.  The same partition $\om$ can arise from two sequences  $\om_{\ell,m}$ and $\om'_{\ell,m}$ if  there is $\ell_0$ so that for all $\ell$ and $m$,  $\om_{\ell,m}=\om'_{\ell+\ell_0,m}$, we then call $\om_{\ell,m}$ and $\om'_{\ell,m}$
equivalent and denote by $\om$ classes of equivalence and by $\Om$ the space of all such equivalent classes.

It is sometimes convenient to describe a configuration $\om$ as the collection $(x_{\ell,m})$ of the positions of the left points of the atoms $\om_{\ell,m}$ of $\om$.  Of course we are interested in equivalent classes, where $(x_{\ell,m})$ is equivalent to $(x'_{\ell,m})$
if there is $n\in \mathbb Z$ so that $x_{\ell,m}=x'_{\ell+n,m}$ for all $\ell$ and $m$.  Local sets in $\Om$ are denoted by $X^*=\{x^*_{\ell,m}, (\ell',m')\le (\ell,m) \le (\ell'',m'')\}$ and are the set of all $(x_{\ell,m})$ such that (for a suitable choice in the equivalence class of  $(x_{\ell,m})$)
$x_{\ell,m} =x^*_{\ell,m}, (\ell',m')\le (\ell,m) \le (\ell'',m'')$.

Calling $\psi$ the map from the space of  spin configurations $\{-1,1\}^{\mathbb Z}$ (such that $\om_{\ell,m}$ is well defined) to the space $\Om$, we define on the local sets $X^*$
   \begin{equation}
        \label{12.7}
P_\ga[X^*] = \mu_\ga \big[ \und s: \psi(\und s) \in X^* \big]
    \end{equation}
where $\mu_\ga$ is the  DLR measure with hamiltonian \eqref{intro.1} at inverse temperature $\beta>1$. $P_\ga$ is then extended to the $\si$-algebra generated by the local events.


\vskip 2cm

\setcounter{equation}{0}

\section{Main results}
\label{sec:13}

The main results in this paper are (1) the proof that the DLR process of the plus, minus and interface intervals is a renewal process and (2) a characterization of the thermodynamics of the system.  We state here the main theorems which will then be proved in the next sections, leaving the more technical details to the appendices.

\vskip.5cm

\subsection{The renewal process}
The space $\Om$ is made of quadruples $(\om_{\ell,1},..,\om_{\ell,4})$ one after the other and indexed by $\ell \in \mathbb Z$.
We will prove that their distribution $P_\ga$ can be realized by suitably clustering together finitely many  quadruples and giving independent  weights to each cluster. Thus the process starts anew every time that a new cluster appears.

We start by looking only at  the lengths of the quadruples without caring about their location.
We denote by $\und u$
finite sequences   of quadruples 
of integers:
$\und u= \{u_{\ell,m}, \ell =1,..,k, m=1,..,4\}$, $k \in \mathbb N$, and define
  \begin{equation}
        \label{13.0}
   \mathcal R=\{\und u:u_{1,1}\ge 3;\;
   u_{\ell,1}\ge 1,\; u_{\ell,2}\ge 2,\; u_{\ell,3}\ge 1,\; u_{\ell,4}\ge 2\}
    \end{equation}
The length of $\und u$ is defined as
  \begin{equation}
        \label{13.0.1}
  |\und u|= \sum_{\ell,m}
u_{\ell,m}
    \end{equation}
 and by \eqref{13.0} if $\und u \in \mathcal R$ then $|\und u| \in [8,\infty)$.

In Section \ref{sec:77.1} we will introduce a probability on $\mathcal R$ (denoted by $w_{\la_\ga}$ for reasons explained in the Remark below) with the following properties (see Theorem \ref{thm77.1} and Theorem \ref{thm77.2}):

\vskip.5cm
\noindent
{\bf Properties of $w_{\la_\ga}$.}

\begin{itemize}

\item $w_{\la_\ga}(\und u)>0$ for all $\und u \in \mathcal R$ and
$\sum_{\und u} w_{\la_\ga}(\und u)=1$ (being a probability).

\item There are $c>0$ and $\delta_\ga>0$ so that for all $R>0$
  \begin{equation}
      \label{77.2}
  \sum_{\und u: |\und u| \ge R}  w_{\la_\ga}(\und u)  \le c e^{-\delta_\ga R}
      \end{equation}

\item
The first moment is finite and denoted by:
 \begin{equation}
        \label{13.1}
\alpha_\ga^{-1} := \sum_{\und u \in \mathcal R} w_{\la_\ga}(\und u) |\und u|
    \end{equation}

\end{itemize}

\medskip

\noindent
{\bf Remark.} Observe that the convergence of the series in \eqref{13.1} follows from \eqref{77.2} because
 \begin{equation}
        \label{13.111}
\sum_{\und u \in \mathcal R} w_{\la_\ga}(\und u) |\und u| = \sum_{n\ge 1}
\sum_{{\und u: |\und u| = n}}  w_{\la_\ga}(\und u)
    \end{equation}
We will see that $\log \alpha_\ga$ and $\log \delta_\ga$ are proportional to $-\ga^{-1}$.

 We will first define weights $w(\und u)$ on $\mathcal R$ which are determined by the statistical weight of the plus, minus and interface intervals.  We will then call $w_\la(\und u):= e^{-\la |\und u|}w(\und u)$, $\la >0$, and prove that there is a unique value $\la_\ga$ of $\la$ for which $w_\la(\und u)$ is a probability.
  We will see that $w_{\la_\ga}(\und u)$ satisfies the properties listed above
 and that  $\la_\ga$ is related to the thermodynamic pressure of the system.

  \medskip

We next denote by $W_\ga$ the probability on $\mathcal R^{\mathbb Z}$ product of the $w_{\la_\ga}$.  The elements of  {$\mathcal R^{\mathbb Z}$
are denoted by $(\und u_i, i \in \mathbb Z)$.  \red{We will often use in the sequel the following classical theorem  (which, for the readers convenience, is proved in Appendix \ref{app-last}):}

\begin{thm}
\label{thm13.1}
There are $c'$ and $\delta'_\ga$ positive so that for any positive integer $n$
  \begin{equation}
        \label{13.1.1}
\Big| W_\ga\Big[\{(\und u_i)_{ i \in \mathbb Z}:\;\text{\rm there is $k$
so that}\;\sum_{i=1}^k |\und u_i| = n\}\Big] - \alpha_\ga \Big| \le
 c'e^{-\delta'_\ga n} 
    \end{equation}
where  $\alpha_\ga$ is defined in \eqref{13.1}.

 \end{thm}

\medskip

We may regard the elements of $\mathcal R^{\mathbb  Z}$ as sequences of rods of lengths $|\und u_i|$ with  internal structure $\und u_i$: our next step is ``to put them'' on $\mathbb Z$.

\medskip

\noindent
{\bf Definitions and notation.}  {\em The space $\Om^*$ and the map $\phi:\Om^*\to \Om$.}

 Consider the sequence of pairs
  \begin{equation}
        \label{13.1.2}
(\und u_i, x_i)_{ i \in \mathbb Z}: \und u_i\in \mathcal R, x_i \in \mathbb Z,
x_{i+1} - x_i = |\und u_i|
    \end{equation}
$x_i$ is interpreted as the position of the rod $\und u_i$, more precisely of its left endpoint; the rods are placed consecutively, one after the other by the last condition in \eqref{13.1.2}.  The labelling is not important and we call equivalent $(\und u_i, x_i)_{ i \in \mathbb Z}$ and $(\und u_i', x'_i)_{ i \in \mathbb Z}$ if there is $n\in \mathbb Z$ such that
$(\und u_i', x'_i) = (\und u_{i+n}, x_{i+n})$ for all $i$.  Notice that an element $(\und u_i, x_i)_{ i \in \mathbb Z}$ is determined by the sequence of the lengths of the rod and by the position of only one of the rods, as the other positions are fixed by the constraints $x_{i+1} - x_i = |\und u_i|$.
We call $\Om^*$ the space of all $(\und u_i, x_i)_{ i \in \mathbb Z}$ (identifying equivalent elements).

There is a ``natural map'' $\phi:\Om^*\to \Om$ defined
by looking in an element of $\Om^*$ only at the sequence $\om_{\ell,m}$ of the internal structures of the rods in $(\und u_i, x_i)_{ i \in \mathbb Z}$. The range of $\phi$ is actually a subset $\Om^{\ge 3}$ of $\Om$ of all  $\om$ such that $\{\ell:|\om_{\ell,1}|\ge 3\}$ is a doubly infinite sequence (the DLR  measure of the complement is equal to 0).
Alternatively we may recover the elements of $\Om^*$ from those of $\Om^{\ge 3}$ by putting ``marks'' in the set $\{\om_{\ell,1}: |\om_{\ell,1}|\ge 3\}$:
then a mark at $(\ell,1)$ selects a site $x$ defined as the left endpoint of $\om_{\ell,1}$. The sets $\{x_i\}$ of such marks are then identified with the sets $\{x_i\}$
 in $(\und u_i, x_i)_{ i \in \mathbb Z}$, the specification $\und u_i$ being then the lengths of the $\om_{\ell,m}$ between $x_i$ and $x_{i+1}$.

In agreement with the notation of Section \ref{sec:12} given an element $(\und u_i, x_i)_{ i \in \mathbb Z}\in \Om^*$ we denote by
$\{x_{\ell,m}, \ell \in \mathbb Z, m=1,..,4\}$ the set of all the left endpoints of the partition  $\om=\phi((\und u_i, x_i)_{ i \in \mathbb Z})$.

\vskip1cm

We define a ``renewal probability''
$P^*_\ga$ on  $\Om^*$ by specifying first the probability of the ``cylinders'' and then extending this to the minimal $\si$-algebra generated by the cylinders.

The specification of a cylinder set is a sequence
$ (\und v_i,y_i)_{i\in [1,k]}$, $k\ge 1$,
such that $\und v_i \in \mathcal R$, $i=1,..,k$ and $y_{i+1}-y_i=|\und v_i|$, $i=1,..,k-1$.  The cylinder with such specification is:
  \begin{equation}
        \label{13.2}
C_{(\und v_i,y_i)_{i\in [1,k]}} = \Big\{(\und u_i, x_i)_{i\in \mathbb Z}\in \Om^*:\;(\und u_i, x_i)=(\und v_i,y_i), i=1,..,k  \Big\}
    \end{equation}
Since $\Om^*$ is defined modulo equivalence this is
the same as
  \begin{equation}
        \label{13.2.1}
C_{(\und v_i,y_i)_{i\in [1,k]}} = \Big\{(\und u_i, x_i)_{i\in \mathbb Z}\in \Om^*:\;(\und u_{n+i}, x_{n+i})=(\und v_i,y_i), i=1,..,k  \Big\}
    \end{equation}
for any $n \in \mathbb Z$. Physically   $C_{(\und v_i,y_i)_{i\in [1,k]}}$ is the event where there is a rod $\und v_1$ at $y_1$ and the next $k-1$ rods have specification $\und v_2,..,\und v_k$.

We next define the $P^*_\ga$ probability of a cylinder as
  \begin{equation}
        \label{13.3}
P^*_{\ga} \Big[C_{(\und v_i,y_i)_{i\in [1,k]}}\Big] = \alpha_\ga \prod_{i=1}^k w_{\la_\ga}(\und u_i)
    \end{equation}
The probability $P^*_{\ga}$ on $\Om^*$ is finally defined as the probability which extends \eqref{13.3} to the minimal $\si$-algebra generated by
the cylinders.
%
%

%
%
%
%

 In Section \ref{app:thm4.1.3-b} we will prove the following theorem about the renewal property of the distribution of the plus, minus and interface intervals:

\medskip

\begin{thm} [Renewal property of the DLR measure]
\label{thm4.1.3-a}

The inverse image under $\phi$ of any local set $X^*$ in $\Om$ is a countable union of disjoint cylinders  $C_i$ in $\Om^*$. It is therefore in the $\si$-algebra where $P^*_\ga$ is defined and
  \begin{equation}
        \label{13.3.0}
P_\ga[ X^*] = \sum_i P^*_\ga [C_i],\quad \phi^{-1}X^*= \bigcup_i C_i
    \end{equation}
As a consequence $P_\ga$ (which is defined on the $\si$-algebra generated by the local sets) satisfies:
 \begin{equation}
        \label{13.3.00}
P_\ga =P^*_\ga \;  \phi^{-1}
    \end{equation}

\end{thm}

%
%
%
%
%

\vskip.5cm

\subsection{Thermodynamics of the model}

Besides the thermodynamic pressure $p_\ga$ (of our Ising model with hamiltonian \eqref{intro.1}) we will introduce several other pressures.  In particular we consider here the pressure $p^{+}_\ga$ relative to plus intervals (namely obtained
by restricting to plus intervals), see the next section for a precise definition. We will prove at the end of Section \ref{sec:77.1} that:

\medskip

\begin{thm}
\label{thm13.3}
The thermodynamic pressure $p_\ga$ is equal to
 \begin{equation}
        \label{13.5.0}
p_\ga = p^{+}_\ga + \frac{\la_\ga}{\beta \ell^+_\ga}
    \end{equation}
where $\la_\ga$ is the positive parameter introduced in the Remark at the beginning of this section (existence of $\la_\ga$ and properties of $w_{\la_\ga}(\und u)$ are proved in Section \ref{sec:77.1}).  Moreover if $Z^{\rm pbc}_{\La_n}$ is the partition function in the region $\La_n=[-n,n]$ (in block spin variables) with periodic boundary conditions, then
%
 \begin{equation}
        \label{13.5.1}
\Big| \frac{Z^{\rm pbc}_{\La_n}}{e^{\beta p_\ga(2n+1)\ell^+_\ga}} -1 \Big| \le c'' e^{-\delta''_\ga n}
    \end{equation}
where $c''$ and $\delta''_\ga$ are positive constants.

\end{thm}

\medskip

Thus by Theorem \ref{thm13.3} the  renewal property which is
strictly related to $\la_\ga$ has a thermodynamic meaning in terms of the pressure difference $p_\ga - p^{+}_\ga$.   $\la_\ga$ however has also a thermodynamic interpretation in terms of ``surface tension''. In fact in \eqref{77.4} it is shown
that $\big|\la_\ga-\eps_\ga\big| \le  c_\ga \eps_\ga^2$ and in \eqref{77.1.0.00.2} that  $\big|\alpha_\ga- \frac{ \eps_\ga}{2}\big| \le  c'\eps_\ga^2$
where $\eps_\ga \approx e^{-c\ga^{-1}}$, $c>0$, while $c_\ga \approx e^{c'\ga^{-b}}$, $b \in (\frac 12,1)$; thus for $\ga$ small $\la_\ga$ is ``essentially'' equal to $\eps_\ga$.  It is  $\eps_\ga$ which
has  a thermodynamic meaning in terms of ``surface tension''. Since there is no phase transition this cannot be taken literally and what we mean by surface tension here is the free energy cost of replacing a plus interval by a plus and a minus interval  separated in the middle by an interface interval.  The precise definition is given in \eqref{111.1} and it is proved in \eqref{111.5} that the surface tension is equal to $-\frac 1\beta \log \eps_\ga$. The physical reason behind the fact that  the  difference $p_\ga-p^+_\ga$ is related to the surface tension is
that
the ``true'' pressure $p_\ga$ is larger than the plus pressure $p^+_\ga$ ($=p^-_\ga$) as it gets an entropic contribution by alternating  plus and minus intervals of all possible lengths and the cost of inserting interface intervals is given by $\eps_\ga$.

 The limit $\ga\to 0$ is also interesting, in fact we have
 \begin{equation}
        \label{13.5.2}
\lim_{\ga\to 0} \ga \log \la_\ga = \lim_{\ga\to 0} \ga \log \eps_\ga = -\beta \bar f
    \end{equation}
$\bar f$, which has been defined in \eqref{intro.7}, is the surface tension of the mesoscopic system derived in the limit $\ga\to 0$ and described by the free energy functional $\mathcal F(m)$, see \eqref{intro.5}.  $\bar f$ is the free energy of the instanton profile which optimizes the free energy of profiles which are asymptotically $\pm m_\beta$.

\vskip2cm

\setcounter{equation}{0}

\section{Outline of proofs}
\label{sec:outline}

In this section we give an outline of how the proof of the statements of Section \ref{sec:13} is organized.

\begin{itemize}

\item
In Section \ref{sec:10} we prove that to leading order the dependence {from the boundary conditions}
of the partition function restricted to plus intervals factorizes into a product
$e^{F_\ga(s_0)}e^{F_\ga(s_{n+1})}$ where $s_0$ is the (block spin) left boundary condition and $s_{n+1}$ the right boundary condition.  By the spin flip symmetry the analogous statement holds for the minus intervals.  The precise statement is given in Theorem \ref{thm10.1} whose proof is quite technical: it is outlined in Appendix \ref{appzero} while the details are  given  in Appendix \ref{appD} and Appendix \ref{appE}.

\item
In Section \ref{sec:20} we use the analysis on the plus and minus intervals to show that  the partition function (with periodic boundary conditions) can be written as a sum of products of weights (denoted by $w(\und u)$) plus remainders which are negligible, as proved in Appendix \ref{BRPF}.

\item
The next step is to control the weights $w(\und u)$ which involve the partition functions restricted to interface intervals.  This is done in Section \ref{sec:11}, more technical details are proved in Appendix \ref{appF}.

\item  In Section \ref{sec:77.1} we go back to the expression for the partition function in terms of products of weights   $w(\und u)$ obtained in Section \ref{sec:20} which is estimated using Laplace transform as in a Tauberian theorem.  We thus write $w_\la(\und u):= e^{-\la |\und u|}w(\und u)$ and prove that there is a  {unique}
    value of $\la$, denoted by $\la_\ga$, such that $w_{\la_\ga}(\und u)$ has the properties stated in Section \ref{sec:13},
    see Theorem \ref{thm77.2}.  Technical details of its proof are given in Appendix \ref{appendix D}.  We then conclude the proof of Theorem \ref{thm13.3} at the end of Section \ref{sec:77.1}.

\item  The proof of the renewal property of the plus, minus and interface interval as stated in Theorem \ref{thm4.1.3-a} is given in Section
\ref{app:thm4.1.3-b}.

\end{itemize}

\vskip2cm

\setcounter{equation}{0}

\section{Statistical weight of plus and minus intervals}
\label{sec:10}

As already mentioned
$w_{\la_\ga}(\und u)$ will be defined in terms of the statistical weight
 of plus, minus and interface intervals. In this section we estimate
 the statistical weight of the plus and minus intervals.

The statistical weight $Z^+_{\ga,n}$ of a plus interval of length $u=n+2 \ge 3$ is
     \begin{equation}
      \label{10.1}
Z^+_{\ga,n}(s_0,s_{n+1}) := \sum_{s_1,..,s_n}
\mathbf 1_{\theta_1=\theta_n=1}\mathbf 1_{\Theta_i \ge 0, i=2,..,n-1} e^{-\beta H_\ga (s_1,..,s_n| s_0,s_{n+1})}
     \end{equation}
where $\theta(s_0)=\theta(s_{n+1})=1$. It is the statistical weight of a plus interval whose endpoints are $0$ and $n+1$, by  translation invariance
the definition applies to all plus intervals and by the spin flip symmetry also to the minus intervals.

\vskip.5cm

\begin{thm}
\label{thm10.1}
For all $\ga$ small enough there are $p^{+}_\ga$, $F^{(k)}_{\ga}(s)$, $k=1,2$, and $G^{(1)}_{\ga,n}(s',s'')$, so that
    \begin{equation}
      \label{10.2}
  Z^+_{\ga,n}(s_0,s_{n+1}) = e^{\beta (\ell^+_\ga n) p^+_\ga + F^{(1)}_{\ga} (s_0)+F^{(2)}_{\ga} (s_{n+1})+
G^{(1)}_{\ga,n}(s_0,s_{n+1})}
     \end{equation}
where  $n\ge 1$ and
    \begin{equation}
      \label{10.3}
F^{(k)}_{\ga}(s)\le c \ga^{-1}, k=1,2; \quad |G^{(1)}_{\ga,n}(s_0,s_{n+1})| \le a
e^{-b_0 \ga \ell_\ga^{+}n}
     \end{equation}
$c$, $a$, $b$ positive constants independent of $\ga$.

\end{thm}

\vskip.5cm

\noindent
{\bf Remarks.}

\begin{itemize}

\item  By the spin flip symmetry the statistical weight of a minus interval is equal to
     \begin{equation}
      \label{10.4}
Z^-_{\ga,n}(s_0,s_{n+1}) =  e^{\beta (\ell_\ga^+n) p^-_\ga + F^{(3)}_{\ga} (s_0)+F^{(4)}_{\ga} (s_{n+1})+
G^{(3)}_{\ga,n}(s_0,s_{n+1})}
    \end{equation}
where $\theta(s_0)=\theta(s_{n+1})=-1$ and
    \begin{eqnarray}
      \label{10.5}
&& p^-_\ga =  p^+_\ga, \quad F^{(3)}_{\ga} (s_0) =  F^{(1)}_{\ga} (-s_0)
, \quad
F^{(4)}_{\ga}(s_{n+1})= F^{(2)}_{\ga}(-s_{n+1}), \nn\\&&
G^{(3)}_{\ga,n}(s_0,s_{n+1}) = G^{(1)}_{\ga,n}(-s_0,-s_{n+1})
   \end{eqnarray}
   \item  The renewal property is a consequence of
   \eqref{10.2}.  In fact if we neglect $G^{(1)}_{\ga,n}(s_0,s_{n+1})$ the dependence of the partition function on $s_0$ and $s_{n+1}$ factorizes.  With this in mind we will write
            \begin{equation}
     \label{10.5bis}
e^{G^{(1)}_{\ga,n}(s_0,s_{n+1})} = \{e^{G^{(1)}_{\ga,n}(s_0,s_{n+1})} - e^{A_{\ga,n}}\}
+ e^{A_{\ga,n}},\quad A_{\ga,n} = \min_{s_0,s_{n+1}} G^{(1)}_{\ga,n}(s_0,s_{n+1})
    \end{equation}
When we take the term $e^{A_{\ga,n}}$ we decouple right and left, when we take the curly bracket we get a small contribution (as $n$ will typically be large).

   \item  While $n$ is the length of the interval in the block spin
   representation $\ell^+_\ga n$ is the length of the same interval in the original
   spin variables ($\ell^+_\ga n=\ga^{-1-\alpha}n$) so that $p^+_\ga$ is the pressure in the plus ensemble.  As we shall see the true pressure has an additional contribution of entropic origin, due to the alternating presence of plus and minus intervals.

   \item   The equality \eqref{10.2} reminds of the expression obtained for the partition function using the transfer matrix method.  The role of our boundary terms $F^{(i)}_\ga$ are played in the  transfer matrix approach by the right and left eigenvectors
       of the  transfer matrix relative to the maximal eigenvalue.  The latter is identified to the exponential of $\beta p$, $p$ the pressure, and it gives rise (as in our case) to a term
       $e^{-\beta p |\La|}$, $|\La|$ the volume of the region where the partition function is computed.  Besides these terms there is an exponentially small correction due to the spectral gap of the transfer matrix which in our case is played by $e^{G^{(i)}_\ga}$.

       The  use of the transfer matrix method with Kac potentials has been central in the  analysis of the van der Waals phase transition, see  the original papers by Kac et al., \cite{KUH1},\cite{KUH2},\cite{KUH3}.
       Its application is however very delicate because in the limit $\ga\to 0$ the maximal eigenvalue becomes degenerate.  In our case however this does not happen because of the restriction to plus intervals, nonetheless the use of   transfer matrix techniques  is not straightforward in our case and we will instead use the Lebowitz-Penrose coarse graining technique.

    \item  The equality \eqref{10.4} also reminds of the
    cluster expansion estimates where the influence  of the boundaries comes from the clusters which touch the boundaries.  The main contribution is due to the small clusters which give rise to surface corrections as our terms  $F^{(i)}_\ga$.  Clusters which connect the two boundaries are long and thus exponentially small, in our case they correspond to the term $e^{G^{(i)}_\ga}$.  We do not know whether cluster expansion works in our case, we have therefore followed a different route based on the Dobrushin uniqueness approach.

\end{itemize}

\vskip.5cm
In Appendix \ref{appzero} we outline the main steps of the proof of Theorem \ref{thm10.1}, details are then given  in Appendix \ref{appD} and Appendix \ref{appE}.

\vskip2cm

\section{The weights $w_{\la_\ga}$}
\label{sec:20}

The pressure $p^+_\ga$ defined in  Theorem \ref{thm10.1} is clearly a lower bound to the true pressure  $p_\ga$ because it is the pressure of restricted  partition functions (to plus intervals). The true pressure $p_\ga$
gets instead an additional
entropic contribution by alternating  plus and minus intervals of all possible lengths.  However each transition from a plus to a minus interval (and viceversa) involves an interface interval whose statistical weight is very small (as $\ga \to 0$) as we shall see in the next section. We will indeed prove that $p_\ga-p^+_\ga=\red{\frac{\la_\ga}{\beta \ell_\ga^+}}>0$, with  ``the correction'' $\la_\ga$  to the pressure  being exponentially small in $\ga^{-1}$.  $\la_\ga$ is the same parameter introduced in Section \ref{sec:13}, this is due to the fact that the
computation of the partition functions will involve the weights $w_{\la_\ga}(\und u)$.

In general the thermodynamic pressure is better approximated by partitions functions with periodic boundary conditions.  Let then $\La_n$ be the interval $[-n,n]$, \red{in block-spin variables, while in the original spin variables it is the interval $[-n\ell_\ga^+,(n+1)\ell_\ga^+)\;$}, $\und s=(s_{-n},...,s_n)$ a block spin configuration in $\La_n$ and
$H_\ga^{\rm pbc}(\und s)$ the hamiltonian with $\La_n$-periodic boundary conditions, our aim is to study the partition function
   \begin{equation}
      \label{20.0}
Z_{\La_n}^{\rm pbc}:= \sum_{\und s}e^{-\beta H_\ga^{\rm pbc}(\und s)}
     \end{equation}
We shall prove that $Z_{\La_n}^{\rm pbc} = e^{\beta p^+_\ga (2n+1)\ell^+_\ga
+\la_\ga (2n+1)}$ plus corrections which are  exponentially small in the sense of \eqref{13.5.1}.

As mentioned above the key point will be to reduce the computation of
$Z_{\La_n}^{\rm pbc}$ to a partition function with the weights $w_{\la_\ga}(\und u)$, to this end we need to re-introduce the partitions into plus-minus and interface intervals in the context of the ``torus'' $\La_n$.

Given $\und s$ we denote by $\und s'$ its $\La_n$-periodic extension to $\mathbb Z$, i.e.\
$s'_{i + k(2n+1)} = s_i$, $i\in \La_n$, $k \in \mathbb Z$.
By restricting $i$ to $\La_n$, $\theta(i,\und s')$ and  $\Theta(i,\und s')$ are the  phase indicators in $\La_n$.
 As a difference with the infinite volume, the sets $\mathcal X^0$ and $\mathcal X^{\pm}$ of configurations $\und s$, where for all $i\in \La_n$: $\Theta_i = 0$, respectively $\Theta_i \ge 0$ (and somewhere  $=1$)
and  $\Theta_i \le 0$ (and somewhere $ =-1$),  have non zero probability.  We denote by $g$ the complement \red{of $\mathcal X^0 \cup \mathcal X^+ \cup \mathcal X^-$} and write
   \begin{equation}
      \label{20.0.222}
Z_{\La_n}^{\rm pbc}=Z_{\La_n}^{\rm pbc, g} + Z_{\La_n}^{\rm pbc,\mathcal X^0}+Z_{\La_n}^{\rm pbc,\mathcal X^+}+Z_{\La_n}^{\rm pbc,\mathcal X^-}
     \end{equation}
where on the right hand side we have written the partition functions restricted to configurations in $g$ and respectively $\mathcal X^0$, $\mathcal X^+$ and $\mathcal X^-$.  We will prove in Appendix \ref{BRPF} that
$ Z_{\La_n}^{\rm pbc,\mathcal X^0}$, $Z_{\La_n}^{\rm pbc,\mathcal X^+}$ and $Z_{\La_n}^{\rm pbc,\mathcal X^-}$ are bounded by $c_\ga n e^{\beta p^+ _\ga \ell^+_\ga (2n+1)}$ and are therefore negligible with respect to the estimate
$e^{\beta p^+_\ga (2n+1)\ell^+_\ga
+\la_\ga (2n+1)}$ that we want to prove for $Z_{\La_n}^{\rm pbc}$.
We  will thus study in the sequel of this section $Z_{\La_n}^{\rm pbc, g}$.
  Since $\und s'$ is periodic
with period $2n+1$ if  $\und s\in g$  then $\und s'$ has infinitely many sites where $\Theta(i) =\pm 1$ and therefore it  defines a partition $\om'$ of $\mathbb Z$ into finite atoms $\om'_{\ell,m}$.  Like $\und s'$  also the partition $\om'$ is periodic, namely if $\om'_{\ell,m}$ is an atom of $\om'$ then also its translates by $2n+1$ are atoms of $\om'$.

Call $\om'_\ell= \om'_{\ell,1}\cup\cdots\cup\om'_{\ell,4}$ and let $\ell_1$ be such that
$\om'_{\ell_1} \ni -n$, let $x'$ be the leftmost point of $\om'_{\ell_1}$ and
$\om'_{\ell_1}\dots \om'_{\ell_k}$ the successive atoms of $\om'$ which cover the interval
$[x',x'')$, $x''= x' + 2n+1$ (by periodicity in fact at $x''$ starts the atom of $\om'$ which is $\om'_{\ell_1}$ shifted by $2n+1$).  We then denote by $\om=\{\om_{i,m} = f(\om'_{\ell_i,m}), m=1,..,4;i=1,..,k\}$ the partition of $\La_n$ image
of the atoms $\om'_{\ell_i,m}$ and call $x=f(x')$ its ``starting point'' (the map $f:\mathbb Z \to \La_n$ has been defined above).  We finally denote by $\und u = \{u_{\ell,m}, \ell=1,..,k; m=1,..,4\}$ the lengths of the intervals $\om_{\ell,m}$ and observe that the pairs $(x,\und u)$ are in one to one correspondence with the partitions $\om$.
Calling $u_\ell= u_{\ell,1}+\cdots+u_{\ell,4}$, $|\und u| = \sum_\ell u_\ell$, the set of possible values of  $(x,\und u)$ is:
   \begin{equation}
      \label{20.0.1}
A= \{(x,\und u): \; x\in [-n,n], \;|\und u|=2n+1; x+ u_1-1 \ge n+1,\;\rm{when}\; x > -n \}
     \end{equation}
the condition $x+ u_1-1 \ge n+1,\;\rm{when}\; x > -n$ corresponds to the condition that $\om'_{\ell_1}\ni -n$.
We remark that the sequences $\und u \in A$ are  not necessarily in $\cR$ \red{namely } $u_{1,1}$ could be smaller than 3, see \eqref{13.0}.

Recalling that
   \begin{equation}
      \label{20.0.2}
Z_{\La_n}^{\rm pbc, g}:= \sum_{\und s\in g}e^{-\beta H_\ga^{\rm pbc}(\und s)}
     \end{equation}
we then have:
   \begin{equation}
      \label{20.0.3}
Z_{\La_n}^{\rm pbc, g}:= \sum_{(x,\und u) \in A}\sum_{\und s\to (x,\und u)}e^{-\beta H_\ga^{\rm pbc}(\und s)}
     \end{equation}
where 
for $(x,\und u) \in A$  we denote by $\und s \to (x,\und u)$ the set of all $\und s\in g$ such that the induced partition
$\om$ gives $(x,\und u)$.
We sum over the interiors of the plus and minus  intervals
using \eqref{10.2} and then over the interface intervals, and get
    \begin{eqnarray}
      \label{20.0.4}
&&Z_{\La_n}^{\rm pbc, g} =  e^{\beta p^+_\ga (2 n+1) \ell^+_\ga} \sum_{(x,\und u) \in A}
\sum_{\{s_{\ell,m}\}} \prod_{\ell=1}^{k(\und u)}\prod_{m=1}^{4} e^{V^m_{u_{\ell,m}}(s_{\ell,m},s_{\ell,m+1})}
     \end{eqnarray}
where $k(\und u)=k$ if $\und u = (u_{\ell,m}, \ell=1,..,k)$. In \eqref{20.0.4}
$\{s_{\ell,m}\}$  are block-spins and the sum  is over all $ s_{\ell,m}, \ell=1,..,k(\und u), m=1,..,4$, such that $\theta(s_{\ell,m})=1$, $m=1,2$; $\theta(s_{\ell,m})=-1$, $m=3,4$;
If $u_{\ell,1}=1$ then $s_{\ell,1}=s_{\ell,2}$ and $s_{\ell,3}=s_{\ell,4}$ if $u_{\ell,3}=1$.
Moreover $s_{\ell,5}:=s_{\ell+1,1}$ and $s_{k(\und u)+1,1}=s_{1,1}$.
\red{$s_{\ell,m}$ , $m=1,2$,  are the block-spins of the leftmost and rightmost blocks  in $\om_{\ell,1}$ whose length is $u_{\ell,1}$.
 Analogously $s_{\ell,m}$, $m=3,4$, are the block-spins of the leftmost and rightmost blocks in $\om_{\ell,3}$ whose length is $u_{\ell,3}$.  }}
It remains to define the two body potentials
$V^m_{u}(s ,s')$.

The potentials $V^m_{u}(s ,s')$, $m=2,4$, $u\ge 2$, take into account the contribution of the interfaces.  Let $\theta(s_0)=1$, $\theta(s_{u+1})=-1$ and
          \begin{eqnarray}
      \label{11.2}
      Z^{+,-}_{\ga,u}(s_0,s_{u+1}) &=&  \sum_{s_{[1,u]}}
\mathbf 1_{\theta_1= 1,\theta_u= -1}
\mathbf 1_{\Theta_i =0, i=1,..,u} e^{-\beta H_\ga (s_{[1,u]}| s_0, s_{u+1})}\nn\\ Z^{-,+}_{\ga,u}(-s_0,-s_{u+1}) &=&  \sum_{s_{[1,u]}}
\mathbf 1_{\theta_1= -1,\theta_u= 1}
\mathbf 1_{\Theta_i =0, i=1,..,u} e^{-\beta H_\ga (s_{[1,u]}| -s_0, -s_{u+1})}
     \end{eqnarray}
Then \red{with $F^{(i)}_\ga(s)$ as in Theorem \ref{thm10.1}}
     \begin{equation}
      \label{20.3.0}
e^{V^2_{u}(s ,s')}  =  \frac{Z^{+,-}_{\ga,u}(s,s')e^{-\beta [H_\ga(s)+H_\ga(s')]+F^{(2)}_\ga(s)+F^{(3)}_\ga(s')}}{e^{\beta p^+_\ga (u+2)\ell^+_\ga}}
     \end{equation}
Similarly
     \begin{equation}
      \label{20.3.1}
e^{V^4_{u}(s ,s')}  =  \frac{Z^{-,+}_{\ga,u}(s,s')e^{-\beta [H_\ga(s)+H_\ga(s')]+F^{(4)}_\ga(s)+F^{(1)}_\ga(s')}}{e^{\beta p^+_\ga (u+2)\ell^+_\ga}}
     \end{equation}
     \red{ When the length $u$ of a plus interval is $\ge 3$ the sum over the spins in its interior is bounded by the r.h.s. of \eqref{10.2}. The factor $e^{\beta p_\ga^+ \ell_\ga^+ (u-2)}$ contributes to the first factor on the r.h.s. of  \eqref{20.0.4}; the factor $e^{\beta (F^{(1)}(s)+F^{(2)}(s))}$ }
     \red{is put  in $V^m_u(\cdot,\cdot)$, $m=2,4$. Then $V^1_u(s,s')= G^{(1)}_{\ga,u}(s,s')$.}

     \red{When $u=1,2$ \eqref{10.2} does not apply and we need to subtract in $V^1_{u}(s ,s') 1_{u\le 2}$ the terms $F^{(1)}_\ga(s),\; F^{(2)}_\ga(s)$ that we have put in  $V^m_u(\cdot,\cdot)$, $m=2,4$. In this case in fact there is no interior in the plus interval: when $u=1$ the contribution of the plus interval is $1$, when $u=2$ it is $e^{-\beta W_\ga(s|s')}$ where
      $W_\ga(s|s')$ is the interaction between two contiguous block spins. The   definition of
      $V^1_u(s,s')$ is therefore:}
         \begin{eqnarray}
      \label{20.3.2}
   V^1_{u}(s ,s') &=&  \Big(\beta H_\ga(s)+\beta p^+_\ga\ell^+_\ga-F^{(1)}_\ga(s)-F^{(2)}_\ga(s)\Big) \red{\mathbf 1_{u=1, s=s'}}\nn\\ &+&   \Big(-\beta W_\ga(s|s') -F^{(1)}_\ga(s)-F^{(2)}_\ga(s')\Big)\red{\mathbf 1_{u=2}} + G^{(1)}_{\ga,u}(s,s')\red{\mathbf 1_{u\ge 3}}
     \end{eqnarray}
     \red{where the extra terms in \eqref{20.3.2}   compensate with  the corresponding factors in the definition of $V^2_u(s,s')$ and $V^4_u(s,s')$}.

     Similarly
             \begin{eqnarray}
      \label{20.3.3}
   V^3_{u}(s ,s') &=&  \Big(\beta H_\ga(s)+\beta p^+_\ga\ell^+_\ga-F^{(3)}_\ga(s)-F^{(4)}_\ga(s)\Big)\red{\mathbf 1_{u=1, s=s'}} \nn\\ &+&   \Big(-\beta W_\ga(s|s') -F^{(3)}_\ga(s)-F^{(4)}_\ga(s')\Big)\red{\mathbf 1_{u=2}} + G^{(3)}_{\ga,u}(s,s')\red{\mathbf 1_{u\ge 3}}
     \end{eqnarray}
Recalling the strategy outlined in the second remark after Theorem \ref{thm10.1} whenever $u \ge 3$ we write each $e^{V^{1}_{u}(s,s')}$ as
           \begin{equation}
    \label{6.11b}
e^{V^1_{u}(s,s')} = \{e^{G^{(1)}_{\ga,u-2}(s ,s') }
- e^{A^{(1)}_{\ga,u-2}}\}
+ e^{A^{(1)}_{\ga, u -2}}
    \end{equation}
By the definition of $A^{(1)}_{\ga,u}$ (see \eqref{10.5bis}) the curly bracket is non negative.  We have
     \begin{equation}
      \label{20.0.7}
Z_{\La_n}^{\rm pbc, g} = Z_{\La_n}^{\rm pbc, gb}+Z_{\La_n}^{\rm pbc, gg}
     \end{equation}
where $Z_{\La_n}^{\rm pbc, gb}$ comes from the case where  for all $\ell$  either  $u_{\ell,1}<3$ or we take the curly bracket.
Thus
    \begin{equation}
      \label{20.0.8}
Z_{\La_n}^{\rm pbc, gb} = e^{\beta p^+_\ga (2n+1) \ell^+_\ga} \sum_{(x,\und u) \in A} w^{(b)}(\und u)
     \end{equation}
where
  \begin{eqnarray}
      \label{20.0.9}
&& w^{(b)}(\und u) =
\sum_{\{s_{\ell,m}\}} \{\prod_{\ell=1}^{k(\und u)}\prod_{m=2}^{4} e^{V^m_{u_{\ell,m}}(s_{\ell,m},s_{\ell,m+1})}\} \prod
_{\ell=1}^{k(\und u)} K_{u_{\ell,1}}(s_{\ell,1},s_{\ell,2})\nn\\&&
K_{u}(s ,s') = e^{V^1_u} \mathbf 1_{u\le 2} +  \{e^{G^{(1)}_{\ga,u-2}(s ,s') }
- e^{A^{(1)}_{\ga,u-2}}\} \mathbf 1_{u\ge 3}
     \end{eqnarray}
      The sum over
$\{s_{\ell,m}\}$ in \eqref{20.0.9} is as in   \eqref{20.0.4}, $s_{k(\und u),5} = s_{1,1}$.  We will prove in Appendix \ref{BRPF}, see \eqref{G9}, \red{that there exists $\zeta_\ga>0$ so that: }
  $$
  \lim_{n\to\infty}    \frac {Z_{\La_n}^{\rm pbc, gb}} {e^{\beta p_\ga (2n+1)\ell^+_\ga -\red{\zeta_\ga(2n+1)}}}{=0}
  $$
so that also the contribution of $Z_{\La_n}^{\rm pbc, gb}$ is negligible \red{in the proof of \eqref{13.5.1}}, we can therefore restrict to  $Z_{\La_n}^{\rm pbc, gg}$.

To write  explicitly $Z_{\La_n}^{\rm pbc, gg}$ we replace $A$ by $A^*$, where
  \begin{equation}
      \label{20.0.10}
A^*= \{(x,\und u_1,..,\und u_k),\; i=1,\dots,k:\; \und u_i \in \mathcal R;\sum_i |\und u_i|=2n+1; x+ |\und u_1|-1 \ge n+1,\;\rm{when}\; x > -n \}
     \end{equation}
and  $\mathcal R$ is defined in  \eqref{13.0}.
Namely we start from a sequence $\und u$ as in \eqref{20.0.1}, and call $\mathcal L$ the set of labels $\ell$ s.t. $u_{\ell,1}\ge 3$ and  the last term in \eqref{6.11b} is taken. We then group together the $u_{\ell',m}$ with $\ell'$ in between successive values of $\ell$ in $\cL$ and each of these groups is a $\und u_i$ in \eqref{20.0.10}.
%
With these notation
    \begin{equation}
      \label{20.0.11}
Z_{\La_n}^{\rm pbc, gg} = e^{\beta p^+_\ga (2n+1) \ell^+_\ga} \sum_k\sum_{(x,\und u_1,,,\und u_k) \in A^*} \prod_{i=1}^k w(\und u_i)
     \end{equation}
where, similarly to \eqref{20.0.9},
  \begin{eqnarray}
      \label{20.70}
&& w(\und u) =
\sum_{\{s_{\ell,m}\}} \{\prod_{\ell=1}^{k(\und u)}\prod_{m=2}^{4} e^{V^m_{u_{\ell,m}}(s_{\ell,m},s_{\ell,m+1})}\} e^{A^{(1)}_{\ga,u_{1,1}}}\prod
_{\ell=2}^{k(\und u)} K_{u_{\ell,1}}(s_{\ell,1},s_{\ell,2})
     \end{eqnarray}

For any $\la>0$ we define:
  \begin{eqnarray}
      \label{20.711}
&& w_{\la}(\und u)= e^{-\la |\und u|}w(\und u), \quad
|\und u| =\sum_{\ell,m} u_{\ell,m}
     \end{eqnarray}
 and get
        \begin{equation}
      \label{20.0.12.00}
Z_{\La_n}^{\rm pbc, gg} = e^{ (2n+1)[\beta p^+_\ga \ell^+_\ga+\la]}
 \sum_k\sum_{(x,\und u_1,,,\und u_k) \in A^*} \prod_{i=1}^k w_\la(\und u_i)
     \end{equation}
While \eqref{20.0.12.00} holds for all $\la>0$ it greatly simplifies if we choose $\la=\la_\ga$, $\la_\ga>0$ such that $ w_{\la_\ga}(\und u)$ satisfies the properties listed in Section \ref{sec:13}.  We will thus continue the estimate of $Z_{\La_n}^{\rm pbc, gg}$ at the end of Section \ref{sec:77.1}, after proving the existence
of  $\la_\ga$ and properties of $ w_{\la_\ga}(\und u)$.

\vskip2cm

\setcounter{equation}{0}

\section{{Statistical} weight of interfaces}
\label{sec:11}

In Theorem \ref{thm10.1} we have estimated the statistical weight
of a plus interval defined as the partition function $Z^+_{\ga,n}(s,s')$.
Analogously the  statistical weight of a $+-$ interface is defined as
the partition function $Z^{+,-}_{\ga,n}(s,s')$ of \eqref{11.2} ($-+$ interfaces are just $+-$ interfaces after spin flip, so that in the sequel we just refer to the former).
In Appendix \ref{appF} we will prove the following bound:

\medskip

\begin{thm}
\label{thm111.2}
There are $c_0>0$, a positive integer $n_0$, $b \in(\frac 12,1)$ and $c_b>0$ so that
 for any $s_0$ and $s_{n+1}$ such that $\theta(s_0)=1=-\theta(s_{n+1})$
         \begin{eqnarray}
      \label{111.6}
  &&    Z^{+,-}_{\ga,n}(s_0,s_{n+1}) \le
     {Z^{++}_{\ga,n}(s_0,-s_{n+1})}\;  e^{-\ga^{-1}\beta \bar f + c_b \ga^{-b} - \ga^{-1}c_0 (n-n_0)\mathbf 1_{n\ge n_0}}
   \end{eqnarray}
$\bar f$ is the free energy of an instanton, see \eqref{intro.7},
and {$Z^{++}_{\ga,n}$ is defined in \eqref{D.2}}.

  Moreover
     \begin{equation}
      \label{111.7}
Z^{+,-}_{\ga,2}(s_0,s_{3}) \ge e^{-\ga^{-1}\beta \bar f - c_b \ga^{-b}}
 Z^{++}_{\ga,2}(s_0,-s_{3})
     \end{equation}

\end{thm}

\medskip
\red{Observe that $Z^{++}_{\ga,n}$ has still a ``dangerous" dependence on the boundary conditions because the interaction with the boundaries is exponential in $\ga^{-1}$, this will be settled in the next subsection.}
\red{In a final } subsection we will relate the statistical weight  of the interface to the surface tension introduced at the end of Section \ref{sec:13}.

\vskip.5cm

\subsection{Bounds on the potentials $V^m_{u}(s ,s')$, $m=2,4$}
\label{subsec:11.1}
We will first bound
   \begin{eqnarray}
      \label{111.4}
  \eps_\ga(n):=  \sum_{s:\theta(s)=1}
    \sum_{s':\theta(s')=-1} e^{V^2_{n}(s ,s')}
     \end{eqnarray}
and
  \begin{equation}
      \label{111.5.0}
\eps_\ga:= \sum_n \eps_\ga(n)
   \end{equation}
 {Recalling \eqref{20.3.0}
     \begin{eqnarray}
      \label{111.5.1}
 \eps_\ga(n) = \sum_{s:\theta(s)=1}
    \sum_{s':\theta(s')=-1}   e^{-\beta p^+_\ga(n+2)\ell^+_\ga}
     Z^{+,-}_{\ga,n}(s,s')
    e^{-\beta[H_\ga(s)+ H_\ga(s')] +F^{(2)}(s)+F^{(3)} (s')}
   \end{eqnarray}
$\eps_\ga$ will be referred to as the statistical weight of the interface.
   \medskip

\begin{thm}
\label{thm111.3.0}

There are  $b \in(\frac 12,1)$ and $c_b>0$ so that for all $\ga$ small enough
    \begin{equation}
      \label{111.5.2}
\eps_\ga(n) \le  e^{-\ga^{-1}\beta \bar f + c_b \ga^{-b} - \ga^{-1}c_0 (n-n_0)\mathbf 1_{n\ge n_0}}
     \end{equation}
     \begin{equation}
      \label{111.5.3}
e^{-\ga^{-1}\beta \bar f - c_b \ga^{-b}}\le \eps_\ga \le  e^{-\ga^{-1}\beta \bar f + c_b \ga^{-b}}
     \end{equation}

\end{thm}

\medskip

\noindent
{\bf Proof.}
 Using \eqref{111.6}   we get from \eqref{111.5.1}
         \begin{eqnarray}
      \label{??}
  &&   \eps_{\ga}(n) \le  \om_{\ga,n} \eps^*_\ga(n), \quad
  \om_{\ga,n} =  e^{-\ga^{-1}\beta \bar f + c_b \ga^{-b} -\ga^{-1}c_0 (n-n_0)\mathbf 1_{n\ge n_0}}
  \\&&\eps^*_\ga(n):=e^{-\beta p^+_\ga(n+2)\ell^+_\ga}
  \{ \sum_{ \theta(s)=\theta(s')=1}  {Z^{++}_{\ga,n}}(s,s')
   e^{-\beta[H_\ga(s)+ H_\ga(s')] +F^{(2)}(s)+F^{(1)} (s')}\}\nn
   \end{eqnarray}
 We are going to prove that
         \begin{eqnarray}
      \label{???}
  &&  \eps^*_\ga(n) \le 1
   \end{eqnarray}
 which then proves \eqref{111.5.2}.

     Let $m>1$, eventually $m\to\infty$, then by Theorem \ref{thm10.1}
        \begin{eqnarray*}
      \label{11.6}
   1 &=& 
 Z^+_{\ga,m}(s_{-m-1},s)e^{-\{\beta p^+_\ga\ell^+_\ga m  + F^{(1)}_{\ga} (s_{-m-1})+F^{(2)}_{\ga} (s)+
G^{(1)}_{\ga,m}(s_{-m-1},s\}} \\&
= &Z^+_{\ga,m}(s',s_{n+m+2})e^{-\{\beta p^+_\ga\ell^+_\ga m + F^{(1)}_{\ga} (s')+F^{(2)}_{\ga} (s_{n+m+2})+
G^{(1)}_{\ga,m}(s',s_{n+m+2})\}} \nn
   \end{eqnarray*}
hence 
           \begin{eqnarray*}
    \eps^*_{\ga}(n) &\le&  e^{-\beta p^+_\ga(n+2m+2)\ell^+_\ga
   -F^{(2)}_{\ga} (s_{n+m+2})-F^{(1)}_{\ga} (s_{-m-1})+
2\|G^{(1)}_{\ga,m}\|_\infty} \\ &\times& \sum_{s_{[-m,n+m+1]}}e^{-\beta H_\ga(s_{[-m,n\red{+m}+1]}|s_{-m-1},s_{m+n+2})}
\\ &\times& \mathbf 1_{\Theta_i\ge 0, i \in [-m,n+m+1]}
\mathbf 1_{\theta_{-m}=\theta_{{n+m+1}}=1}
 {\mathbf 1_{\Theta_{i}=1, i\in [0,n+1]}}
   \end{eqnarray*}
The last factor is bounded by $Z^+_{\ga,n+2m+2}(s_{-m-1},s_{m+n+2})$,
(having  dropped the   characteristic function ${\mathbf 1_{\Theta_{i}=1, i\in [0,n+1]}}$). Then using again Theorem \ref{thm10.1}:
          \begin{equation*}
    \eps^*_{\ga}(n)  \le
  e^{2\|G^{(1)}_{\ga,m}\|_{\infty}
+\|G^{(1)}_{\ga,2m+n+2}\|_{\infty}}
   \end{equation*}
Letting $m\to\infty$ we prove \eqref{???}
so that the upper bound in \eqref{111.5.2} and \eqref{111.5.3} is proved.

By \eqref{111.7}
         \begin{eqnarray*}
  &&   \eps_{\ga}(2) \ge e^{-\beta p^+_\ga 4\ell^+_\ga} \om'_{\ga,2}
  \red{\times }\{ \sum_{ \theta(s)=\theta(s')=1} Z^{++}_{\ga,2}(s,s')
   e^{-\beta[H_\ga(s)+ H_\ga(s')] +F^{(2)}(s)+F^{(1)} (s')}\}
   \end{eqnarray*}
with $\om'_{\ga,2}$ obtained from $\om_{\ga,2}$ by changing the sign of the term $c_b\ga^{-b}$.
As before and with $n\equiv 2$ below,
          \begin{eqnarray*}
    \eps_{\ga}(2) &\ge& \om'_{\ga,2}e^{-\beta p^+_\ga(n+2m+2)\ell^+_\ga
   -F^{(2)}_{\ga} (s_{n+m+2})-F^{(1)}_{\ga} (s_{-m-1})-
2\|G^{(1)}_{\ga,m}\|_\infty} \\ &\times& \sum_{s_{[-m,n+m+1]}}e^{-\beta H_\ga(s_{[-m,{m+n+1}]}|s_{-m-1},s_{m+n+2})}\mathbf 1_{\Theta_i\ge 0, i \in [-m,n+m+1]}
\mathbf 1_{\theta_{-m}=\theta_{{n+m+1}}=1}
 \mathbf 1_{\Theta_{i}=1, i\in [0,3]}
   \end{eqnarray*}
Call $\mu^+(s_{[-m,{m+n+1}]}|s_{-m-1},s_{m+n+2})$ the Gibbs measure in the interval $[-m,{m+n+1}]$ conditioned to
$\{\Theta_i\ge 0, i \in [-m,n+m+1];
 \theta_{-m}=\theta_{n+m+1} =1\}$ and with boundary conditions
 $s_{-m-1},s_{m+n+2}$.  Then the last sum is equal to
          \begin{eqnarray*}
  Z^+(s_{-m-1},s_{m+n+2}) \mu^+\Big[\{\Theta_{i}=1, i\in [0,3]\}
  \Big]
   \end{eqnarray*}
 and
          \begin{eqnarray*}
 \mu^+\Big[\{\Theta_{i}=1, i\in [0,3]\}
  \Big] \ge 1 - 4 e^{-3 b'\ga^{-1}},\quad b'>0
   \end{eqnarray*}
 which follows from \eqref{D.8}.  The lower bound is proved \red{ because $\eps_\ga\ge \eps_\ga(2)$} .  \qed

\medskip



\vskip .5cm

\medskip

\red{We shall see, in  Appendix F (see \eqref{DD.15}--\eqref{DD.16} and Subsection \ref{subsect-F3}), that  the factors $e^{V^3_{u}(s ,s')}$ and $K_{u}(s ,s')$ in \eqref{20.0.9} can  be bounded uniformly in $s$ and $s'$ if $u\ge 3$ (but not when $u<3$).  Thus if the intervals to the right and left of an interface have both $u\ge 3$, then, after using the above uniform bounds, we are in the setup of  Theorem \ref{thm111.3.0} and get an estimate for the contribution of the interface. In the other cases we proceed as follows.}

\red{We call two interfaces ``connected'' if they are separated by an interval of length $u<3$.  We then consider a maximal sequence of connected interfaces which starts on the left from the interface interval $(\ell',m')$ and ends on the right with $ (\ell'',m'')$, calling $(\ell,m)$ the intervals (interface or not) in between the two extremal ones. Thus the lengths $u_{\ell,m}$, $m\in\{1,3\}$,  of the intervals between  $(\ell',m')$ and  $ (\ell'',m'')$ have all length $<3$ while  the two intervals one to the left of $(\ell',m')$ and the other  to the right of $(\ell'',m'')$ have both length $\ge 3$. By using the above uniform bounds for  these latter we are then left with the case considered in the following theorem:}

\medskip

\begin{thm}
\label{thm111.12}

\red{There is $c>0$ so that for all $\ga$ small enough the following holds. Fix any maximal connected  sequence of interfaces  with $(\ell',m')$, $(\ell'',m'')$ and $\{u_{\ell,m}\}$   as above. Call $k$ the number of the interfaces in the sequence,
then}
         \begin{eqnarray}
      \label{111.12}
  && \sum_{\{u_{\ell,m}\}} \sum_{\{s_{\ell,m}\}} \prod_{\ell,m}  e^{V^m_{u_{\ell,m}}(s_{\ell,m} ,s_{\ell,m+1})} \le c^k\eps_\ga^k e^{2k  c_b \ga^{-b}}
   \end{eqnarray}
where $\theta(s_{\ell,m})=1$ when $m=1,2$ and $\theta(s_{\ell,m})=-1$
when $m=3,4$.

\end{thm}}

\medskip

\noindent
{\bf Proof.} Fix $\{u_{\ell,m}\}$ supposing for the sake of definiteness that $m'=2$ and $m''=4$. Call
     $$
     n := \sum_{\ell,m} u_{\ell,m}
     $$
and shorthand $s_0 = s_{\ell',m'}$, $s_{n+1} = s_{\ell'',m''}$.  We write
$s_{[1,n]} \in \mathcal S_{[1,n]}(\{u_{\ell,m}\})$ for the configurations which give $\{u_{\ell,m}\}$, then
        \begin{eqnarray}
             \label{111.13}
\sum_{\{s_{\ell,m}\}} \prod_{\ell,m}  e^{V^m_{u_{\ell,m}}(s_{\ell,m} ,s_{\ell,\red{m+1}})}
  &=& e^{-\beta p^+_\ga(n+2)\ell^+_\ga}\sum_{s_{[0,n+1]}} e^{-\beta H_\ga(s_{[0,n+1]})}e^ {F^{(2)}_\ga(s_0)+F^{(1)}_\ga(s_{n+1})}\nn\\ &\times& \mathbf 1_{ s_{[1,n]} \in \mathcal S_{[1,n]}(\{u_{\ell,m}\})}
  \mathbf 1_{ \theta(s_0)=\theta(s_{n+1})=1}
          \end{eqnarray}
\red{Noticing that for $n\le2$  $Z^+_{\ga,n}=Z^{++}_{\ga,n}$, } by the spin flip symmetry, \red{ by \eqref{111.6} and  \eqref{??}} 
 this is bounded by
        \begin{eqnarray*}
  && [e^{-\ga^{-1}\beta \bar f + c_b \ga^{-b}}]^k
  e^{ - \ga^{-1}c_0 \sum_{\ell}\sum_{ m\in\{2,4\}}(u_{\ell,m}-n_0)\mathbf 1_{u_{\ell,m}\ge n_0}} \eps^*_{\ga}(n)
   \end{eqnarray*}
where $ \eps^*_{\ga}(n)$ is defined in  \eqref{??} and it is  bounded by 1 \red{ (see \eqref{???}  ). By \eqref{111.5.3}:}
   $$
   e^{-\ga^{-1}\beta \bar f + c_b \ga^{-b}} \le \eps_\ga
   e^{2c_b \ga^{-b}}
   $$
 hence \eqref{111.12} recalling that $u_{\ell,m} \le 2$ when $m\in \{1,3\}$.
  \qed

\vskip2cm

\subsection{Free energy of an interface}
\label{subsec:11.2}

We define the free energy $\phi_\ga$ of an interface as the free energy cost of splitting a plus interval into a plus and a minus interval with an interface in between them.  More precisely
         \begin{eqnarray}
      \label{111.1}
  \phi_\ga:=- \frac 1 \beta\log \sum_{n\ge 2}\liminf_{m\to \infty}\frac {A^{+-}_{m,n}(s_{-m-1},s_{m+n+2})}{Z^{+}_{\ga,2m+n}(s_{-m-1},-s_{m+n+2})}
     \end{eqnarray}
where $s_{-m-1}$ and $s_{m+n+2}$ are arbitrary with the only condition that $\theta(s_{-m-1})=-\theta(s_{m+n+2})=1$. In \eqref{111.1}
         \begin{eqnarray}
      \label{111.2}
A^{+-}_{m,n}(s_{-m-1},s_{m+n+2})&=&  \sum_{s_{[-m,n+m+1]}}e^{-\beta H_\ga(s_{[-m,n+1]}|s_{-m-1},s_{m+n+2})}\mathbf 1_{\Theta_i\ge 0, i \le 0}
\mathbf 1_{\Theta_i\le 0, i\ge n+1} \nn\\&\times& \mathbf 1_{\Theta_i= 0, i\in[1,n]}
\mathbf 1_{\Theta_0=1, \Theta_{n+1}=-1}
     \end{eqnarray}
is the partition function with the constraint that $[-m-1,0]$ is a plus interval,
$[1,n]$ is a $+-$ interface and $[n+1,n+m+2]$ is a minus interval.
The liminf in \eqref{111.1} is actually a limit:

\begin{thm}
\label{thm111.1}
For any sequence $s_{-m-1}$ and $s_{m+n+2}$ as above
    \begin{equation}
      \label{111.3}
\lim_{m\to \infty}\frac {A^{+-}_{m,n}(s_{-m-1},s_{m+n+2})}{Z^{+}_{\ga,2m+n}(s_{-m-1},-s_{m+n+2})}
=       \eps_\ga(n)
\end{equation}
{where $\eps_\ga(n)$ is defined in \eqref{111.4}}

\end{thm}

\medskip

\noindent
{\bf Proof.}  Let $\theta(s)=1=-\theta(s')$ then by Theorem \ref{thm10.1}
%
%
%
%

   \begin{eqnarray*}
A^{+-}_{m,n}(s_{-m-1},s_{m+n+2}) &=&  \{ \sum_{s:\theta(s)=1}
\sum_{s':\theta(s')=-1} e^{V^2_{n}(s ,s')} 
    e^{
    G^{(1)}_{m}(s_{-m-1},s)+G^{(3)}_{m}(s',s_{m+n+2})}\} \\&\times&
 e^{\beta p^+_\ga (2m+n+2)\ell^+_\ga}  e^{F^{(1)}(s_{-m-1})+F^{(4)} (s_{m+n+2}))}
\end{eqnarray*}
The denominator $Z^{+}_{\ga,2m+n}(s_{-m-1},-s_{m+n+2})$ in \eqref{111.1} is equal to
   \begin{eqnarray*}
Z^{+}_{\ga,2m+n}(s_{-m-1},-s_{m+n+2}) &=&   e^{\beta p^+_\ga(2m+n+2)\ell^+_\ga
+F^{(1)}(s_{-m-1})+F^{(2)} (-s_{m+n+2}) +
    G^{(1)}_{2m+2}(s_{-m-1},-s_{m+n+2}) }
\end{eqnarray*}
Since $F^{(2)} (-s_{m+n+2})=F^{(4)} (s_{m+n+2})$
   $$
\frac {A^{+-}_{m,n}(s_{-m-1},s_{m+n+2})}{Z^{+}_{\ga,2m+n}(s_{-m-1},-s_{m+n+2})} \le
\{e^{\|G^{(1)}_{m}\|_\infty+\|G^{(3)}_{m}\|_\infty+ \| G^{(1)}_{2m+2}\|_\infty}\}\eps_\ga(n)
  $$
  Analogously
    $$
\frac {A^{+-}_{m,n}(s_{-m-1},s_{m+n+2})}{Z^{+}_{\ga,2m+n}(s_{-m-1},-s_{m+n+2})} \ge
\{e^{-\|G^{(1)}_{m}\|_\infty-\|G^{(3)}_{m}\|_\infty- \| G^{(1)}_{2m+2}\|_\infty}\}\eps_\ga(n)
  $$

By \eqref{10.3} the curly bracket converges to $1$ when $m\to\infty$ and
\eqref{111.3} then follows.  \qed

%
%
%
%
%
%
%

\medskip

By Theorem \ref{thm111.1} {and by Theorem \ref{thm111.3.0}} we then have: 

\medskip

\begin{thm}
\label{thm111.3}

The free energy $\phi_\ga$ defined in \eqref{111.1} is well defined, the liminf on the right hand side is a limit and
  \begin{equation}
      \label{111.5}
 e^{-\beta \phi_\ga} = \eps_\ga
   \end{equation}
Then, recalling \eqref{111.5.3},
     \begin{equation}
      \label{111.8}
 |\phi_\ga-\ga^{-1} \bar f| \le c_b \ga^{-b}
     \end{equation}
Thus
         \begin{eqnarray}
      \label{111.9}
\lim_{\ga\to 0}  \ga \phi_\ga =\bar f
     \end{eqnarray}
\end{thm}

\vskip2cm

\setcounter{equation}{0}

\section{Normalization of the weights $w(\und u)$}
\label{sec:77.1}

In this section we prove the existence of $\la_\ga$  and  other properties of $w_{\la_\ga}(\und u)$  \red{defined in \eqref{20.711}}, including those stated in Section \ref{sec:13}.  The  following theorem is proved in Appendix \ref{appendix D}:

\medskip

\begin{thm}

\label{thm77.1}
\red{Taking $c_b$ as in in Theorem \ref{thm111.3.0},  there is a constant $c>0$} 
such that for all $\ga$ small enough and $\frac 12{\eps_\ga}  \le \la \le  \frac 32 {\eps_\ga}$
   \begin{eqnarray}
      \label{77.1.0}
&& \Big|\sum_{\und u} w_{\la}(\und u) -(\frac{\eps_\ga}{\la})^2\Big|
\le c_\ga \, \eps_\ga ,   \quad c_\ga= c e^{8c_b\ga^{-b}} 
     \end{eqnarray}

\end{thm}

\medskip
%

%
%
%
%

\medskip

\begin{corollary}
\label{coro12}
{In the context of Theorem \ref{thm77.1} for $\ga$ small enough} there is $c>0$ so that for any $\la \in (\frac 34 \eps_\ga,\frac 54 \eps_\ga)$ and any $R>0$
  \begin{equation}
      \label{77.2.07}
  \sum_{\und u: |\und u| \ge R}  w_{\la}(\und u)  \le c e^{-\eps_\ga R/4},\quad \sum_{\und u \in \mathcal R} w_{\la}(\und u) |\und u| \le c \frac{1}{1-e^{-\eps_\ga /4}}
     \end{equation}
As a consequence the function $\la \to  \sum_{\und u \in \mathcal R} w_{\la}(\und u)$ is in $C^1$ when $\la \in (\frac 34 \eps_\ga,\frac 54 \eps_\ga)$ .

\end{corollary}

\vskip.5cm

\noindent
{\bf Proof.}
\red{We obviously have for any $\la \in (\frac 34 \eps_\ga,\frac 54 \eps_\ga)$ }:
\begin{equation}
      \label{77.2.0}
  w_{\la}(\und u)  \le  e^{- |\und u|\frac{\eps_\ga}4}w_{\frac 12 \eps_\ga}(\und u)
     \end{equation}
 \red{and the } first inequality \red{in \eqref{77.2.07} } follows from \eqref{77.2.0} with
$c=  \sum_{\und u}  w_{ \frac 12 \eps_\ga}(\und u)$. The second inequality follows from
 \eqref{13.111} and the first inequality.  \qed

%
%
\medskip

\begin{thm}

\label{thm77.2}
For any $\ga$ small enough there is a unique $\la_\ga>0$ such that
  \begin{equation}
      \label{77.3}
\sum_{\und u} w_{\la_\ga}(\und u)=1
     \end{equation}
Moreover there is $c$ so that for all $\ga$    small enough
      \begin{equation}
      \label{77.4}
\big|\la_\ga-\eps_\ga\big| \le  c \eps_\ga^2
     \end{equation}
Finally the probability $ w_{\la_\ga}(\und u)$ on $\mathcal R$ satisfies the properties listed in Section \ref{sec:13}.


\end{thm}

\medskip

\noindent
{\bf Proof.}
  Uniqueness follows because $\sum_{\und u} w_{\la}(\und u)$ is a strictly decreasing function of $\la$ when it is finite.  We will next prove
  \eqref{77.3}.
We postpone the (elementary) proof that
   \begin{equation}
      \label{77.6}
(\frac{\eps_\ga}{\eps_\ga-c_\ga\eps_\ga^2})^2 \ge 1 + c_\ga\eps_\ga,\quad
(\frac{\eps_\ga}{\eps_\ga+c_\ga\eps_\ga^2})^2 \le 1 - c_\ga\eps_\ga
     \end{equation}
Then, by \eqref{77.1.0}
  $$
 \sum_{\und u} w_{\eps_\ga-c_\ga\eps_\ga^2}(\und u) \ge (\frac{\eps_\ga}{\eps_\ga-c_\ga\eps_\ga^2})^2
  -c_\ga\eps_\ga \ge 1
  $$
Similarly
  $$
  \sum_{\und u} w_{\eps_\ga+c_\ga\eps_\ga^2}(\und u) \le (\frac{\eps_\ga}{\eps_\ga+c_\ga\eps_\ga^2})^2
  +c_\ga\eps_\ga \le 1
  $$

By Corollary \ref{coro12} $\sum_{\und u}w_\la(\und u)$ is a $C^1$ function of $\la$ in $(\frac 34{\eps_\ga} ,\frac 54{\eps_\ga})$, hence by continuity
there is a value $\la_\ga$ in $[\eps_\ga-c_\ga{\eps_\ga}^2,\eps_\ga+c_\ga{\eps_\ga}^2]$ where it is equal to 1.

The proof of \eqref{77.6} follows from the inequalities:
   $$
(\frac{1}{1-c_\ga\eps_\ga})^2 \ge ( 1+c_\ga\eps_\ga)^2 \ge  1+c_\ga\eps_\ga
   $$
     $$
(\frac{1}{1+c_\ga\eps_\ga})^2 \le ( 1-c_\ga\eps_\ga+(c_\ga\eps_\ga)^2)^2 \le 1 - c_\ga\eps_\ga
   $$
In the last inequality we have used that $4c_\ga\eps_\ga \le 1$.

The proof of \eqref{77.4} follows from
  \eqref{77.1.0} with $\la=\la_\ga$.
Finally
\eqref{77.2} is proved by  \eqref{77.2.07}.
The existence of the first moment, as stated in
\eqref{13.1}, follows also from \eqref{77.2.07}.

  \qed

\medskip

By \eqref{77.1.0} $\sum_{\und u} w_{\la}(\und u)$ as a function of $\la$ is to first order approximated by $(\frac{\eps_\ga}{\la})^2$, it is actually true that also its derivative with respect to $\la$ is well approximated by the derivative of $(\frac{\eps_\ga}{\la})^2$.  Indeed the following theorem holds (we omit for brevity its proof which is based on the estimates obtained in Appendix \ref{appendix D}):

\medskip

\begin{thm}

\label{thm77.1.00}
There is $c>0$
such that for all $\ga$ small enough
   \begin{eqnarray}
      \label{77.1.0.00}
&& \Big|\sum_{\und u} w_{\la_\ga}(\und u)|\und u| -\frac 2{ \eps_\ga} \Big|
\le c 
     \end{eqnarray}
As a consequence there is $c'>0$ so that
   \begin{eqnarray}
      \label{77.1.0.00.2}
&& |\alpha_\ga - \frac{ \eps_\ga} 2 |
\le c' \eps_\ga^2
     \end{eqnarray}

\end{thm}

\vskip.5cm

\noindent
{\bf Proof of Theorem \ref{thm13.3}}.

We have shown in Section \ref{sec:20} that
   \begin{equation}
      \label{20.0.12?}
Z_{\La_n}^{\rm pbc} = Z_{\La_n}^{\rm pbc, gg} + Z_{\La_n}^{\rm pbc, \mathcal X^0}+ Z_{\La_n}^{\rm pbc, \mathcal X^+}+  Z_{\La_n}^{\rm pbc, \mathcal X^-}
+ Z_{\La_n}^{\rm pbc, gb}
     \end{equation}
      In Appendix \ref{BRPF} it is proved that \red{there exists $\zeta_\ga>0$ so that}
     \begin{equation}
     \label{G9.00}
  \lim_{n\to\infty}    \frac {1} {e^{\beta p_\ga (2n+1)\ell^+_\ga\red{-\zeta_\ga(2n+1) }}}
  \Big(Z_{\La_n}^{\rm pbc, \mathcal X^0}+ Z_{\La_n}^{\rm pbc, \mathcal X^+}+  Z_{\La_n}^{\rm pbc, \mathcal X^-}
+ Z_{\La_n}^{\rm pbc, gb}\Big)= 0
    \end{equation}

\eqref{20.0.12.00} with $\la=\la_\ga$ becomes
    \begin{equation}
      \label{20.0.12}
Z_{\La_n}^{\rm pbc, gg} = e^{ (2n+1)[\beta p^+_\ga \ell^+_\ga+\la_\ga]}
\sum_{(x,\und u_1)\in A'}w_{\la_\ga}(\und u_1)  \sum_k\sum_{\und u_2,,,\und u_k: \sum |\und u_i| = 2n+1-|\und u_1|} \prod_{i=2}^k w_{\la_\ga}(\und u_i)
     \end{equation}
where $A'= \{(x,\und u_1): x +|\und u_1|-1 \ge n+1 \;{\rm when}\; x>-n, |\und u_1| \le 2n+1\}$ (if $|\und u_1| = 2n+1$ then the sum over $k$ in \eqref{20.0.12} is absent).  We have:
   \begin{equation}
      \label{20.0.13}
\sum_{(x,\und u_1)\in A', |\und u_1|\le n/2}w_{\la_\ga}(\und u_1) =
\sum_{|\und u_1|\le n/2}|\und u_1| w_{\la_\ga}(\und u_1) =
\frac{1}{\alpha_\ga} - \sum_{|\und u_1|> n/2}|\und u_1| w_{\la_\ga}(\und u_1)
     \end{equation}
with the last term exponentially small in $n$ (by the properties of $\la_\ga$ stated in Section \ref{sec:13}).  {By \eqref{13.1.1}}  for any $|\und u_1| \le n/2$  the quantity 
    \begin{equation}
      \label{20.0.14}
\sum_k\sum_{\und u_2,,,\und u_k: \sum |\und u_i| = 2n+1-|\und u_1|} \prod_{i=2}^k w_{\la_\ga}(\und u_i) -  \alpha_\ga
     \end{equation}
decays exponentially in $n$.  This proves \eqref{13.5.1} and therefore Theorem \ref{thm13.3}.

\vskip2cm

%
%
%
%
%
%
%
%
%

\vskip2cm

\section{Proof of Theorem \ref{thm4.1.3-a} }
\label{app:thm4.1.3-b}

In this section we will prove \eqref{13.3.0}
and hence Theorem \ref{thm4.1.3-a}. We thus fix a local event
 $X^*=\{x^*_{\ell,m}, (\ell',m')\le (\ell,m) \le (\ell'',m'')\}$ (see the end of Section \ref{sec:12}).  As a general result for one dimensional Gibbs measure we have
    \begin{equation}
    \label{G1}
 \mu_{\ga}[\und s:\psi(\und s)\in X^*] = \lim_{n\to \infty}
  \mu_{\ga,\La_n}^{\text{pbc}} [\und s':\psi(\und s')\in X^*]
   \end{equation}
where  (using the notation in Section \ref{sec:20}) $\La_n=[-n,n]$, $\und s'$ is the $\La_n$-periodic extension of the configuration in $\La_n$
and  $\mu_{\ga,\La_n}^{\text{pbc}}$ is the Gibbs measure with periodic boundary conditions.
   \begin{equation}
   \label{G1.0}
  \mu_{\ga,\La_n}^{\text{pbc}} [\und s':\psi(\und s')\in X^*] =
  \frac{Z_{\La_n}^{\text{pbc}}[\und s':\psi(\und s')\in X^*]}{Z_{\La_n}^{\text{pbc}}}=  \frac{Z_{\La_n}^{\text{pbc}}[\und s':\psi(\und s')\in X^*]}{e^{\beta p_\ga (2n+1)\ell^+_\ga}} \frac{e^{\beta p_\ga (2n+1)\ell^+_\ga}}{Z_{\La_n}^{\text{pbc}}}
   \end{equation}
where $Z_{\La_n}^{\text{pbc}}[\und s':\psi(\und s')\in X^*]$ is the partition function restricted to  $\{\und s':\psi(\und s')\in X^*\}$.

 By \eqref{13.5.1} the last factor converges to 1 as $n\to \infty$.
 To estimate the first one we proceed as in Section \ref{sec:20} with the role of the point $-n$ replaced by $x^*_{\ell',1}$.  We then have
     \begin{equation}
     \label{G2}
  \frac{Z_{\La_n}^{\text{pbc}}[\und s':\psi(\und s')\in X^*]}{e^{\beta p_\ga (2n+1)\ell^+_\ga}} = Z_{\La_n}^{\text{pbc,*}}
  +    \frac {Z_{\La_n}^{\rm pbc, gb}} {e^{\beta p_\ga (2n+1)\ell^+_\ga}}
    \end{equation}
where  $Z_{\La_n}^{\rm pbc, gb}$ is defined in \eqref{20.0.7} and the
last term in \eqref{G2} vanishes by \eqref{G9} below ; $ Z_{\La_n}^{\text{pbc,*}}$ is:
    \begin{eqnarray}
     \label{G3}
  &&Z_{\La_n}^{\text{pbc,*}}:=
   \sum_{(x,k,\und u_1,\dots \und u_k)\in A^*}\Ii_{\{(x,k,\und u_1,\dots \und u_k) \to X^*\}} \prod_{i=1}^{k}w_{\la_\ga}(\und u_i)
    \end{eqnarray}
where $A^*=\{(x,k,\und u_1,\dots \und u_k): x^*_{\ell',1}\in [x,x+|\und u_1|-1] ;
|\und u_1|+\dots +|\und u_k| = 2n+1\}$.  Moreover $(x,k,\und u_1,\dots \und u_k)$ defines a sequence $x_{\ell,m}, \ell=1,..,k;m=1,..,4$ where $x_{1,1}=x$,
$x_{1,2}=x+ u_{1,1}$, $x_{\ell,m} = x_{\ell,m-1} + u_{\ell,m-1}$ (with $x_{\ell,0}=  x_{\ell-1,4}$).  With these notation $ \{(x,k,\und u_1,\dots \und u_k) \to X^*\} $ is the set of elements in $A^*$ such that the corresponding $(x_{\ell,m})$ verifies $x_{\ell+\ell_0,m} = x^*_{\ell,m}, \ell=\ell',..,\ell''$ for some $\ell_0$.

Let $C_1= [x, x+|\und u_1|-1]=:[x'_1,x''_1]$, $C_i= [x''_{i-1}+1,  x''_{i-1}+ |\und u_i|]=:[x'_{i},x''_{i}]$, $i \le k$.  By the definition of $A^*$
the point $x^*_{\ell',1} \in C_1$ and we call $C_h$ the last set $C_i$ which has non empty intersection with  $[x^*_{\ell',1},x^*_{\ell'',4}]$ so that
 $C_1,...,C_h$ is the collection of all $C_i$ which have non empty intersection with $[x^*_{\ell',1},x^*_{\ell'',4}]$.
Since the $C_i$ are disjoint their number $h \le x^*_{\ell'',4}-x^*_{\ell',1}+1$.

We fix a positive number $R>x^*_{\ell'',4}-x^*_{\ell',1}+1$, and take $n$ so large that $R< 2n+1$.
We then split the sum in \eqref{G3}, over configurations $(x,k,\und u_1, \dots \und u_k)$ such that
$|C_{i}|\le R, i=1,..,h$ and the other ones.  The contribution of the latter ones is bounded by:
  \begin{equation}
        \label{G.3.0}
2\sum_{|\und u|>R} w_{\la_\ga}(\und u)|\und u|
    \end{equation}
 In fact  if  $|C_{i}|> R$ then necessarily either $i=1$ or $i=h$ (or both) which are bounded in the same way, hence the factor 2 in \eqref{G.3.0}.
 The quantity in \eqref{G.3.0} is bounded by:
    \begin{equation}
        \label{G.3.0.1}
 e^{- R\frac{\eps_\ga}4} \sum_{|\und u|>R} w_{\frac 12 \eps_\ga}(\und u)|\und u| \le c''_\ga e^{- R\frac{\eps_\ga}4}
    \end{equation}
 where we have used \eqref{77.2.0}

It remains to estimate
   \begin{eqnarray}
     \label{G3.0.2}
  &&Z_{\La_n}^{\text{pbc,**}}:=
   \sum_{(x,k,\und u_1,\dots \und u_k)\in A^*}\Ii_{\{(x,k,\und u_1,\dots \und u_k) \to X^*\}} \Ii_{\{|C_i| \le R, i=1,..,h \}}\prod_{i=1}^{k}w_{\la_\ga}(\und u_i)
    \end{eqnarray}
This can be written as (see \eqref{13.1.1} for notation):
  \begin{eqnarray}
     \label{G3.0.3}
  &&
   \sum_{(x,h,\und u_1,\dots \und u_h)} \Ii_{\{|C_i| \le R, i=1,..,h \}}\Ii_{\{(x,h,\und u_1,\dots \und u_h) \to X^*\}}\prod_{i=1}^{h}w_{\la_\ga}(\und u_i)\nn\\&&
   \times   W_\ga\Big[ \;\text{\rm there is $k$
so that}\;\sum_{i=1}^k |\und u_i| = 2n+1 -\big(|C_1|+\cdots +|C_h|\big)\Big]
   \end{eqnarray}
  We next let $n\to \infty$, since $|C_1|+\cdots +|C_h| \le Rh$ (recall that $h \le x^*_{\ell'',4}-x^*_{\ell',1}+1$), then by \eqref{13.1.1} the last factor converges to $\alpha_\ga$.  Then we let $R\to \infty$ and finally get
 \begin{eqnarray}
     \label{G3.0.4}
  &&
 \alpha_\ga  \sum_{(x,h,\und u_1,\dots \und u_h)\in B}
 \Ii_{\{(x,h,\und u_1,\dots \und u_h) \to X^*\}}
 \prod_{i=1}^{h}w_{\la_\ga}(\und u_i)
   \end{eqnarray}
where $B$ is the collection of all $(x,h,\und u_1,\dots \und u_h)$ such that
the corresponding sets $C_1,..,C_h$ cover $[x^*_{\ell',1},x^*_{\ell'',4}]$ and each one has non empty intersection with $[x^*_{\ell',1},x^*_{\ell'',4}]$.

\section*{Acknowledgement}
We are grateful to Roberto Fernandez and Pierre Picco for many useful discussions.
\newpage
\appendix

\setcounter{equation}{0}

\section{Outline of the proof of Theorem \ref{thm10.1}}
\label{appzero}

The advantage of describing the spin configurations in terms of plus-minus and interface intervals is to split difficulties. In fact in a plus (or minus) interval the system sees only one phase and therefore one can exploit the local stability of single phases.  The interface intervals are dealt with large deviation estimates as   in Theorem \ref{thm111.2}. What we said above about the plus  (or minus) interval is not entirely true because the definition of the plus interval allows for values of $\Theta$ which are not always equal to 1, as $\Theta=0$ is also allowed, this spoils a direct applications of the previous stability statements.

\medskip

\subsection{Reduction to the restricted ensemble}

The problem is settled in the next theorem where we show that the computation of $Z^+_{\ga,n}$ can be reduced to that of a partition function in the ``restricted plus ensemble'' where $\Theta \equiv 1$ (i.e.\ without fluctuations).  The price is an additional hamiltonian which however is proved to be ``small''.

\vskip.5cm

\begin{thm}
\label{thm10.2}
There are
$\ga^*$,  $c$ and $b$ all positive so that for all $\ga \le \ga^*$ and $n\ge 1$.
   \begin{eqnarray}
      \label{10.6}
&&Z^{+}_{\ga,n}(s_0,s_{n+1}) =  \sum_{(s_1,..,s_n): \theta(s_i)= 1, i=1,..,n}
 e^{-\beta[ H_\ga (s_1,..,s_n|s_0,s_{n+1})+   U_\ga(s_1,..,s_n|s_0,s_{n+1})]}\\&&
 U_\ga(s_1,..,s_n|s_0,s_{n+1})=\sum_{\Delta \subset [0,n+1]:N_\Delta \ge 5} u_{\Delta,\ga}(s_\Delta)\nn
     \end{eqnarray}
The latter is the energy in  $[1,n]$ in interaction with    $\{0,n+1\}$ of the hamiltonian $U_\ga$ with many body potentials $\{u_{\Delta,\ga}(s_\Delta)\}$ which are bounded by

\begin{equation}
      \label{10.7}
| u_{\Delta,\ga}(s_\Delta)| \le ce^{-{b}\ga^{-1}N_\Delta}, \quad \text{\rm
$N_\Delta$ the cardinality of $\Delta$}
     \end{equation}
when $n=1,2$ the additional hamiltonian $U_\ga$ is absent.
\end{thm}

\medskip

The proof of Theorem \ref{thm10.2} is reported in Appendix \ref{appD}.
The strategy of the proof is like in a typical step of the Pirogov-Sinai analysis of  Ising systems at low temperatures where the energy is reduced to the ground state energy (i.e.\  with all plus, or all minus) and the partition function becomes the partition function of a gas of polymers whose activities are the weights of the contours. In our case we reduce to the restricted ensemble and the contribution of the contours is expressed (using cluster expansion) in terms of the additional hamiltonian $U_\ga$.

\medskip

\subsection{An interpolation hamiltonian}

It is now convenient  to go back to the original spins $\si(x)$ and,
following the proof in Chapter 11 in \cite{presutti},
we introduce a reference hamiltonian
      \begin{equation}
      \label{10.8}
H^{\rm free}   = -\sum_{x } m_\beta \si(x)
     \end{equation}
     \red{where $m_\beta$ is defined in \eqref{intro.4}}.

Recalling from \eqref{12.2} that $C^{\pm}_i= [ \ell^{\pm}_\ga i,\ell^{\pm}_\ga(i+1))$, we  write
  \begin{equation}
      \label{10.16}
\La_n = \bigcup_{i=1}^{n} C^+_i,
 \quad \La_{n,-} = \bigcup_{i=-\infty}^{ n} C^+_i, \quad\La_{n,+} = \bigcup_{i=1}^\infty C^+_i
      \end{equation}
Namely     $\La_{n,-}$ and $\La_{n,+}$ are obtained from $\La_{n}$ by adding all the intervals $C_i^+$ to the left and respectively to the right of $\La_n$.  We also write:
          \begin{equation}
      \label{10.17}
\La^*_n = \bigcup_{i=0}^{n+1} C^+_i,
 \quad \La^*_{n,-} = \bigcup_{i=-\infty}^{ n+1} C^+_i, \quad\La^*_{n,+} = \bigcup_{i=0}^\infty C^+_i
      \end{equation}
We denote by $\mu^0_{t, \La_n,\si_{\La_n^c}}$  the  Gibbs measure on the plus ensemble $\Theta \equiv 1$  with the interpolating hamiltonian
     \begin{eqnarray}
      \label{10.9}
&&t[H_\ga (\si_{\La_n}| \si_{\La_n^c})+ U_\ga(\si_{\La_n}|\si_{\La^*_n\setminus \La_n})] +(1-t) H^{\rm free}_{ \La}(\si_{\La_n}),\quad
t\in[0,1]\\&& U_\ga(\si_{\La_n}|\si_{\La^*_n\setminus \La_n})
= \sum_{\Delta \subset \La^*_n, \Delta \cap \La_n \ne \emptyset } u_{\Delta,\ga}\nn
     \end{eqnarray}
$U_\ga(\si_{\La_n}|\si_{\La^*_n\setminus \La_n})$ is equal to the expression in \eqref{10.6} once written in the original spins $\si(x)$, the condition
$\Delta \cap \La_n \ne \emptyset$ is redundant because $\Delta \subset \La^*_n$ and $N_\Delta\ge 5$.
Then
    \begin{equation}
      \label{10.10}
\frac{1}{\beta} \log \frac{Z^{+}_{\ga,\La_n}(\si_{\La_n^c})}{[2\cosh (\beta m_\beta)]^{|\La_n|}} = \int_{0}^{1}dt E_{\mu_{t, \La_n,\si_{\La_n^c}}}
\Big[ H^{\rm free}_{ \La_n}(\si_{\La_n})-H_\ga (\si_{\La_n}| \si_{\La_n^c})-U_\ga(\si_{\La_n}|\si_{\La^*_n\setminus \La_n})\Big]
     \end{equation}
$|\La_n| = n\ell^+_\ga$, i.e.\ the number of original spins in $\La_n$.

\medskip

The key point in what follows is that the interpolating hamiltonian restricted to the plus ensemble satisfies the Dobrushin conditions for uniqueness and the corresponding Gibbs measures have exponential decay of correlations.  We will exploit all that in the analysis of the right hand side of \eqref{10.10}.  We will write the integrand as sum of terms which are divided into three categories: those localized away from the boundaries, those close to the right boundary and finally those close to the left boundary.  The first ones will contribute to the pressure, the second ones to $F^{(2)}_\ga$ and the last ones to  $F^{(1)}_\ga$.  The  term $G^{(1)}_{\ga,n}$ takes into account the small dependence on the far away boundaries.  We outline below how we implement this strategy.

\bigskip

\subsection{The infinite volume measures $\mu_t$ and the pressure $p^{+}_\ga$}.

We will prove in Appendix \ref{appE} that for $\ga$ small enough and any $t\in [0,1]$ there is a unique DLR measure $\mu_t$ on the plus ensemble $\Theta\equiv 1$ with hamiltonian $t[H_\ga+U_\ga]+(1-t) H^{\rm free}$ where $U_\ga=\{u_{\Delta,\ga}\}$.
%
%
%
We
will prove
Theorem  \ref{thm10.1} with
    \begin{equation}
      \label{10.11}
p^+_{\ga} =\frac 1\beta\log\{2\cosh (\beta m_\beta)\}+ \int_{0}^{1}dt \frac 1{\ell^+_\ga}\sum_{x\in [0,\ell^+_\ga)} E_{\mu_{t}}
 [ k_{x}  ]
     \end{equation}
where
        \begin{equation}
      \label{10.12}
 k_{x} = -m_\beta \si(x) + \frac 12\sum_y J_\ga(x,y)\si(x)\si(y)
 -\sum_{\Delta \ni x}\frac 1{|\Delta|}u_{\Delta,\ga}(\si_\Delta)
     \end{equation}
By \eqref{10.7} the series in the last term is absolutely convergent.

\eqref{10.10} can  be written in a similar way
         \begin{equation}
      \label{10.13}
\frac{1}{\beta} \log \frac{Z^{+}_{\ga,\La_n}(\si_{\La_n^c})}{[2\cosh (\beta m_\beta)]^{|\La_n|}} = \sum_{x\in \La_n^*} \int_{0}^{1}dt E_{\mu^0_{t, \La_n,\si_{\La_n^c}}}
\Big[k_{\La_n,x} \Big]
     \end{equation}
     where for $x\in \La_n^*$
                \begin{equation}
      \label{10.14}
 k_{\La_n,x}  = -m_\beta \si(x)\mathbf 1_{x\in \La_n} + \frac 12\sum_{y: \{x,y\} \cap \La_n \ne \emptyset} J_\ga(x,y)\si(x)\si(y)
 -\sum_{\Delta\subseteq \La_n^*, \Delta\ni x}\frac 1{|\Delta|}u_{\Delta,\ga}(\si_\Delta)
     \end{equation}
Observe that if above $\Delta\subseteq \La_n^*$ then necessarily \red{$\Delta \cap \La_n \ne \emptyset$} because $N_\Delta \ge 5$. Thus
         \begin{equation}
      \label{10.15}
\frac{1}{\beta} \log Z^{+}_{\ga,\La_n}(\si_{\La_n^c})-|\La_n|p^+_{\ga}  = \int_{0}^{1}dt\{ \sum_{x\in \La_n^*} E_{\mu^0_{t, \La_n,\si_{\La_n^c}}}
\Big[ k_{\La_n,x}\Big] - \sum_{x \in \La_n} E_{\mu_{t}}
\Big[ k_{x} \Big]\}
     \end{equation}

\medskip

\subsection{Semi-infinite measures and the surface terms $F^{(k)}_\ga$}.

We will prove in Appendix \ref{appE} that for $\ga$ small enough and any $t\in [0,1]$ for any configuration $\si_{\La_{n,-}^c}$ in \red{$\Theta\equiv 1$}
there is a unique semi-infinite DLR measure $\mu^0_{t,\La_{n,-}, \si_{\La_{n,-}^c}}$ on the plus ensemble $\Theta\equiv 1$ with hamiltonian $t[H_\ga+U'_\ga]+(1-t) H^{\rm free}$ where $U'_\ga=\{u_{\Delta,\ga}, \Delta \subset \La_{n,-}^*\}$, analogous statement holding for $\mu^0_{t,\La_{n,+}, \si_{\La_{n,+}^c}}$.


For any $x \in \La_{n,\mp}^*$ we first define
           \begin{equation}
      \label{10.14.1}
 k_{\La_{n,\mp},x}  = -m_\beta \si(x)\mathbf 1_{x\in \La_{n,\mp}} + \frac 12\sum_{y: \{x,y\} \cap \La_{n,\mp} \ne \emptyset} J_\ga(x,y)\si(x)\si(y)
 -\sum_{\Delta\subseteq \La_{n,\mp}^*, \Delta\ni x}\frac 1{|\Delta|}u_{\Delta,\ga}(\si_\Delta)
     \end{equation}
and then
        \begin{eqnarray}
      \label{10.18}
&& F^{(2)}_\ga(\si_{\La_{n,-}^c})  = \int_{0}^{1}dt\{ \sum_{x\in \La_{n,-}}\Big( E_{\mu^0_{t, \La_{n,-}; \si_{\La_{n,-}^c}}}
\Big[ k_{\La_{n,-},x}\Big] -  E_{\mu_{t}}
\Big[ k_{x} \Big] \Big)\nn\\&&\hskip5cm  +\sum_{x \in \La_{n,-}^*\setminus\La_{n,-}} E_{\mu^0_{t, \La_{n,-}; \si_{\La_{n}^c}}}
\Big[ k_{\La_{n,-},x}\Big]\}
     \end{eqnarray}
 Notice that    $F^{(2)}_\ga(\cdot)$ does not depend on $n$.
  Analogously
         \begin{eqnarray}
      \label{10.19}
&&F^{(1)}_\ga(\si_{\La_{n,+}^c})  = \int_{0}^{1}dt\{ \sum_{x\in \La_{n,+}}\Big( E_{\mu^0_{t, \La_{n,+}; \si_{\La_{n,+}^c}}}
\Big[ k_{\La_{n,+},x}\Big] -  E_{\mu_{t}}
\Big[ k_{x} \Big] \Big) \nn
\\&&\hskip5cm+\sum_{x \in \La_{n,+}^*\setminus \La_n} E_{\mu^0_{t, \La_{n}; \si_{\La_{n}^c}}}
\Big[ k_{\La_{n,+},x}\Big]\}
     \end{eqnarray}

     \medskip

     \subsection{The terms $G^{(i)}_{\ga,n}(s_0,s_{n+1})$}

We are going to show that the difference between the right hand side of \eqref{10.15} and the sum
$F^{(1)}_\ga(\si_{\La_{n,+}^c})+ F^{(2)}_\ga(\si_{\La_{n,-}^c})$, which thus defines the remainder term
$G^{(1)}_{\ga,n}(s_0,s_{n+1})$, is
         \begin{equation}
      \label{10.20}
G^{(1)}_{\ga,n}(s_0,s_{n+1}) = \sum_{i=1}^6\Psi^{(i)}
     \end{equation}
where, calling $x^*:= \ell^+_\ga [\frac n2]$, $[\frac n2]$ the integer part of $\frac n2$,
         \begin{equation}
      \label{10.21}
\Psi^{(1)}  = \int_{0}^{1}dt  \sum_{x\in \La_n, x< x^*}\Big(E_{\mu^0_{t, \La_{n}; \si_{\La_{n}^c}}}
\Big[ k_{\La_{n},x}\Big] - E_{\mu^0_{t, \La_{n,+}; \si_{\La_{n,+}^c}}}
\Big[ k_{\La_{n,+},x}\Big]\Big)
     \end{equation}
        \begin{equation}
      \label{10.22}
\Psi^{(2)}  = \int_{0}^{1}dt  \sum_{x\in \La_n, x \ge x^*}\Big(E_{\mu^0_{t, \La_n; \si_{\La_n^c}}}
\Big[ k_{\La_n,x}\Big] - E_{\mu^0_{t, \La_{n,-}; \si_{\La_{n,-}^c}}}
\Big[ k_{\La_{n,-},x}\Big]\Big)
     \end{equation}
         \begin{equation}
      \label{10.23}
\Psi^{(3)}  =- \int_{0}^{1}dt  \sum_{x\in \La_n, x \ge x^*}\Big( E_{\mu^0_{t, \La_{n,+}; \si_{\La_{n,+}^c}}}
\Big[ k_{\La_{n,+},x}\Big] -  E_{\mu_{t}}
\Big[ k_{x} \Big] \Big)
     \end{equation}
        \begin{equation}
      \label{10.24}
\Psi^{(4)}  =- \int_{0}^{1}dt  \sum_{x\in \La_{n,-}, x < x^*}\Big( E_{\mu^0_{t, \La_{n,-}; \si_{\La_{n,-}^c}}}
\Big[ k_{\La_{n,-},x}\Big] -    E_{\mu_{t}}
\Big[ k_{x} \Big] \Big)
     \end{equation}
                      \begin{equation}
      \label{10.25}
\Psi^{(5)}  =  \int_{0}^{1}dt  \sum_{x \in \La_{n,+}^*\setminus \La_{n,+}} \{E_{\mu^0_{t, \La_{n}; \si_{\La_{n}^c}}}
\Big[ k_{\La_n,x}\Big]-E_{\mu^0_{t, \La_{n,+}; \si_{\La_{n}^c}}}
\Big[ k_{\La_{n,+},x}\Big]\}
     \end{equation}
             \begin{equation}
      \label{10.26}
\Psi^{(6)}  =  \int_{0}^{1}dt  \sum_{x \in \La_{n,-}^*\setminus \La_{n,-}} \{E_{\mu^0_{t, \La_{n}; \si_{\La_{n}^c}}}
\Big[ k_{\La_n,x}\Big]-E_{\mu^0_{t, \La_{n,-}; \si_{\La_{n}^c}}}
\Big[ k_{\La_{n,-},x}\Big]\}
     \end{equation}

In Appendix \ref{appE} we will prove that the series in the definition of the
$\Psi^{(i)}$
are absolutely convergent.  To prove \eqref{10.20} we observe that
       \begin{eqnarray*}
F^{(1)}_\ga(\si_{\La_{n,+}^c}) +\Psi^{(1)} +\Psi^{(3)}+\Psi^{(5)} &=&  \int_{0}^{1}dt\Big(\sum_{x\in \La_n,x < x^*} \{ E_{\mu^0_{t, \La_n,\si_{\La_n^c}}}
\Big[ k_{\La_n,x}\Big] -E_{\mu_{t}}
\Big[ k_{x} \Big]\}\nn\\ &+&\sum_{x \in \La_{n,+}^*\setminus \La_n} E_{\mu^0_{t, \La_{n}; \si_{\La_{n}^c}}}
\Big[ k_{\La_n,x}\Big]\Big)
     \end{eqnarray*}
           \begin{eqnarray*}
F^{(2)}_\ga(\si_{\La_{n,+}^c}) +\Psi^{(2)} +\Psi^{(4)} +\Psi^{(6)}&=&  \int_{0}^{1}dt\Big(\sum_{x\in \La_n,x \ge x^*} \{ E_{\mu^0_{t, \La_n,\si_{\La_n^c}}}
\Big[ k_{\La_n,x}\Big] -E_{\mu_{t}}
\Big[ k_{x} \Big]\}\nn\\ &+&\sum_{x \in \La_{n,-}^*\setminus \La_n} E_{\mu^0_{t, \La_{n}; \si_{\La_{n}^c}}}
\Big[ k_{\La_n,x}\Big]\Big)
     \end{eqnarray*}
Summing   the last two equations we see that $F^{(1)}_\ga+F^{(2)}_\ga+\Psi^{(1)}+\cdots + \Psi^{(6)}$
is equal to
the right hand side of \eqref{10.15}, hence \eqref{10.20}.

As we shall see in Appendix \ref{appE}, the logic behind the above manipulations is that
the terms in the $\Psi^{(i)}$ can be reduced to difference of conditional Gibbs expectations of cylindrical functions localized away from where the conditioning of the Gibbs measures differ from each other and thus exploit the exponential decay of correlations.

\vskip2cm

\setcounter{equation}{0}

\section{Contours, cluster expansion and proof of Theorem \ref{thm10.2}}
\label{appD}

\eqref{10.6} shows that $Z^+_{\ga,n}(s_0,s_{n+1})$ can be written as a partition function where the spin configurations are restricted to the plus ensemble with all $\Theta_i =1$.  This is automatically true when $n=1,2$, when $n\ge 3$ it can still be done but the energy gets an additional term.  This is what we will prove in the sequel using contours.
Contours are denoted by $\Ga= \{\rm{sp}(\Ga), \theta_\Ga\}$, where $\rm{sp}(\Ga)$ is the spatial support of $\Ga$, namely an interval of length $N_\Ga\ge 3$ ($N_\Ga\ge 3$
tacitly in the sequel); $\theta_\Ga$, called the specification of $\Ga$, is a function on $\rm{sp}(\Ga)$, namely a sequence $\theta_1,..,\theta_n$, $n=N_\Ga$, with values $0,\pm 1$ and such that, setting $\theta_0=\theta_{n+1}=1$,
     \begin{equation}
      \label{D.1}
\theta_1=\theta_n= 1,\;\:  \Theta_i=0,\; i=1,..,n
     \end{equation}
The weight of a contour $\Ga$ with $\rm{sp}(\Ga)=\{1,..,n\}$ and specification   $(\theta_1,..,\theta_n)$ is
     \begin{eqnarray}
      \label{D.2}
&& W_{\ga}(\Ga| s_0,s_{n+1}) =\frac{ \sum_{s_1,..,s_n }
\mathbf 1_{\theta(s_i) =\theta_i \; i=1,2,.., n }
e^{-\beta H_\ga (s_1,..,s_n| s_0,s_{n+1})}}{Z^{++}_{\ga,n}(s_0,s_{n+1}) } \\&&
Z^{++}_{\ga,n}(s_0,s_{n+1}) = \sum_{s_1,..,s_n}
\mathbf 1_{\theta_i= 1,\; i=1,2,..,n}
e^{-\beta H_\ga (s_1,..,s_n| s_0,s_{n+1})}\nn
     \end{eqnarray}
with $\theta(s_0)=\theta(s_{n+1})=1$.  By translations the definition is then extended to all contours $\Ga$.  Denoting by $\und \Ga$ any collection of ``compatible'' contours
$\Ga$
we have:
    \begin{equation}
      \label{D.3}
Z^{+}_{\ga,n}(s_0,s_{n+1}) = \sum_{(s_1,..,s_n): \theta(s_i)= 1, i=1,..,n}
\Xi_{\ga,[1,n],s_0,..,s_{n+1}} e^{-\beta H_\ga (s_1,..,s_n| s_0,s_{n+1})}
     \end{equation}
     where
        \begin{equation}
      \label{D.4}
\Xi_{\ga,[1,n],s_0,..,s_{n+1}}:= \sum_{\und \Ga:\rm{sp}(\Ga) \subseteq\{1,..,n\}} \prod_{\Ga \in \und \Ga} W_{\ga}(\Ga|s_0,..,s_{n+1} )
     \end{equation}
$W_{\ga}(\Ga| s_0,..,s_{n+1} )$  actually depends only on $s_m$ and $s_{m+k+1}$ if sp$(\Ga)=(m+1,..,m+k)$.
Theorem \ref{10.2} will follow by proving that
       \begin{equation}
      \label{D.5}
\log \Xi_{\ga,[1,n], s_0,..,s_{n+1}}=  -\beta\sum_{\Delta \subset [0,n+1]:N_\Delta \ge 5} u_{\Delta,\ga}(s_\Delta)
     \end{equation}
($N_\Delta$ the number of sites in the interval $\Delta$) with $u_{\Delta,\ga}$ satisfying \eqref{10.7}.
$\Xi_{\ga,[1,n], s_0,..,s_{n+1}}$ is
the partition function of a gas of polymers with weights $ W_{\ga}(\Ga|s_0,..,s_{n+1} )$ and \eqref{D.5} will follow from cluster expansion whose validity depends on the smallness of the weights
$W_{\ga}(\Ga|s_0,..,s_{n+1} )$.

\medskip

\begin{thm}
\label{thmD1}
There are $c$ and $\ga^*$ positive so that for all $\ga \le \ga^*$  and  $n\ge 3$,
     \begin{equation}
      \label{D.6}
W_{\ga}(\Ga|s_0,s_{n+1})\le w_\ga(\Ga):=\exp\{- c \delta \zeta^2 \ga^{-1} N_\Ga\},\quad N_\Ga=n
     \end{equation}

\end{thm}

\noindent{\bf Proof.}
In Theorem 9.2.5.1 of \cite{presutti} it is proved that
    \begin{equation}
      \label{D.7}
W_{\ga}(\tilde\Ga|s_0,s_{n+1})\le \exp\{- c' \delta \zeta^2 \ga^{-1} N_\Ga\}
     \end{equation}
where the contour $\tilde\Ga$ is defined by specifying the values of $\eta$ on ${\rm sp}(\Ga)$ rather than the values of $\theta$ as in our case (under the smallness conditions on $\delta$ and $\zeta$ that we have required). It should also be remarked
 that
 $\delta$ and $\zeta$ in
Theorem
9.2.5.1 depend suitably on $\ga$ but the proof works also in our case.
To use \eqref{D.7} we observe that the number of specifications $\eta$
which give a same specification $\theta$  is bounded by
$3^{N_\Ga \ell^+_\ga/\ell^-_\ga}$, $\ell^-_\ga:=\delta\ga^{-1}$, so that
    \begin{equation*}
W_{\ga}(\Ga|s_0,s_{n+1})\le 3^{N_\Ga \ell^+_\ga/\ell^-_\ga}e^{- c' \delta \zeta^2 \ga^{-1} N_\Ga} 
     \end{equation*}
which for $\alpha$ small yields
\eqref{D.6} (for   $\ga$ small enough).

 \qed

\vskip.5cm

The proof of \eqref{D.5}  follows from cluster expansion whose validity relies on the Peierls estimates proved in Theorem \ref{thmD1}.
An immediate consequence of  \eqref{D.6} is the validity of a strengthened  K-P (Kotecki-Preiss) condition: there is $b'>0$ so that for all $\ga$ small enough
    \begin{equation}
      \label{D.8}
\sum_{\Ga: \rm{sp}
(\Ga)\ni x }e^{b'\ga^{-1}N_\Ga}W_{\ga}(\Ga)\le 1 
 \end{equation}
where $N_\Ga$ 
is the number of {$\ell_\ga^+$-blocks} in the spatial support of $\Ga$.

The K-P condition
allows to exponentiate $\Xi_{\ga,[1,n],s}$ in a convergent series. Referring to the literature for a proof we just state the result in Theorem \ref{thmD2} below.  We need some extra notation: we regard the space of all contours $\Ga$ (denoted by $\{\Ga\}$) as a graph with nodes $\Ga$ and connections $(\Ga,\Ga')$ if
${\rm sp}(\Ga)\cap {\rm sp}(\Ga')\ne \emptyset$.  $I(\Ga)$ denotes any integer valued function on $\{\Ga\}$ such that $\{\Ga: I(\Ga)>0\}$ is a connected set.

\medskip

\begin{thm} [Cluster expansion]
\label{thmD2}
Let $\ga$ be so small that the K-P condition (see \cite{kotpreis}) \eqref{D.8} holds. Then the sum on the right hand side of \eqref{D.9} below
(over all  functions $I$ as above) is absolutely convergent and
     \begin{equation}
      \label{D.9}
\log \Xi_{\ga,[1,n],s}=
\sum_{I}a_I W_\ga^I, \quad W_\ga^I :=\prod_{\Ga}W_\ga(\Ga)^{I(\Ga)}
     \end{equation}
The coefficients $a_I$ are combinatorial (signed) factors whose explicit definition is not needed here.  Moreover
given any $\Ga$  and any subset $\mathcal I$ in $\{I\}$ such that
$I(\Ga)\ge 1$ for all $I\in \mathcal I$ and a non negative function $f(I)$ on  $\mathcal I$,
\begin{equation}
    \label{D.10}
\sum_{I\in \mathcal I}f(I)I(\Ga)|a_I|W_\ga^I \le W_\ga(\Ga) e^{(1+\ga^{-1}b')N_\Ga} \sup_{I\in \mathcal I} f(I)
e^{-\ga^{-1}b'\|I\|},\quad \|I\| = \sum_{\Ga} N_\Ga I(\Ga)
   \end{equation}

\end{thm}

\medskip
The convergence of the series on the right hand side of \eqref{D.9} follows from \eqref{D.10}. Indeed, calling $ \mathcal I_\Ga=\{I: I(\Ga)\ge 1\}$ and using \eqref{D.10} and  \eqref{D.8}
     \begin{eqnarray*}
\sum_{I}|a_I| W_\ga^I &\le & \sum_{i=1}^n \sum_{\Ga: {\rm sp}(\Ga)\ni i}\sum_{I\in \mathcal I_\Ga} |a_I|W_\ga^I  \\ &\le &\sum_{i=1}^n \sum_{\Ga: {\rm sp}(\Ga)\ni i}
W_\ga(\Ga) e^{(1+\ga^{-1}b')N_\Ga} \le en
     \end{eqnarray*}

     \medskip

We can now complete the proof of Theorem \ref{thm10.2}.
Calling $\Delta_0$ the interior of an interval $\Delta$, \eqref{10.6} holds with
\begin{equation}
    \label{D.11}
   -\beta u_{\Delta,\ga}=
\sum_{I\in \mathcal I_\Delta}a_I W_\ga^I,\quad
  \mathcal I_\Delta = \{I: \bigcup_{\Ga: I(\Ga)\ge 1} {\rm sp}(\Ga) = \Delta_0\}
   \end{equation}
and by \eqref{D.10} with $\mathcal I= \mathcal I_\Ga\cap
\mathcal I_\Delta$,
for any $i\in \Delta_0$
     \begin{eqnarray*}
|\beta u_{\Delta,\ga}|   &\le & 
\sum_{\Ga: {\rm sp}(\Ga)\ni i,
{\rm sp}(\Ga)\subset \Delta_0}\sum_{I\in \mathcal I_\Ga\cap
\mathcal I_\Delta} |a_I|W_\ga^I  \\ &\le &  
\sum_{\Ga: {\rm sp}(\Ga)\ni i,
{\rm sp}(\Ga)\subset \Delta_0}  W_\ga(\Ga) e^{(1+\ga^{-1}b')|\Ga|} e^{-\ga^{-1}b'N_{\Delta_0}}
 \\ &\le & 
 e^{1-\ga^{-1}b'N_{\Delta_0}}
%
     \end{eqnarray*}
because $\|I\| = \sum_{\Ga} N_\Ga I(\Ga) \le N_{\Delta_0}$ when
$I\in \mathcal I_\Ga\cap
\mathcal I_\Delta$.   Observe that the minimal value of $N_\Delta$ is achieved when $N_{\Delta_0}=3$ is minimal and   therefore by \eqref{D.8} $N_\Delta\ge 5$.

\vskip2cm

\setcounter{equation}{0}

\section{Dobrushin uniqueness and proof of Theorem \ref{thm10.1}}
\label{appE}

In the proof of  Theorem \ref{thm10.1} we will need to compare expectations of Gibbs measures where we change both the boundary conditions and the potential $U_\ga$.  With this in mind we call
$U'$ and $U''$  energies with many body potentials
$\{u'_\Delta\}$ and $\{u''_\Delta\}$ which verify the bound \eqref{10.7}.
Denote by  $A$   an interval
union of intervals $C_i^+$ and by $\mu'_{t,A,\si'_{A^c}}$  the  Gibbs measure  with hamiltonian
     \begin{eqnarray}
      \label{E.1}
&&t[H_\ga (\si_A| \si'_{A^c})+ U'  (\si_A|\si'_{A^c})] +(1-t) H^{\rm free}_{ \La}(\si_A),\quad
t\in[0,1]\\&&
U'  (\si_A|\si'_{A^c}) = \sum_{\Delta: \Delta \cap A \ne \emptyset}
u'_\Delta
     \end{eqnarray}
restricted to the plus ensemble where $\Theta \equiv 1$ ($\Theta_x(\si_A,\si'_{A^c})=1$ identically).
$\mu''_{t,A,\si''_{A^c}}$ is defined analogously with $U'$ replaced by $U''$.
The key ingredient in the sequel is that for all $\ga$ small enough and for all $t\in[0,1]$ these hamiltonians
satisfy the Dobrushin uniqueness criterion
with respect to a suitable Vaserstein distance as stated below.

\medskip

\begin{thm} [Dobrushin uniqueness]
\label{thmE1}

There are
$\ga^*$,  $c_1$ and $b_1$ all positive so that for all $\ga \le \ga^*$, $t \in [0,1]$, $A$, {$B\supseteq A$}, (unions of $C_i^+$ intervals) $\si'_{A^c}$, $\si''_{A^c}$, $U'$
and $U''$ as above and with  $u'_\Delta=u''_\Delta$ when $\Delta \subseteq B$  the following holds. There is a joint representation $\mathcal P$ (its expectation denoted by  $\mathcal E$) of $\mu'_{t,A,\si'_{A^c}}$ and $\mu''_{t,A,\si''_{A^c}}$ such that for all $C_i^- \subset A$:
   \begin{equation}
      \label{E.1}
\mathcal E \Big[ d_{C_i^-}(\si'_A,\si''_A) \Big] \le \sum_{C^-_j \subset B\setminus A}c_1 e^{-b_1 \;\ga\;{\rm dist}(C^-_i,C^-_j)}
d_{C_j^-}(\si'_{A^c},\si''_{A^c})+\sum_{C^-_j \subset B^c}c_1 e^{-b_1 \;\ga\;{\rm dist}(C^-_i,C^-_j)}
     \end{equation}
where
   \begin{equation}
      \label{E.2}
d_{C_i^-}(\si'_A,\si''_A) = \sum_{x\in C_i^-} |\si'_A(x)-\si''_A(x)|
     \end{equation}

\end{thm}

\medskip

The theorem is proved in Chapter 11.5 of \cite{presutti}, it would take too long to enter into its proof and we just refer to \cite{presutti}.

\medskip

%
Recalling \eqref{10.16} and \eqref{10.17} for notation we
state and prove a first corollary of Theorem \ref{thmE1}:

\begin{corollary}
  \label{coroE2}
For any $\ga \le \ga^*$ (see Theorem \ref{thmE1})
and $t\in [0,1]$  there is a unique DLR measure $\mu_t$ on $\Theta \equiv 1$ with hamiltonian $t[H_\ga+U] +(1-t) {H_\ga^{\rm free}}$, $U =\{u_{\Delta,\ga}\}$. Moreover for any $\si_{\La_{n,\pm}^c}$ with $\Theta\equiv 1$ there is a unique
semi-infinite DLR measures    $\mu^0_{t,\La_{n,\pm},\si_{\La_{n,\pm}^c}}$
on the plus ensemble $\Theta \equiv 1$.

\end{corollary}

\noindent
{\bf Proof.}  We will first prove the statement relative to $\mu_t$.
Let   $f$ be any cylindrical function so that there \red{exists a constant } $c_f\le 2\|f\|_\infty$ and  a set $B_f$ (union of intervals $C_i^-$)   such that
     \begin{equation}
   \label{E.13.0}
   | f(\si')-f(\si'')| \le c_f \sum_{C^-_{i}\subset B_f} d_{C^-_{i}}(\si',\si'')
     \end{equation}
Then uniqueness of $\mu_t$ follows from
   \begin{equation*}
\lim_{A \nearrow \mathbb Z}\sup_{\si'_{A^c},\si''_{A^c}}| E_{\mu_{t,A,\si'_{A^c}}}[f]
-E_{\mu_{t,A,\si''_{A^c}}}[f]|=0
     \end{equation*}
the sup being on configurations with $\Theta\equiv 1$ and
$\mu_{t,A,\si_{A^c}}$ the Gibbs measure in $A$, boundary conditions
$\si_{A^c}$ and energy $U =\{u_{\Delta,\ga}\}$.
%
Let $A =[-n\ell^+_\ga, n\ell^+_\ga)$ with $n$ so large that $B_f \subset A$, then by Theorem \ref{thmE1}
        \begin{equation*}
E_{\mu_{t,A ,\si'_{A ^c}}}[f]
-E_{\mu_{t,A ,\si''_{A ^c}}}[f] = \mathcal E [ f(\si')-f(\si'')]
     \end{equation*}
and by \eqref{E.13.0}
    \begin{equation}
      \label{E.5}
| E_{\mu_{t,A ,\si'_{A^c}}}[f]
-E_{\mu_{t,A ,\si''_{A^c}}}[f]|\le  c_f  \sum_{C_i^-\subset B_f}
\sum_{C^-_j \subset A ^c}c_1 e^{-b_1 \;\ga\;{\rm dist}(C^-_i,C^-_j)}
2\delta\ga^{-1}
     \end{equation}
which vanishes as $A \nearrow \mathbb Z$.  Translation invariance by $k\ell^+_\ga$ follows from uniqueness.

The  proof  for $\mu^0_{t,\La_{n,-}}$ is
similar.   In this case $A=A_k=[-k\ell^+_\ga,n\ell^+_\ga)$, $U'=U''$ with many body potentials $\{u_{\Delta,\ga}\mathbf 1_{\Delta \subseteq \La_{n,-}^*}\}$, $u_{\Delta,\ga}$ as in \eqref{10.6}.  We then have to bound
        \begin{equation*}
E_{\mu^0_{t,A_k,\si'_{(-\infty,-k\ell^+_\ga)}, \si_{[n\ell^+_\ga,(n+1)\ell^+_\ga)}}}[f]
-E_{\mu^0_{t,A_k,\si''_{(-\infty,-k\ell^+_\ga)}, \si_{[n\ell^+_\ga,(n+1)\ell^+_\ga)}}}[f]
     \end{equation*}
The important point is that the boundary spins differ only when $x \le -k\ell^+_\ga$
as they coincide for $x \ge n\ell^+_\ga$.  Then the same argument used for $\mu_t$ applies also in this case.  The proof for $\mu^0_{t,\La_{n,+}}$ is completely analogous and
the corollary is proved.  \qed

\vskip.5cm

We will prove  that the series defining $F^{(2)}_\ga(\si_{\La_{n,-}^c})$ is absolutely convergent and that $F^{(2)}_\ga(\si_{\La_{n,-}^c})$ satisfies the inequality \eqref{10.3}, the proof of the same statement for  $F^{(1)}_\ga(\si_{\La_{n,-}^c})$ is analogous and omitted.  Let $C_i^- \subset \La_n$ and
      \begin{equation}
      \label{E.6}
g_i(\si)  = \sum_{x \in C_i^-} \si(x), \quad
h_i (\si) = \frac 12
 \sum_{x \in C_i^-, y \in \La_{n,-}} J_\ga(x,y)\si(x)\si(y)
     \end{equation}
     Recalling the definition of $k_{x}$ and $k_{\La_{n},x}$,
     {(see \eqref{10.12}, \eqref{10.14}) in \eqref{10.18}, } we define
\begin{eqnarray}
      \label{E.7}
R_{1,t}&=& \sum_{C_i^- \subset \La_{n,-}} \{|E_{\mu^0_{t, \La_{n,-}; \si_{\La_{n,-}^c}}}
[ g_i ]-E_{\mu_{t}}[g_i]|+  | E_{\mu^0_{t, \La_{n,-}; \si_{\La_{n,-}^c}}}
[ h_i]-E_{\mu_{t}}[h_i]|\}
\nn\\&+& \sum_{\Delta \subseteq \La_{n,-} }|  E_{\mu^0_{t, \La_{n,-}; \si_{\La_{n,-}^c}}}
[ u_{\Delta,\ga}]-E_{\mu_{t}}[ u_{\Delta,\ga}] |
     \end{eqnarray}
\begin{eqnarray}
      \label{E.8}
R_{2,t}&=& \sum_{x \in \La_{n,-},y \notin \La_{n,-}} J_\ga(x,y)\{|E_{\mu^0_{t, \La_{n,-}; \si_{\La_{n,-}^c}}}
[ \si(x)\si(y)]|+ |E_{\mu_{t}}[ \si(x)\si(y)]|\}
     \end{eqnarray}
  \begin{eqnarray}
      \label{E.9}
&& \hskip-3cm R_{3,t} = \sum_{\Delta:\Delta\subset \La_{n,-}^*;
\Delta\cap  \La_{n,-}^* \setminus \La_{n,-} \ne \emptyset}
 \|u_{\Delta,\ga}\|_\infty +
  \sum_{\Delta:\Delta\cap  \La_{n,-}\ne \emptyset;
\Delta\cap  \La_{n,-}^c\ne \emptyset}\| u_{\Delta,\ga}\|_\infty
     \end{eqnarray}
We will prove that
       \begin{equation}
       \label{E.10}
 |F^{(2)}(\si_{\La_{n,-}^c}) | \le  \int_{0}^{1}dt  \sum_{i=1}^3 R_{i,t} < \infty
    \end{equation}
The first inequality follows directly from the definition  \eqref{10.18} of $F^{(2)}(\si_{\La_{n,-}^c})$ and the   definition of the terms $R_{i,t}$, we thus have to bound the latter.

 We have
    $$
     R_{2,t}\le 2\sum_{x \in \La_{n},y \notin \La_{n}} J_\ga(x,y)
     \le c \ga^{-1}
     $$
By \eqref{10.7} $R_{3,t}$ is bounded by
     $$
   2  \sum_{\Delta:\Delta\cap  \La_{n,-}\ne \emptyset;
\Delta\cap  \La_{n,-}^c\ne \emptyset}|u_{\Delta,\ga}|
\le 2 ce^{-(b/2)\ga^{-1}5}2\sum_{k>0,k'>0}e^{-(b/2)\ga^{-1}(k+k')}
\le c'e^{-(b/2)\ga^{-1}5}
     $$
To bound $R_{1,t}$ we will use the following lemma:

 \medskip

 \begin{lemma}
   \label{lemmaE3.0}
Let $f$ be a cylindrical function on $\La_{n,-}$ and let $c_f$ and $B_f$ as in \eqref{E.13.0}.
Then:
        \begin{equation}
   \label{E.14.0}
| E_{\mu^0_{t, \La_{n,-}; \si_{\La_{n,-}^c}}}[f] -  E_{\mu_{t}}[f]|
\le c_f \sum_{C^-_{i}\subset B_f} \sum_{C^-_{j}\subset \La_{n,-}^c} 2\delta \ga^{-1}c_1e^{-b_1 \ga {\rm dist}(C^-_{i},C^-_{j})}
    \end{equation}

\end{lemma}

\medskip

\noindent
{\bf Proof.}   Let $A_k= [-k \ell^+_\ga,n\ell^+_\ga)$ (eventually $k\to \infty$), then
        \begin{eqnarray*}
 &&  E_{\mu^0_{t, \La_{n,-}; \si_{\La_{n,-}^c}}}[f] -  E_{\mu_{t}}
 [f] = \int \mu_{t, \La_{n,-}; \si_{\La_{n,-}^c}}(d\si')
 \mu_{t}(d\si'')\\&& \hskip2cm \times
 \Big( E_{\mu'_{t,A_k,\si'_{A_k^c}}}[f]
-E_{\mu''_{t,A_k,\si''_{A_k^c}}}[f]\Big)
          \end{eqnarray*}
where $\mu'$ is defined with the hamiltonian $U'=\{u_{\Delta,\ga} \mathbf 1_{\Delta \subset \La_{n,-}^*}\}$ while $\mu''$ is defined with the hamiltonian $U''=\{u_{\Delta,\ga} \}$.  Let $\mathcal E$ be the expectation relative to the coupling of Theorem \ref{thmE1}, then
        \begin{eqnarray*}
 && E_{\mu'_{t,A_k,\si'_{A_k^c}}}[f]
-E_{\mu''_{t,A_k,\si''_{A_k^c}}}[f] = \mathcal E \Big[f(\si'_{A_k})-f(\si''_{A_k})\Big]
          \end{eqnarray*}
Since $d_{C_i^-}\le 2 \delta \ga^{-1}$,  by Theorem \ref{thmE1}
     \begin{equation*}
| E_{\mu'_{t, \La_{n,-}; \si'_{A_k^c}}}[f] -  E_{\mu''_{t, \La_{n,-}; \si''_{A_k^c}}}[f]|
\le c_f \sum_{C^-_{i}\subset B_f} \sum_{C^-_{j}\subset A_k^c} 2\delta \ga^{-1}c_1e^{-b_1 \ga {\rm dist}(C^-_{i},C^-_{j})}
    \end{equation*}
 which proves \eqref{E.14.0} by letting $k\to \infty$. \qed

\bigskip

We will apply Lemma \ref{lemmaE3.0} with $f$ equal to $g_i$, $h_i$ (see
\eqref{E.6}) and $\Delta$.

 \medskip

 \begin{lemma}
   \label{lemmaE3.00}
For $f=g_i$ the coefficient $c_f$ is equal to 1 and $B_f=C_i^-$.
For $f=h_i$,  $c_f =1 $ and
    $$
    B_f= \bigcup_{{C_j^- \subset \La_{n,-}},{\rm dist}(C_j^-,C_i^-) \le \ga^{-1}}C_j^-
    $$
For $f=u_{\Delta,\ga}$, $c_f = 2\| u_{\Delta,\ga}\|_\infty$ and
$B_f=\Delta$.

\end{lemma}

\medskip

\noindent
{\bf Proof.} The statements for $g_i$ and $u_\Delta$ are obviously true.  The proof for $h_i$ is as follows. By
\eqref{E.6}
      \begin{equation*}
h_i (\si') -h_i (\si'')=
 \frac 12
 \sum_{x \in C_i^-, {y \in \La_{n,-}}}J_\ga(x,y) \Big(\si'(x)[\si'(y)-\si''(y)]
 +[\si'(x)-\si''(x)]\si''(y)\Big)
     \end{equation*}
 Since   $J_\ga(x,y) =0$ if $|x-y| > \ga^{-1}$ and $\sum_x J_\ga(x,y)=1$
          \begin{equation*}
| \sum_{x \in C_i^-, y \in \La_{n,-}}J_\ga(x,y)  \si'(x)[\si'(y)-\si''(y)]|
 \le  \sum_{{C_j^- \subset \La_{n,-}}}\mathbf 1_{{\rm dist}(C_j^-,C_i^-) \le \ga^{-1}}
  d_{C^-_{j}}(\si',\si'')
     \end{equation*}
     Thus
          \begin{equation*}
|h_i (\si') -h_i (\si'')| \le
 \frac 12 \Big( d_{C^-_{i}}(\si',\si'') +\sum_{{C_j^- \subset \La_{n,-}}}\mathbf 1_{{\rm dist}(C_j^-,C_i^-) \le \ga^{-1}}
  d_{C^-_{j}}(\si',\si'')\Big)
     \end{equation*}
hence the thesis.  \qed

\bigskip

As a corollary of Lemma \ref{lemmaE3.0} and Lemma \ref{lemmaE3.00} we have:

 \medskip

 \begin{corollary}
   \label{lemmaE3}
There are $c'$ and $b'$ so that for any $\ga \le \ga^*$ (see Theorem \ref{thmE1})
and $t\in [0,1]$ 
         \begin{eqnarray}
      \label{E.11}
 &&    | E_{\mu^0_{t, \La_{n,-}; \si_{\La_{n,-}^c}}}[g_i] -  E_{\mu_{t}}
 [g_i] | \le 2\delta \ga^{-1}\sum_{C_j^- \subset \La_{n,-}^c}  c_1  e^{-b_1  \ga\, {\rm dist}(C^-_i,C_j^-)}\\&&
     | E_{\mu^0_{t, \La_{n,-}; \si_{\La_{n,-}^c}}}[h_i] -  E_{\mu_{t}}
 [h_i] | \le \sum_{C^-_{i'}\subset \La_{n,-}:{\rm dist}(C^-_{i'},C^-_{i})\le \ga^{-1}}
 2\delta \ga^{-1} \sum_{C_j^- \subset \La_{n,-}^c}  c_1  e^{-b_1  \ga\, {\rm dist}(C^-_{i'},C_j^-)}  \nn
          \end{eqnarray}
where $h_i$ and $g_i$ are defined in \eqref{E.6}.
Moreover for any $\Delta \subseteq    \La_{n,-}$
      \begin{equation}
   \label{E.12}
   | E_{\mu^0_{t, \La_{n,-}; \si_{\La_{n,-}^c}}}
[u_{\Delta,\ga}] -  E_{\mu_{t}}
 [ u_{\Delta,\ga}] | \le 2\| u_{\Delta,\ga}\|_\infty \sum_{C^-_{i}\subset \Delta}
 2\delta \ga^{-1}\sum_{C_j^- \subset \La_{n,-}^c}  c_1  e^{-b_1  \ga\, {\rm dist}(C^-_i,C_j^-)}
     \end{equation}
\end{corollary}

%
%

\medskip

 \medskip

\noindent
{\bf Proof of the first inequality in \eqref{10.3}}.

We will next prove that $R_{1,t} \le c \ga^{-1}$ which together with the analogous bounds already proved for $R_{2,t}$ and $R_{3,t}$ yields via \eqref{E.10}
that $|F^{(2)}(\si_{\La_{n,-}^c}) | \le c \ga^{-1}$.

Let $m$ be the integer such that $(n+1)\ell^+_\ga = m\ell^-_\ga$.  Then in \eqref{E.11} $i\le m-1$,
$j\ge m$ and  ${\rm dist}(C^-_i,C_j^-)= \delta \ga^{-1}(j-i-1)$ so that
          \begin{equation*}
\sum_{C_i^- \subset \La_{n,-}} | E_{\mu^0_{t, \La_{n,-}; \si_{\La_{n,-}^c}}}[g_i] -  E_{\mu_{t}}
 [g_i] | \le 2\delta \ga^{-1}\sum_{j\ge m, i\le m-1}  c_1  e^{-b_1 \delta(j-i-1)} \le c'\ga^{-1}
     \end{equation*}
The sum over $C_i^- \subset \La_{n,-}$ of the right hand side in the second inequality in \eqref{E.11} is bounded by
         \begin{equation}
      \label{E.14bis}
 2\delta \ga^{-1}\sum_{i'\le m-1}[c''\delta^{-1}]
 \sum_{j\ge m}  c_1  e^{-b_1  \delta(j-i'-1)}
     \end{equation}
where $c''\delta^{-1}$ bounds the number of intervals $C^-_i$ which have distance $\le \ga^{-1}$ from $C_{i'}^-$.  Thus
         \begin{equation*}
\sum_{C_i^- \subset \La_{n,-}}  | E_{\mu^0_{t, \La_{n,-}; \si_{\La_{n,-}^c}}}[h_i] -  E_{\mu_{t}}
 {[h_i]} |  \le c'\ga^{-1}
     \end{equation*}
Finally by \eqref{E.12}
        \begin{equation}
          \label{E.15bis}
 \sum_{\Delta \subseteq \La_{n,-} }|\{E_{\mu^0_{t, \La_{n,-}; \si_{\La_{n,-}^c}}}
[ u_{\Delta,\ga}]-E_{\mu_{t}}[ u_{\Delta,\ga}]\}| \le K
2\delta \ga^{-1}\sum_{i\le m-1,j\ge m}  c_1  e^{-b_1   \delta(j-i-1)}
   \end{equation}
where, given $ C^-_{i}$,  $K \ge 2\sum_{ \Delta \supset  C^-_{i}}
\| u_{\Delta,\ga}\|_\infty $. We are going to show that $K$ is finite:
by \eqref{10.7}
   $$
2\sum_{ \Delta \supset  C^-_{i}}
\| u_{\Delta,\ga}\|_\infty \le 2 \sum_{n \ge 5} n ce^{-b \ga^{-1}n}
   $$
The sum is finite and vanishing as $\ga\to 0$ hence the existence of $K$, so that by \eqref{E.15bis}
        \begin{equation*}
 \sum_{\Delta \subseteq \La_{n,-} }|\{E_{\mu^0_{t, \La_{n,-}; \si_{\La_{n,-}^c}}}
[ u_{\Delta,\ga}]-E_{\mu_{t}}[ u_{\Delta,\ga}]\}| \le c' \ga^{-1}
   \end{equation*}

 \qed

\vskip.5cm

We shall next prove that   $G^{(1)}_{\ga,n}(s_0,s_{n+1})$ as given in \eqref{10.20} satisfies the bound  \eqref{10.3} and thus complete the proof of Theorem \ref{10.1}.  We will prove the bound for $\Psi^{(2)}+\Psi^{(4)}$, the proof for $\Psi^{(1)}+\Psi^{(3)}$ is similar and omitted.  As in the proof of $F^{(2)}$ we write

      \begin{equation}
      \label{E.15}
\Psi^{(2)} \le \int_0^1 dt \sum_{i=1}^3 S_{i,t}
     \end{equation}
where {recalling that $x^*=\ell^+_\ga\left[\frac{n}{2}\right]$, (see after \eqref{10.20})}.

\begin{eqnarray}
      \label{E.16}
&&S_{1,t}= \sum_{C_i^- \subset \La_n, i\ell^-_\ga \ge x^*} \{|E_{\mu^0_{t, \La_{n}; \si_{\La_{n}^c}}}
[ g_i ]-E_{
\mu^0_{t,\La_{n,-}; \si_{\La_{n,-}^c}}}[g_i]|
+  | E_{\mu^0_{t, \La_{n}; \si_{\La_{n}^c}}}
[h_i ]-
E_{\mu^0_{t,\La_{n,-}; \si_{\La_{n,-}^c}}}
[h_i]|\}
\nn\\&& \hskip2cm + \sum_{\Delta \subseteq \La_{n}, \Delta \cap [x^*,\infty)\ne \emptyset}|E_{\mu_{t, \La_{n}; \si_{\La_{n}^c}}}[ u_{\Delta,\ga}]- E_{\mu_{t, \La_{n,-}; \si_{\La_{n,-}^c}}}
[ u_{\Delta,\ga}] |
 \\
      \label{E.17}
&&S_{2,t}= \sum_{x \in \La_{n},y \ge (n+1)\ell^+_\ga} J_\ga(x,y)|E_{\mu^0_{t, \La_{n}; \si_{\La_{n}^c}}}[ \si(x)]-E_{\mu^0_{t, \La_{n,-}; \si_{\La_{n,-}^c}}}
[ \si(x)]| \nn\\&& \hskip2cm + \sum_{\Delta \subseteq \La_n^*, \Delta \cap [(n+1)\ell^+_\ga,(n+2)\ell^+_\ga)\ne \emptyset}|E_{\mu^0_{t, \La_{n}; \si_{\La_{n}^c}}}[ u_{\Delta,\ga}]- E_{\mu^0_{t, \La_{n,-}; \si_{\La_{n,-}^c}}}
[ u_{\Delta,\ga}] |
 \\
      \label{E.18}
&& S_{3,t} =   2\sum_{\Delta:\Delta\subset  \La^*_{n};
\Delta\cap [0,\ell^+_\ga)\ne \emptyset,\Delta\cap [x^*,\infty)\ne \emptyset} \| u_{\Delta,\ga}\|_\infty  +\sum_{\Delta \subset \La_{n,-}^*:
\Delta\cap [-\ell^+_\ga,0)\ne \emptyset,\Delta\cap [x^*,\infty)\ne \emptyset} \| u_{\Delta,\ga}\|_\infty
     \end{eqnarray}

Let $f$ satisfy \eqref{E.14.0}, then proceeding as in the proof of Lemma \ref{lemmaE3.0} we write
        \begin{eqnarray*}
 &&  E_{\mu^0_{t, \La_{n}; \si_{\La_{n}^c}}}[f] -E_{\mu^0_{t, \La_{n,-}; \si_{\La_{n,-}^c}}}[f]
 = \int \mu^0_{t, \La_{n,-}; \si_{\La_{n,-}^c}}(d\si''_{(-\infty, \ell_\ga^+)})
 \\&& \hskip2cm \times
 \Big( E_{\mu'_{t, \La_{n}; \si_{(-\infty, \ell_\ga^+)}, \si_{[(n+1)\ell^+_\ga,\infty)}}}[f]
-E_{\mu''_{t,\La_n;\si''_{(-\infty, \ell_\ga^+)}, {\si_{[(n+1)\ell^+_\ga,\infty)}}}}[f]\Big)
          \end{eqnarray*}
where $\mu'$ is defined with the hamiltonian $U'=\{u_{\Delta,\ga} \mathbf 1_{\Delta \subseteq \La_{n}^*}\}$ while $\mu''$ is defined with the hamiltonian $U''=\{u_{\Delta,\ga} 1_{\Delta \subset \La_{n,-}^*}\}$.  To simplify notation we just write
   $$
   \mu'_t \equiv \mu'_{t, \La_{n}; \si_{(-\infty, \ell_\ga^+)}, \si_{[(n+1)\ell^+_\ga,\infty)}},\quad \mu''_t\equiv \mu''_{t,\La_n;\si''_{(-\infty, \ell_\ga^+)}, {\si_{[(n+1)\ell^+_\ga,\infty)}}}
   $$
By \eqref{E.14.0} and \eqref{E.1}
      \begin{eqnarray}
   \label{E.19}
| E_{\mu'_{t}}[f]
-E_{\mu''_{t}}[f] |
\le c_f \sum_{C^-_{i}\subset B_f} \sum_{C^-_{j}\subset (-\infty, \ell^+_\ga)} 2\delta \ga^{-1}c_1e^{-b_1 \ga {\rm dist}(C^-_{i},C^-_{j})}
    \end{eqnarray}
Then, by Lemma \ref{lemmaE3.00},
         \begin{eqnarray}
      \label{E.20}
 &&    \hskip-1cm |  E_{\mu'_{t}}[g_i] -E_{\mu''_{t}}[g_i] | \le 2\delta \ga^{-1}\sum_{C_j^- \subset (-\infty,\ell^+_\ga)}  c_1  e^{-b_1  \ga\, {\rm dist}(C^-_i,C_j^-)}\\&&
  \hskip-1cm   |E_{\mu'_{t}}[h_i] -E_{\mu''_{t}}[h_i] | \le \sum_{{C^-_{i'}\subset \La_n}:{\rm dist}(C^-_{i'},C^-_{i})\le \ga^{-1}}
 2\delta \ga^{-1}  \sum_{C_j^- \subset (-\infty,\ell^+_\ga)}  c_1  e^{-b_1  \ga\, {\rm dist}(C^-_{i'},C_j^-)}  \label{E.21}
          \end{eqnarray}
Let $m^*$, $m$  and $m'$ be the integers such that $m^*\ell^-_\ga= x^*=\ell^+_\ga\left[\frac{n}{2}\right]$, $m\ell^-_\ga= (n+1)\ell^+_\ga$ and $m'\ell^-_\ga = \ell^+_\ga$.
Then
          \begin{equation*}
\sum_{ i\in [m^*,m-1]} | E_{\mu^0_{t, \La_{n,-}; \si_{\La_{n,-}^c}}}[g_i] -  E_{\mu_{t}}
 [g_i]| \le 2\delta \ga^{-1}\sum_{j<m', i\in [m^*,m-1]}  c_1  e^{-b_1 \delta(i-j-1)}
     \end{equation*}
Calling $i=m^*+k$, $k\ge 0$, and $j= m'+k'$ the right hand side is bounded by
          \begin{equation*}
 2\delta \ga^{-1}c_1  e^{-b_1 \delta (m^*-m')}\sum_{k'<0, k\ge 0}    e^{-b_1 \delta(k-k'-1)}
 \le c' e^{-b_1 \ga[(n/2)-2]\ell^+_\ga}
     \end{equation*}
because   $\delta (m^*-m') = \ga (\delta\ga^{-1}) (m^*-m')= \ga (x^*-\ell^+_\ga)\ge \ga[(n/2)-2]\ell^+_\ga$.

The values of $i'$ in \eqref{E.21} are bounded from below by $i'\ge m^*-k_0$,
$k_0 = \delta^{-1} + 1$ (because of the condition ${\rm dist}(C^-_{i'},C^-_{i})\le \ga^{-1}$). Thus calling $i'=m^*+k$, $k\ge -k_0$,
$j=m'+k'$,
          \begin{eqnarray*}
\sum_{ i\in [m^*,m-1]} | E_{\mu^0_{t, \La_{n,-}; \si_{\La_{n,-}^c}}}[h_i] -  E_{\mu_{t}}
 [h_i]| &\le& 2\delta \ga^{-1}[c''\delta^{-1}] e^{-b_1 \delta (m^*-m')}
 \sum_{k'<0, k\ge -k_0}  c_1  e^{-b_1 \delta(k-k'-1)}\\ &\le &
 c \ga^{-1} e^{-b_1 \ga[(n/2)-2]\ell^+_\ga}
     \end{eqnarray*}
see \eqref{E.14bis} for the term $[c''\delta^{-1}]$.

We split the sum over $\Delta$ in the last term of \eqref{E.16} by distinguishing whether
$ \Delta$ is or is not in $ [x^*,\infty)$.  In the former case we proceed as for $F^{(2)}_\ga$ and get
 \begin{eqnarray*}
&& \sum_{\Delta \subset  \La_{n} \cap [x^*,\infty)}|E_{\mu^0_{t, \La_{n}; \si_{\La_{n}^c}}}[ u_{\Delta,\ga}]- E_{{\mu^0_{t, \La_{n,-}; \si_{\La_{n,-}^c}}}}
[ u_{\Delta,\ga}] | \le  K
2\delta \ga^{-1}\sum_{i\ge m^*,j<0}  c_1  e^{-b_1   \delta(j-i-1)}
     \end{eqnarray*}
which is then bounded by
 \begin{eqnarray*}
&&  K
2\delta \ga^{-1}c_1  e^{-b_1   \delta m^*}\sum_{k\ge0,j<0}  e^{-b_1   \delta(k-j-1)}
     \end{eqnarray*}
see \eqref{E.15bis}. It remains to consider the case where $\Delta = [k \ell^+_\ga,k'\ell^+_\ga)$, $0\le k < [\frac n2]$, $[\frac n2] \le k' < n$.  We tacitly suppose below that $k$ and $k'$ satisfy the above bounds, then calling
$\Delta=[k\ell^+_\ga,k'\ell^+_\ga)$
 \begin{eqnarray*}
&& \sum_{k,k', \Delta} |E_{\mu'_{t}}[ u_{\Delta,\ga}]- E_{\mu''_{t}}
[ u_{\Delta,\ga}] | \\&& \hskip2cm \le \sum_{k,k'}ce^{-b\ga^{-1}|k'-k|} \sum_{C_i^- \subset [k\ell^+_\ga,k'\ell^+_\ga)}
2\delta \ga^{-1}\sum_{j<0}  c_1  e^{-b_1   \delta(i-j-1)}
     \end{eqnarray*}
which is bounded by
\begin{eqnarray*}
&& \{ \sum_{k,k'}ce^{-(b/2)\ga^{-1}|k'-k|} e^{-(b/2)\ga^{\alpha} (x^*-k\ell^+_\ga)}\} \{\frac{ |k'-k|\ell^+_\ga}{ \ell^-_\ga}\}
2\delta \ga^{-1}\sum_{j<0}  c_1  e^{-b_1 [\ga k\ell^+_\ga +\delta(- j-1)]}
     \end{eqnarray*}
For $\ga$ small enough $(b/2)\ga^{\alpha} (x^*-k\ell^+_\ga) > b_1 \ga (x^*-k\ell^+_\ga)$, so that the above is bounded by
\begin{eqnarray*}
&& 2\delta \ga^{-1} c_1 e^{-b_1 \ga x^*} \{ \sum_{k,k'}\frac{ c|k'-k|\ell^+_\ga}{ \ell^-_\ga}e^{-(b/2)\ga^{-1}|k'-k|}
\} \sum_{j<0}  c_1  e^{-b_1 \delta(- j-1)]} \le  c'\ga^{-1}  e^{-b_1 \ga x^*}
     \end{eqnarray*}

Calling $\Om(x):=|E_{\mu^0_{t, \La_{n}; \si_{\La_{n}^c}}}[ \si(x)]-E_{\mu^0_{t, \La_{n,-}; \si_{\La_{n,-}^c}}}
[ \si(x)]|$ we bound
   $$
   \Om(x) \le \mathcal E  \Big[ d_{C_i^-}(\si'_A,\si''_A) \Big]
   $$
Recalling that
$\si_{\La_{n}^c}(x)={\si_{\La_{n,-}^c}(x)}$ for $x \in \La_{n,-}^c$ we get from \eqref{E.1}
\begin{eqnarray*}
 \sum_{x \in \La_{n},x \ge n\ell^+_\ga-\ga^{-1}} J_\ga(x,y)\Om(x)
\le \sum_{i=m-k_0}^m +\sum_{C^-_j \subset B^c}c_1 e^{-b_1 \;\ga\;{\rm dist}(C^-_i,C^-_j)} \le c' e^{-b_1 \ga (n\ell^+_\ga)}
    \end{eqnarray*}
For the second sum we use \eqref{E.19} with  $f=u_{\Delta,\ga}$.  We have
\begin{eqnarray*}
&& |u_{\Delta, \ga}(\si')-u_{\Delta, \ga}(\si'')| \le \mathbf 1_{\si'_{\Delta
\cap \La_n} \ne \si''_{\Delta \cap \La_n}}
ce^{-b \ga |\Delta|} \le ce^{-b \ga |\Delta|}\sum_{C_i^- \subset \Delta \cap \La_n} d_{C_i^-}(\si',\si'')
     \end{eqnarray*}
so that
\begin{eqnarray*}
&&\sum_{\Delta \subseteq [0,(n+1)\ell^+_\ga), \Delta \cap [n\ell^+_\ga,(n+1)\ell^+_\ga)\ne \emptyset}|E_{\mu^0_{t, \La_{n}; \si_{\La_{n}^c}}}[ u_{\Delta,\ga}]- E_{\mu^0_{t, \La_{n,-}; \si_{\La_{n,-}^c}}}
[ u_{\Delta,\ga}] |\\&& \hskip2cm \le\sum_{k\in [0,n)} 2\delta \ga^{-1} c_1 c\{e^{-(b/2)\ga^{-1}|n-k|- b_1 \ga k\ell^+_\ga}\}e^{-(b/2)\ga^{-1}|n-k|}
\frac{|n+1-k|\ell^+_\ga}{\ell^-_\ga}\\
&&\hskip3cm \times \sum_{j<0}  c_1  e^{-b_1 \delta(- j-1)]}\le c' \delta \ga^{-1} e^{-b_1 \ga^{\alpha} n \ell^+_\ga}
     \end{eqnarray*}

Finally
\begin{eqnarray*}
&&S_{3,t} \le 2\sum_{k'\ge n; k<0} c e^{-b \ga^{-1}|k'-k| } \le 2c e^{-b\ga^{\alpha} n \ell^+_\ga} \{ \sum_{k'>0,k<0} e^{-b\ga^{-1}(k'-k)}\}
\le c'  e^{-b \ga^{\alpha} n \ell^+_\ga}
     \end{eqnarray*}

$\Psi^{(4)}$ is like $F^{(2)}$ but with the sum restricted to $x<x^*$.  The analysis
of the many terms which contribute to $\Psi^{(4)}$ are therefore like the corresponding ones for $F^{(2)}$ but we need to exploit the fact that the conditioning where the Gibbs measures differ is far away.  This is done with the same ideas used so for $\Psi^{(2)}$, we omit the details.  Also the analysis of
$\Psi^{(6)}$ is similar to that of terms already considered and it is omitted.

\vskip2cm

\setcounter{equation}{0}

\section{Bounds on restricted partition functions}
\label{BRPF}

In this appendix we will bound the restricted partition functions introduced in Section \ref{sec:20} namely: $Z_{\La_n}^{\rm pbc,\mathcal X^0}$, $Z_{\La_n}^{\rm pbc,\mathcal X^+}$, $Z_{\La_n}^{\rm pbc,\mathcal X^-}$ see \eqref{20.0.222} and
$Z_{\La_n}^{\rm pbc, gb}$ see \eqref{20.0.8}.  We start from $Z_{\La_n}^{\rm pbc,\mathcal X^+}$.

\medskip

\subsection{Bounds on $Z_{\La_n}^{\rm pbc,\mathcal X^{\pm}}$}

By  the spin flip symmetry it is sufficient to bound $Z_{\La_n}^{\rm pbc,\mathcal X^+}$.  Let
\begin{equation}
     \label{U.1}
Z_{\La_n;0}^{\rm pbc}  := \sum_{\Theta_i \ge 0, \Theta_0=1} \sum_{s_{\La_n}} e^{-\beta {H^{\rm pbc}_\ga(s_{\La_n})}}
    \end{equation}
then
\begin{equation}
     \label{U.2}
Z_{\La_n}^{\rm pbc,\mathcal X^+} \le (2n+1)Z_{\La_n;0}^{\rm pbc}
    \end{equation}
Recalling that $H^{\rm pbc}_\ga(s_{\La_n})$ is defined with periodic boundary conditions
so that $s_0= s_{2n+1}$
\begin{equation}
     \label{U.3}
Z_{\La_n;0}^{\rm pbc}  := \sum_{s_0: \theta(s_0)=1}e^{-\beta H^{\rm pbc}_\ga(s_0)}
\sum_{s_{[1,2n]}}\mathbf 1_{{\theta_1=\theta_{2n}=1}; \Theta_i \ge 0}
e^{-\beta H^{\rm pbc}_\ga(s_{[1,2n]}|s_0,s_{2n+1})}
    \end{equation}
By Theorem \ref{thm10.1} there is a constant $c>0$ so that
\begin{equation}
     \label{U.4}
Z_{\La_n;0}^{\rm pbc}  \le \sum_{s_0: \theta(s_0)=1}e^{-\beta H^{\rm pbc}_\ga(s_0)}
e^{\beta p^+_\ga(2n)\ell^+_\ga+c\ga^{-1}}
    \end{equation}
so that $Z_{\La_n;0}^{\rm pbc}  \le e^{\beta p^+_\ga(2n)\ell^+_\ga+c' \ell^+_\ga}$ ($c'$ a constant).  In conclusion:
\begin{equation}
     \label{U.5}
Z_{\La_n}^{\rm pbc,\mathcal X^+} \le \{(2n+1)e^{c' \ell^+_\ga}\}  e^{\beta p^+_\ga(2n+1)\ell^+_\ga}
    \end{equation}

\medskip

\subsection{Bound on $Z_{\La_n}^{\rm pbc,\mathcal X^{0}}$}

We will reduce the bound of $Z_{\La_n}^{\rm pbc,\mathcal X^{0}}$   to that of
$Z_{\La_n}^{\rm pbc,\mathcal X^+}$.  A better bound could be derived by using the Lebowitz-Penrose coarse graining techniques and estimates on the associated free energy functional but the bound below is faster and simpler.

We call $\La'_n$ the set $\La_n$ without the points $-1,0,1$.  We then have
\begin{equation}
     \label{U.6}
Z_{\La_n}^{\rm pbc,\mathcal X^{0}}= \sum_{s'_0,s'_{\pm 1}}e^{-\beta H^{\rm pbc}_\ga(s'_0,s'_{\pm 1})}
\sum_{s_{\La'_n}}
e^{-\beta H^{\rm pbc}_\ga(s_{\La'_n}|s'_0,s'_{\pm 1})} {\mathbf 1_{ \Theta_i = 0, i=1,..,2n}}
    \end{equation}
with $\Theta_i$ computed on the configuration $s_{\La'_n},s'_0,s'_{\pm 1}$.
Let $s_0,s_{\pm 1}$ be a configuration with $\theta_i=1$, $i=0,\pm 1$.
We have
\begin{equation}
     \label{U.7}
|H^{\rm pbc}_\ga(s_{\La'_n}|s'_0,s'_{\pm 1})-H^{\rm pbc}_\ga(s_{\La'_n}|s_0,s_{\pm 1})| \le  c_1 \ga^{-1}
    \end{equation}
\begin{equation}
     \label{U.8}
H^{\rm pbc}_\ga(s_0,s_{\pm 1}) \ge - c_2 \ell^+_\ga, \quad  \sum_{s'_0,s'_{\pm 1}}e^{-\beta H^{\rm pbc}_\ga(s'_0,s'_{\pm 1})} \le e^{c_3 \ell^+_\ga}
    \end{equation}
We thus get from \eqref{U.6}
\begin{equation}
     \label{U.9}
Z_{\La_n}^{\rm pbc,\mathcal X^{0}} \le e^{c_3 \ell^+_\ga}e^{\beta c_1 \ga^{-1}}
 e^{ \beta c_2 \ell^+_\ga}
{Z_{\La_n}^{\rm pbc,\mathcal X^{+}}}
    \end{equation}
and using the bound for ${Z_{\La_n}^{\rm pbc,\mathcal X^{+}}}$ proved before,
we get for a suitable constant $c''$:
\begin{equation}
     \label{U.10}
Z_{\La_n}^{\rm pbc,\mathcal X^{0}} \le e^{c'' \ell^+_\ga}
 e^{\beta p^+_\ga(2n+1)\ell^+_\ga}
    \end{equation}

\medskip

\subsection{Bound on $Z_{\La_n}^{\rm pbc, gb}$}

{Recalling \eqref{20.0.8}:}
\begin{equation}
     \label{G9-0}
     \frac {Z_{\La_n}^{\rm pbc, gb}} {e^{\beta p_\ga (2n+1)\ell^+_\ga}}=
   {\frac{1}{e^{\la_\ga(2n+1)}}}\sum_{(x,\und u) \in A} w^{(b)}(\und u)
    \end{equation}
We will next prove that
\red{that there exists $\zeta_\ga>0$ so that: }

\begin{equation}
     \label{G9}
   \lim_{n\to\infty} {\frac{1}{e^{\la_\ga(2n+1)-\red{\zeta_\ga(2n+1)}}}}\sum_{{(x,\und u) \in A}} w^{(b)}(\und u) =0
    \end{equation}
{where $w^{(b)}(\und u)$ is defined in \eqref{20.0.9} and $A$ in \eqref{20.0.1}} {($A$ depends on $n$)}

We write:
\begin{equation}\label{G9.0}
  \sum_{(x,\und u) \in A} w^{(b)}(\und u)=\sum_{(x,\und u) \in A^{\ge 3}} w^{(b)}(\und u)+\sum_{(x,\und u) \in A^{<3}} w^{(b)}(\und u)
\end{equation}
where:
\begin{equation}\label{Age3}
A^{\ge 3}:= \{(x,\und u): |\und u|=2n+1; \;  \exists \bar\ell: u_{\bar\ell,1}\ge 3 ;\; x+ u_1-1 \ge n+1,\;\rm{when}\; x > -n \}
\end{equation}
and $A^{<3}$ the complementary set.

\vskip 1cm
Let $u\ge 3$, recalling \eqref{10.3} and notation in \eqref{20.0.9} we write:
\begin{eqnarray}
      \label{G10}
\frac{K_{u}(s ,s') e^{A^{(1)}_{\ga,u-2}}}{e^{A^{(1)}_{\ga,u-2}}}\mathbf 1_{u\ge 3}&\le& e^{A^{(1)}_{\ga,u-2}} \Big(  \{e^{G^{(1)}_{\ga,u-2}(s ,s') }
- e^{A^{(1)}_{\ga,u-2}}\} e^{-A^{(1)}_{\ga,u-2}}\Big)\mathbf 1_{u\ge 3}
\\
&\le& 4  a {e^{-b_0\ga\ell_\ga^+}  } \cdot e^{A^{(1)}_{\ga,u-2}} \mathbf 1_{u\ge 3}
     \end{eqnarray}
We then use this estimate in the first sum in \eqref{G9.0} on $u_{\bar\ell,1}$ where $\bar\ell$ is defined in \eqref{Age3}:
\begin{equation}\label{ge3}
  w^{(b)}(\und u)\le  4  a {e^{-b_0\ga\ell_\ga^+}  }  w(\und u)
\end{equation}
Substituting in the first term of \eqref{G9.0}, we get by \eqref{77.2}:
\begin{equation}\label{ge3f}
 {\frac{1}{e^{\la_\ga (2n+1)}}}\sum_{(x,\und u) \in A^{\ge 3}} w^{(b)}(\und u)\le 4  a {e^{-b_0\ga\ell_\ga^+}  } (2n+1)
  \sum_{\und u : |\und u|\ge 2n+1 } {w_{\la_\ga}(\und u)}\le  4  a {e^{-b_0\ga\ell_\ga^+}  }(2n+1)e^{-\delta_\ga (2n+1)}
\end{equation}
\red{which vanishes exponentially } when $n\to \infty$.

Let consider other term of \eqref{G9.0}. In this case $u_{\ell,1}<3$ for any $\ell$.
We consider $u_{1,1}$ and bound $K_{u_{1,1}}(s ,s') $ as follows:
\begin{eqnarray}
      \label{G11}
\frac{K_{u}(s ,s') e^{A^{(1)}_{\ga,u-2}}}{e^{A^{(1)}_{\ga,1}}}\mathbf 1_{u<3}&\le& e^{A^{(1)}_{\ga,1}} \Big( e^{V_u^1} e^{-A^{(1)}_{\ga,1}}\Big)\mathbf 1_{u\le 2}
\\
&\le& {e^{-b_0\ga\ell_\ga^+}  } \cdot e^{A^{(1)}_{\ga,1}} \mathbf 1_{u\le 2}
     \end{eqnarray}

     Since $\sum_{u\le 2}\le 2$
\begin{equation}\label{le3f}
{\frac{1}{e^{\la_\ga (2n+1)}}}  \sum_{(x,\und u) \in A^{< 3}} w^{(b)}(\und u)\le 2 {e^{-b_0\ga\ell_\ga^+}  }\cdot 2\cdot
  \sum_{\und u : |\und u|\ge 2n+1 } {w_{\la_\ga}(\und u)}\le  4 {e^{-b_0\ga\ell_\ga^+}  } e^{-\delta_\ga (2n+1)}
\end{equation}
Collecting \eqref{le3f} and \eqref{ge3f} we get
\eqref{G9}.

\vskip2cm

\setcounter{equation}{0}
\section{Instanton  and proof of { Theorem \ref{thm111.2}}}
\label{appF}
\subsection{Coarse graining}
We define the coarse graining maps
$\phi_{\rm cg}: \{-1,1\}^{\mathbb Z} \to [-1,1]^{\mathbb Z}$ and $\psi_{\rm cg}: [-1,1]^{\mathbb Z} \to L^{\infty}(\mathbb R)$
by setting
       \begin{equation}
   \label{F.1}
\phi_{\rm cg}(\si) = \und m=(m_i)_{i\in \mathbb Z},\quad m_i = \ga^{1/2}\sum_{x\in [i\ga^{1/2},(i+1)\ga^{1/2})}\si(x)
    \end{equation}
       \begin{equation}
   \label{F.2}
\psi_{\rm cg}(\und m)(r) =  m_i,\quad \text{when $r\in [i\ga^{1/2},(i+1)\ga^{1/2})$ }
    \end{equation}
We denote by $\mathcal X_\La$ the subspace of $L^\infty(\La, [-1,1])$ made of functions on $\La$
which are constant on the intervals $[i\ga^{1/2},(i+1)\ga^{1/2})$, we tacitly suppose that $\La$ is union of such intervals and by default any function $m$ that we will consider in the sequel is (unless otherwise stated) in $\mathcal X_\La$ for some $\La$.  When $\La= \mathbb R$ we simply write $\mathcal X$.

Let $\und m$ be in the range of $\phi_{\rm cg}$ and, recalling \eqref{12.3} for notation,
       \begin{equation}
   \label{F.3}
Z_{n,\und m,s_{[1,n]^c}} = \sum_{s_{[1,n]}}\mathbf 1_{\phi_{\rm cg}(s_{[1,n]},s_{[1,n]^c})=\und m}
e^{-\beta H_\ga(s_{[1,n]}|s_{[1,n]^c})}
\end{equation}
Then (see for instance Theorem 4.2.2.2 in \cite{presutti}) there is a constant $c$ so that
       \begin{eqnarray}
   \label{F.4}
\log Z_{n,\und m,s_{[1,n]^c}} &\le& - \ga^{-1}\beta F(m_{[\ga^{-\alpha},(n+1)\ga^{-\alpha})} | m_{[\ga^{-\alpha},(n+1)\ga^{-\alpha})^c }) \nn\\&+& c n\ell^+_\ga\ga^{1/2 } \log \ga^{-1}
 \end{eqnarray}
where  $m(\cdot)= \psi_{\rm cg}(\und m)$ and
$ F(m_\La | m_{\Delta})$, $\La\cap \Delta= \emptyset$, is the Lebowitz-Penrose functional
       \begin{eqnarray}
   \label{F.5}
&&F(m_\La | m_{\Delta}) = F(m_\La)- \int_\La dr\int_{\Delta} dr'
 J(|r-r'|)m(r)m(r')\\&& F(m_\La)=\int_{\La} \frac{-1}{\beta} S(m) dr -\frac 12 \int_\La dr \int_{\La} dr'
 J(|r-r'|)m(r)m(r')\nn
\end{eqnarray}
The entropy $S(m)$ is defined after \eqref{intro.3}. The proof of \eqref{F.4} simply follows from the smoothness of the interaction kernel $J(|r-r'|)$ and the Stirling formula.

In analogy with \eqref{12.4} we  define new phase indicators for functions
$m\in L^{\infty}(\mathbb R, [-1,1])$ by setting for any $i\in \mathbb Z$:
     \begin{eqnarray}
      \label{F.6}
&&\eta^*_i =
  \pm 1 \;\;\text{if}\;\;   \big| \delta^{-1} \int_{i\delta}^{(i+1)\delta}dr\,[ m(r)\mp m_\beta]\big| \le \zeta,\quad
  \text{$\eta^*_i=0$ otherwise}
    \end{eqnarray}
and then, like in \eqref{12.5}--\eqref{12.6},

      \begin{eqnarray}
      \label{F.7}
&&\theta^*_i   =
  \pm 1 \;\;\text{if}\;\; \eta^*_j \equiv \pm 1 \;\;\text{for all}\;\; j: [\delta j,\delta(j+1))  \subset [\ga^{-\alpha}i,\ga^{-\alpha}(i+1))\nn\\&&\hskip3cm \text{$\theta^*_i =0$ otherwise}
  \\&&
  \Theta^*_i =
  \pm 1 \;\;\text{if}\;\; \theta^*_j  \equiv \pm 1,\; j=i-1,i, i+1,\quad \text{$\Theta^*_i (\und s)=0$ otherwise}
    \label{F.8}
  \end{eqnarray}
Since $\eta^*_i$, $\theta^*_i$ and $\Theta^*_i$  {are respectively equal to $\eta_i$, $\theta_i$ and $\Theta_i$  when $m = \psi_{\rm cg}(\phi_{\rm cg}(\si))$}, then the statistical weight $Z^{-,+}_{\ga,n}$ of a $-+$ interface is bounded by
         \begin{eqnarray}
      \label{F.9}
&&  \log   Z^{-,+}_{\ga,n}(s_0,s_{n+1}) \nn\\&&  \hskip1cm
\le
- \ga^{-1}\beta   \inf_{\theta^*_n= 1=-\theta^*_1; \Theta^*_i =0, i=1,..,n} F(m_{[\ga^{-\alpha},(n+1)\ga^{-\alpha})} | m_{[0,\ga^{-\alpha})},m_{[(n+1)\ga^{-\alpha},(n+2)\ga^{-\alpha})})\nn
\\&& \hskip1cm+
{\frac{n\ell^+_\ga}{\ga^{-1/2}}\log (2\ga^{-1/2}+1)}
  + c n\ell^+_\ga\ga^{1/2 } \log \ga^{-1}
     \end{eqnarray}
the inf being on functions in $\mathcal X_{[\ga^{-\alpha},(n+1)\ga^{-\alpha})}$ (defined at the beginning of this appendix) and which verify the constraint stated in the argument of the inf.  To get \eqref{F.9} we have used that the cardinality of all $\und m$ is $(2\ga^{-1/2} +1)^{n\ell^+_\ga \ga^{1/2}}$

\vskip.5cm

\subsection{An equilibrium variational problem}

Since the inf in \eqref{F.9} is over $m$ which have
$\theta^*_1=-1$ and $\theta^*_n=1$
in the intervals $[\ga^{-\alpha},2\ga^{-\alpha})$ and  respectively $[n\ga^{-\alpha},(n+1)\ga^{-\alpha})$, then $m$ in the above intervals
is in the minus, plus, equilibrium phase.  In Theorem 6.3.3.1 in \cite{presutti} it is proved that in an equilibrium phase the minimizer of the free energy approaches exponentially fast (away from the boundary conditions) the equilibrium value  $\pm m_\beta$, respectively.  Thus

\medskip

\noindent
{\bf Theorem.} There are $a$ and $c$ positive so that
         \begin{eqnarray}
      \label{F.10}
&& \inf_{\theta^*_n= 1=-\theta^*_1; \Theta^*_i =0, i=1,..,n} F(m_{[\ga^{-\alpha},(n+1)\ga^{-\alpha})} | m_{[0,\ga^{-\alpha})},m_{[(n+1)\ga^{-\alpha},(n+2)\ga^{-\alpha})})\nn
\\&& \hskip1cm \ge \text{$\inf\,^*$}
F(m_{[\ga^{-\alpha},(n+1)\ga^{-\alpha})} | m_{[0,\ga^{-\alpha})},m_{[(n+1)\ga^{-\alpha},(n+2)\ga^{-\alpha})})
- c e^{-a \ga^{-\alpha}}
     \end{eqnarray}
where $\inf^*$ shorthands the inf over $m\in L^\infty([\ga^{-\alpha},(n+1)\ga^{-\alpha}),[-1,1])$ such that:
\begin{itemize}

\item $\theta^*_1 = -1, \theta^*_n= 1 ; \Theta^*_i =0, i=1,..,n$;

\item  $m(r) = - m_\beta$ for all $r\in [\frac 32 \ga^{-\alpha}  -1, \frac 32 \ga^{-\alpha} +1]$;

\item   $m(r) = m_\beta$ for all $r\in [(n+\frac 12)\ga^{-\alpha}  -1, (n+\frac 12)\ga^{-\alpha}  +1]$.

\end{itemize}

\vskip.5cm

\noindent
We split the interval $[\ga^{-\alpha},(n+1)\ga^{-\alpha})$ into three intervals:
        \begin{eqnarray}
      \label{F.11}
&&  A=[\ga^{-\alpha},\frac 32\ga^{-\alpha}), \quad A'=[(n+\frac 12) \ga^{-\alpha},(n+1)\ga^{-\alpha})\nn\\&&
\La=[\frac 32 \ga^{-\alpha},(n+\frac 12)\ga^{-\alpha})
     \end{eqnarray}

\medskip

\begin{lemma}
Let $m_{[\ga^{-\alpha},(n+1)\ga^{-\alpha})}(r)$ be equal to $- m_\beta$ when $r\in [\frac 32 \ga^{-\alpha}, \frac 32 \ga^{-\alpha} +1]$ and to $m_\beta$ when
$r\in [(n+\frac 12)\ga^{-\alpha}, (n+\frac 12)\ga^{-\alpha}  +1]$.  Then
       \begin{eqnarray}
   \label{F.12}
 &&  F(m_{[\ga^{-\alpha},(n+1)\ga^{-\alpha})} | m_{[0,\ga^{-\alpha})},m_{[(n+1)\ga^{-\alpha},(n+2)\ga^{-\alpha})})
   = F_1+F_2+F_3\nn\\&&\hskip2cm
  + 2   m_\beta^2\int_0^1 dr\int_{-1}^0 dr'J(|r-r'|)
    \end{eqnarray}
 where
       \begin{eqnarray}
      \label{F.13}
  && F_1:= F(m_A|m_{[0,\ga^{-\alpha})}, -m_\beta \mathbf 1_{r \in [\frac 32\ga^{-\alpha}-1,\frac 32\ga^{-\alpha}]
  })\nn\\&&
    F_2:= F(m_{A'}|m_{[(n+1)\ga^{-\alpha},(n+2)\ga^{-\alpha})},
  m_\beta \mathbf 1_{[(n+\frac 12) \ga^{-\alpha},(n+\frac 12) \ga^{-\alpha}+1]})\nn
\\&&
F_3:=
F(m_\La |-m_\beta \mathbf 1_{r \in [\frac 32\ga^{-\alpha}-1,\frac 32\ga^{-\alpha}]
  }, m_\beta \mathbf 1_{[(n+\frac 12) \ga^{-\alpha},(n+\frac 12) \ga^{-\alpha}+1]})
     \end{eqnarray}

\end{lemma}

\medskip
\noindent
{\bf Proof.}   The l.h.s.\ of \eqref{F.12} differs from $F_1+F_2+F_3$ because
the interactions across  $\frac 32 \ga^{-\alpha}$ and across $(n+\frac  1 2 ) \ga^{-\alpha}$ are counted twice.  The last term on the r.h.s.\ of \eqref{F.12} corrects such overcounting.  \qed

\medskip

Since $F_1$ and $F_2$ do not depend on $ m_\La$ we can minimize  separately the three $F_i$.

\medskip

\noindent
Minimization of $F_1$.  We need to minimize $F_1$ over functions $m_A$ such that
$\eta^*_i$ is equal to $-1$ for all $i: [\delta i, \delta(i+1)) \subset A$ and with boundary conditions which have the same property (in particular the boundary condition at $[\frac 32\ga^{-\alpha}-1,\frac 32\ga^{-\alpha}]$ is identically $-m_\beta$.  In Theorem 6.3.3.1 in \cite{presutti}  it is  proved that there is a unique minimizer $m^*_A$ which is the unique solution of
       \begin{eqnarray}
      \label{F.14}
  && m^*_A(r) = \tanh\{ J* m(r)\}, \quad r \in A
     \end{eqnarray}
where $m$ in the argument of $\tanh$ is equal to $m^*_A$ in $A$, to $m_{[0,\ga^{-\alpha})}$ in $[0,\ga^{-\alpha})$ and to $-m_\beta$ in $[\frac 32\ga^{-\alpha}-1,\frac 32\ga^{-\alpha}]$.
By the symmetry under change of sign:
       \begin{eqnarray}
      \label{F.15}
  && \inf_{\eta^*_i \equiv -1}F_1 = F(-m^*_A|-m_{[0,\ga^{-\alpha})}, m_\beta \mathbf 1_{r \in [\frac 32\ga^{-\alpha}-1,\frac 32\ga^{-\alpha}]
  })
     \end{eqnarray}
where $\eta^*_i(-m^*_A) \equiv 1$.

\medskip

\noindent
Minimization of $F_2$.  The argument is completely analogous to that for $F_1$, hence there
is $m^*_{A'}$ with $\eta^*_i(m^*_{A'}) \equiv 1$ in $A'$ which satisfies the analogue of
\eqref{F.14} in $A'$ and such that
       \begin{eqnarray}
      \label{F.16}
  && \inf_{\eta^*_i \equiv  1}F_2 = F(m^*_{A'}|m_{[(n+1)\ga^{-\alpha},(n+2)\ga^{-\alpha})}, m_\beta \mathbf 1_{r \in [
  \mathbf 1_{[(n+\frac 12) \ga^{-\alpha},(n+\frac 12) \ga^{-\alpha}+1]})
  })
     \end{eqnarray}

\medskip

\noindent
Minimization of $F_3$.  Let  $\mathcal G$ be the set of $m \in L^\infty(\mathbb R, [-1,1])$
with the following properties:

\begin{itemize}

\item $m (r) = m_\beta$ for $r> (n+1/2)\ga^{-\alpha}$ and
$m (r) = -m_\beta$ for $r\le 3/2\ga^{-\alpha}$.

\item $\theta^*_1 (m) = -1$, $\theta^*_n(m)= 1$.

\item $\Theta^*_i (m) = 0$, $i=1,..,n$.

\end{itemize}
Call  $\mathcal G_\La$   the set of functions which are the restriction to $\La$
of functions in  $\mathcal G$.
\medskip
\begin{lemma}
With the above notation
        \begin{eqnarray}
      \label{F.17}
&& \inf_{m\in \mathcal G_\La}  F_3= \inf_{m\in \mathcal G} \mathcal F( m)  + |\La| f_\beta(m_\beta) +
 m_\beta^2 \int_{r>0} dr \int_{r'<0} dr' J(|r-r'|)
     \end{eqnarray}

\end{lemma}

\medskip

\noindent
{\bf Proof.}  Let $m\in \mathcal G$, then
       \begin{eqnarray*}
 \mathcal F( m) &=& F(m_\La |m_{\La^c})-|\La| f_\beta(m_\beta)
+\int_{-\infty}^{3/2\ga^{-\alpha}}dr \Big( \frac 12 m_\beta^2  - \frac 12
\int_{-\infty}^{3/2\ga^{-\alpha}}dr' m_\beta^2J(|r-r'|)\Big)\\&+&
\int_{(n+1/2)\ga^{-\alpha}}^{\infty}dr \Big( \frac 12 m_\beta^2  - \frac 12
\int_{(n+1/2)\ga^{-\alpha}}^{\infty}dr' m_\beta^2J(|r-r'|)\Big)
    \end{eqnarray*}
    hence \eqref{F.17}. \qed

    \medskip
The $ \inf_{m\in \mathcal G} \mathcal F( m)$  will be studied in the following two subsections, here we proceed by analyzing  the other terms.
Going back to \eqref{F.10} we get
        \begin{eqnarray}
      \label{F.10.0}
&&   \text{$\inf\,^*$}
F(m_{[\ga^{-\alpha},(n+1)\ga^{-\alpha})} | m_{[0,\ga^{-\alpha})},m_{[(n+1)\ga^{-\alpha},(n+2)\ga^{-\alpha})})\ge \inf_{m\in \mathcal G} \mathcal F( m)\nn\\&&
+ F(-m^*_A|-m_{[0,\ga^{-\alpha})}, m_\beta \mathbf 1_{r \in [\frac 32\ga^{-\alpha}-1,\frac 32\ga^{-\alpha}]
  })\nn \\&&+  F(m^*_{A'}|m_{[(n+1)\ga^{-\alpha},(n+2)\ga^{-\alpha})}, m_\beta \mathbf 1_{r \in [
  \mathbf 1_{[(n+\frac 12) \ga^{-\alpha},(n+\frac 12) \ga^{-\alpha}+1]})
  })\nn\\&& + |\La| f_\beta(m_\beta)  + 3
 m_\beta^2 \int_{r>0} dr \int_{r'<0} dr' J(|r-r'|)
     \end{eqnarray}
We claim that 
       \begin{eqnarray}
      \label{F.10.00}
&&   \text{$\inf\,^*$}
F(m_{[\ga^{-\alpha},(n+1)\ga^{-\alpha})} | m_{[0,\ga^{-\alpha})},m_{[(n+1)\ga^{-\alpha},(n+2)\ga^{-\alpha})})\ge \inf_{m\in \mathcal G} \mathcal F( m)\nn\\&&
+  F(\hat m_{[\ga^{-\alpha},(n+1)\ga^{-\alpha})} |-m_{[0,\ga^{-\alpha})},m_{[(n+1)\ga^{-\alpha},(n+2)\ga^{-\alpha})})
     \end{eqnarray}
where $\hat m_{[\ga^{-\alpha},(n+1)\ga^{-\alpha})}$ is equal to $-m^*_A$ and
$m^*_{A'}$ in $A$ and $A'$ and elsewhere it is equal to $m_\beta$.

To prove the claim \red{we need} to show that the sum of the last four terms in \eqref{F.10.0}  is equal to the last term in \eqref{F.10.00}.  To this end
we define  $\mathcal G^*$ as the set of $m \in L^\infty(\mathbb R, [-1,1])$
with the following properties:

\begin{itemize}

\item $m (r) = m_\beta$ for $r> (n+1/2)\ga^{-\alpha}$ and
 for $r\le 3/2\ga^{-\alpha}$.

\item $\theta^*_1 (m) = 1$, $\theta^*_n(m)= 1$.

\item $\Theta^*_i (m) = 0$, $i=1,..,n$.

\end{itemize}

proceed as before with the $\inf^*$ over $m\in \mathcal G^*$ we  get the following analogue of \eqref{F.10.0}
         \begin{eqnarray*}
&&   \text{$\inf\,^*$}
F(\hat m_{[\ga^{-\alpha},(n+1)\ga^{-\alpha})} |-m_{[0,\ga^{-\alpha})},m_{[(n+1)\ga^{-\alpha},(n+2)\ga^{-\alpha})})\ge \inf_{m\in \mathcal G^*} \mathcal F( m)\\&&
+ F(-m^*_A|-m_{[0,\ga^{-\alpha})}, m_\beta \mathbf 1_{r \in [\frac 32\ga^{-\alpha}-1,\frac 32\ga^{-\alpha}]
  }) \\&&+  F(m^*_{A'}|m_{[(n+1)\ga^{-\alpha},(n+2)\ga^{-\alpha})}, m_\beta \mathbf 1_{r \in [
  \mathbf 1_{[(n+\frac 12) \ga^{-\alpha},(n+\frac 12) \ga^{-\alpha}+1]})
  })\\&& + |\La| f_\beta(m_\beta)  + 3
 m_\beta^2 \int_{r>0} dr \int_{r'<0} dr' J(|r-r'|)
     \end{eqnarray*}
 and the claim follows because $\inf_{m\in \mathcal G^*} \mathcal F( m) =0$
 as $m(r)\equiv m_\beta$ is in $\mathcal G^*$.

 \vskip.4cm

 By the smoothness in $A \cup A'$ it follows that the last term in \eqref{F.10.00} is bounded from below by
        \begin{eqnarray*}
&&   F(\tilde m_{[\ga^{-\alpha},(n+1)\ga^{-\alpha})} |-m_{[0,\ga^{-\alpha})},m_{[(n+1)\ga^{-\alpha},(n+2)\ga^{-\alpha})}) - c n \ga^{-\alpha}
     \end{eqnarray*}
where $\tilde m = \psi (\und m)$ with $\und m$ in the range of $\phi$ (see the notation at the beginning of this appendix).  The converse of \eqref{F.4} holds namely (see for instance Theorem 4.2.2.2 in \cite{presutti})
       \begin{eqnarray}
   \label{F.18}
 &&-\ga^{-1}  \beta F(\tilde m_{[\ga^{-\alpha},(n+1)\ga^{-\alpha})} |-m_{[0,\ga^{-\alpha})},m_{[(n+1)\ga^{-\alpha},(n+2)\ga^{-\alpha})})
 \nn \\&&
\hskip2cm \le
\log Z^+_{n,\und m,(-s_0, s_{n+1})} + c n\ell^+_\ga\ga^{-1/2 } \log \ga^{-1}
 \nn \\&&
\hskip2cm \le \log Z^+_{\ga,n}(-s_0, s_{n+1}) + c n\ell^+_\ga\ga^{-1/2 } \log \ga^{-1}
 \end{eqnarray}
{Thus \eqref{111.6}}
will be proved once we show that
   \begin{equation}
      \label{F.19}
\inf_{m\in \mathcal G} \mathcal F( m) \ge  \bar f + \ga^{-1}c_0 (n-n_0)\mathbf 1_{n\ge n_0}
%
     \end{equation}

\vskip.5cm

\subsection{The instanton}

We will prove \eqref{F.19} separately  for ``small and large $n$'', the former in this subsection, latter in the next one.

Recalling the definition \eqref{intro.7} of $\bar f$ we have
       \begin{eqnarray}
      \label{F.20}
&& \inf_{m\in \mathcal G} \mathcal F( m) = \bar f =\int dr \,\bar m(r)
     \end{eqnarray}
where     $\bar m(r)$, called the instanton, solves
      \begin{eqnarray*}
&&     \bar m(r) = \tanh \Big( \int dr' J(|r-r'|)  \bar m(r')\Big)
     \end{eqnarray*}
 with     $ \bar m(r) \to \pm m_\beta$ as $r\to \pm \infty$. $ \bar m(r)$ is modulo translations the only function with such properties, see for instance  Theorem 8.1.2.1 in \cite{presutti}.

 \eqref{F.20} proves \eqref{F.19} for $n$   small, i.e.\ $n<n_0$ (which will be defined in the next subsection).

\vskip.5cm

\subsection{A large deviation estimate}

 The bound \eqref{F.20} being independent of $n$ becomes inadequate when $n$ is large, for that we use a large deviation estimate proved in Theorem 6.4.2.3 in \cite{presutti}.  Let $\mathcal I_0$ be the collection of intervals $[i\delta,(i+1)\delta)$ where $\eta*_i=0$ and
$\mathcal I_{\ne}$ a collection of consecutive pairs of such intervals with $\eta*_i= - \eta*_{i+1}$ and such that no interval appears twice in $\mathcal I_{\ne}$.  One can check that there is $b>0$ so that $|\mathcal I_0| +|\mathcal I_{\ne}| \ge bn$. Then by  Theorem 6.4.2.3 in \cite{presutti} there is a positive constant $c$ so that
        \begin{eqnarray}
      \label{F.22}
&& \inf F_3 \ge c \zeta^2 \delta bn
     \end{eqnarray}
 \eqref{F.20} and \eqref{F.22} yield
        \begin{eqnarray}
      \label{F.23}
&& \inf F_3 \ge  \bar f + c \zeta^2 \delta b(n-n_0) \mathbf 1_{n > n_0},\quad
n_0: c \zeta^2 \delta b n_0> \bar f
     \end{eqnarray}
 thus completing the proof of  \eqref{F.19} and hence \eqref{111.6}.

\vskip2cm

\setcounter{equation}{0}

\section{Proof of Theorem \ref{thm77.1}}
\label{appendix D}

We  split
  \begin{equation}
      \label{DD.1}
\sum_{{\und u\in \mathcal R}} w_{\la}(\und u)= \sum_{k\ge 1} \psi_k(\la),\quad
 {\psi_k(\la)}= \sum_{{\und u\in \mathcal R}: k(\und u)=k} w_{\la}(\und u)
     \end{equation}
and define
      \begin{equation}
      \label{DD.5}
\psi_0(\la){: =} \sum_{\und u\in \mathcal R: k(\und u)=1; u_3\ge 3} w(\und u)
      e^{-\la |\und u|}
     \end{equation}
We will
prove that there is $c$ so that for all $\ga$ small enough and all $\la\in [\frac{\eps_\ga}2,\frac{3\eps_\ga}2]$,
  \begin{eqnarray}
      \label{DD.2}
&& |\psi_0(\la)- (\frac{\eps_\ga}{\la})^2| \le c\eps_\ga
e^{2 c_b \ga^{-b}},\quad|\phi_0(\la)| \le c\eps_\ga e^{4c_b\ga^{-b}},\quad
\psi_0(\la)+ \phi_0(\la){\equiv }\psi_1(\la)
   \nn\\&&
\psi_2(\la) \le c\eps_\ga e^{8c_b\ga^{-b}},\quad \sum_{k>2}\psi_k(\la)  \le c\eps_\ga
    \end{eqnarray}
\red{By \eqref{111.5.3}} $\eps_\ga e^{8c_b\ga^{-b}}$ is infinitesimal as $\ga\to 0$ because $b\in(1/2,1)$.
We start by proving the first inequality in \eqref{DD.2}:


     \medskip
     \subsection{The term $\psi_0$}

We rewrite $\psi_0(\la)$ more explicitly; recalling \eqref{20.70},
        \begin{eqnarray}
      \label{DD.6}
&& \hskip-1cm \psi_0(\la) = \sum_{u_1,u_3 \ge 3} \sum_{u_2,u_4\ge 2} e^{-\la(u_1+\cdots+u_4)} e^{A^{(1)}_{\ga,u_1}}
\sum_{\{s_1,s_2,s_3,s_4\}} e^{G^{(3)}_{\ga,u_3}(s_2,s_4)}e^{V^2_{u_2}(s_2 ,s_3)}
e^{V^4_{u_4}(s_4 ,s_1)}
     \end{eqnarray}
where     $V^2$ and $V^4$ are defined in \eqref{20.3.0} and \eqref{20.3.1}
and the sum over $\{s_1,s_2,s_3,s_4\}$ is restricted to $\theta(s_1)=\theta(s_2)=1$, $\theta(s_3)=\theta(s_4)=-1$.

We write
        \begin{eqnarray}
      \label{DD.7}
&&  \psi_0(\la) = \sum_{u_1,u_3 \ge 3} \sum_{u_2,u_4\ge 2} e^{-\la(u_1+\cdots+u_4)}
\sum_{\{s_1,s_2,s_3,s_4\}} e^{V^2_{u_2}(s_2 ,s_3)}
e^{V^4_{u_4}(s_4 ,s_1)} + \psi'_0(\la)
     \end{eqnarray}
By \eqref{10.3}
       \begin{eqnarray}
      \label{DD.8}
&& | \psi'_0(\la)| \le   \sum_{u_1,u_3 \ge 3} \sum_{u_2,u_4\ge 2} e^{-\la(u_1+\cdots+u_4)} \Big(|1-e^{a e^{-b_0 \ga \ell_\ga^+ u_3}}|
\sum_{\{s_1,s_2,s_3,s_4\}} e^{V^2_{u_2}(s_2 ,s_3)}
e^{V^4_{u_4}(s_4 ,s_1)}  \nn\\
&&\hskip2cm + |1-e^{a e^{-b_0 \ga \ell_\ga^+u_1}}|e^{a e^{-b_0 \ga \ell_\ga^+ u_3}}\sum_{\{s_1,s_2,s_3,s_4\}} e^{V^2_{u_2}(s_2 ,s_3)}
e^{V^4_{u_4}(s_4 ,s_1)}\Big)
     \end{eqnarray}
We then have
      \begin{eqnarray}
      \label{DD.9}
&& | \psi'_0(\la)| \le  \frac{1}{1-e^{-\la}}  \sum_{u_3 \ge 3}  |1-e^{a e^{-b_0 \ga \ell_\ga^+ u_3}}| \sum_{u_2,u_4\ge 2}
\sum_{\{s_1,s_2,s_3,s_4\}} e^{V^2_{u_2}(s_2 ,s_3)}
e^{V^4_{u_4}(s_4 ,s_1)}  \nn\\
&& + \{\sum_{u_1\ge 3}|1-e^{a e^{-b_0 \ga \ell_\ga^+ u_1}}|\}
\{ \frac{1}{1-e^{-\la}}
e^{a e^{-b_0 \ga \ell_\ga^+ 3}}\}\sum_{u_2,u_4\ge 2}\sum_{\{s_1,s_2,s_3,s_4\}} e^{V^2_{u_2}(s_2 ,s_3)}
e^{V^4_{u_4}(s_4 ,s_1)}
     \end{eqnarray}
By \eqref{111.4} and \eqref{111.5}
      \begin{eqnarray}
      \label{DD.10}
&& \sum_{u_2,u_4\ge 2}\sum_{\{s_1,s_2,s_3,s_4\}} e^{V^2_{u_2}(s_2 ,s_3)}
e^{V^4_{u_4}(s_4 ,s_1)} = \eps_\ga^2
     \end{eqnarray}
so that
     \begin{eqnarray}
      \label{DD.11}
&& | \psi'_0(\la)| \le  c\la^{-1}\eps_\ga^2  = c' \eps_\ga
     \end{eqnarray}
when $\la\in [\frac{\eps_\ga}2,\frac{3\eps_\ga}2]$.

Call $\psi_0^*(\la)$ the first term on the right hand side of \eqref{DD.7}, then
        \begin{eqnarray}
      \label{DD.12}
&&  \psi^*_0(\la) = \sum_{u_1,u_3 \ge 3}  e^{-\la(u_1+ u_3)}\Big(
\eps_\ga^2 + \sum_{u_2,u_4\ge 2}[e^{-\la(u_2+ u_4)}-1]\eps_\ga(u_2)\eps_\ga(u_4)\Big)
     \end{eqnarray}
We have
       \begin{eqnarray}
      \label{DD.13}
&& | \sum_{u_1,u_3 \ge 3}  e^{-\la(u_1+ u_3)}
\eps_\ga^2 -(\frac{\eps_\ga}{\la})^2| \le c \eps_\ga
     \end{eqnarray}
In the last term in \eqref{DD.12} we write
  $$
|e^{-\la(u_2+ u_4)}-1| \le |e^{-\la u_2}-1| +  |e^{-\la u_4}-1|
  $$
Then by \eqref{111.5.2}-\eqref{111.5.3} 
  the last term in \eqref{DD.12} is bounded by
 $$
(\frac{1}{1-e^{-\la}})^2 2\eps_\ga [\eps_\ga c\la e^{2 c_b \ga^{-b}}]
\le c' \eps_\ga e^{2 c_b \ga^{-b}}
  $$
which completes the proof of the first inequality in \eqref{DD.2}.

     \medskip
     \subsection{The term $\phi_0$}

By \eqref{DD.2} $\phi_0=\psi_1-\psi_0$, hence
        \begin{eqnarray}
      \label{DD.14}
&& \hskip-1.5cm \phi_0(\la) = \sum_{u_1\ge 3,u_3\le 2} \sum_{u_2,u_4\ge 2} e^{-\la(u_1+\cdots+u_4)} e^{A^{(1)}_{\ga,u_1}}
\sum_{\{s_1,s_2,s_3,s_4\}} e^{V^{3}_{\ga,u_3}(s_2,s_4)}e^{V^2_{u_2}(s_2 ,s_3)}
e^{V^4_{u_4}(s_4 ,s_1)}
     \end{eqnarray}
thus $\phi_0>0$. We bound
    $$
     e^{-\la(u_2+\cdots+u_4)} e^{A^{(1)}_{\ga,u_1}} \le 2
     $$
     so that
             \begin{eqnarray}
      \label{DD.15}
&& \hskip-1.5cm \phi_0(\la) \le \frac{{2}}{1-e^{-\la}} \sum_{u_3\le 2 ; u_2,u_4\ge 2}
\sum_{\{s_1,s_2,s_3,s_4\}} e^{V^{3}_{\ga,u_3}(s_2,s_4)}e^{V^2_{u_2}(s_2 ,s_3)}
e^{V^4_{u_4}(s_4 ,s_1)}
     \end{eqnarray}
By \eqref{111.12} { and \eqref{20.3.3}}
             \begin{eqnarray}
      \label{DD.16}
&& \hskip-1.5cm \phi_0(\la) \le \frac{2}{1-e^{-\la}}
c^2\eps_\ga^2 e^{4  c_b \ga^{-b}}
     \end{eqnarray}
which proves the second inequality in \eqref{DD.2}.

     \medskip
     \subsection{The terms $\psi_k, k\ge 2$}
     \label{subsect-F3}
Fix $k\ge 2$. In the sum over $u_{\ell,m}$, $\ell=1,..,k$, $m=1,..,4$ we distinguish the cases when $u_{\ell,1}$ and $u_{\ell,3}$ are $\ge 3$ or $\le 2$, \red {here} $\ell \ge 2$  because $u_{1,1}\ge 3$ by the definition of $\psi_k(\la)$.
We thus have $2^{2k-1}$ terms and when $u\equiv u_{\ell,1}\ge 3$ we bound $e^{-\la u}\le 1$ and use \eqref{10.3}
to bound (for $\ga$ small enough)
    $$
 \sup_{s,s'}   |e^{G^{(1)}_{\ga,u}(s,s')} - e^{A^{(1)}_{\ga,u}}| \le 2\{
  \sup_{s,s'}|G^{(1)}_{\ga,u}(s,s')|+ |A^{(1)}_{\ga,u}|\}
 \le  4a e^{-b \ga\ell_\ga^+ u}
    $$
Thus the sum over any such  $u_{\ell,1}$ is bounded by a constant.
When $u\equiv u_{\ell,3}\ge 3$ we bound
    $$
\sup_{s,s'} e^{G^{(3)}_{\ga,u}(s,s')} \le 2
    $$
and  the sum over any such  $u_{\ell,3}$ is bounded by $\frac{2}{1-e^{-\la}}$.
For the remaining terms we use \eqref{111.12} and get the bound, ($d$ below is a suitable constant)
             \begin{eqnarray}
      \label{DD.16b}
&& \hskip-1.5cm \psi_k(\la) \le c^k (\frac{1}{1-e^{-\la}})^{k+1}
\eps_\ga^{2k} e^{4k  c_b \ga^{-b}} \le  d^k \eps_\ga^{k-1}  e^{4k  c_b \ga^{-b}}
     \end{eqnarray}
(recall that $\eps_\ga/\la \le 3/2$).  When $k=2$:
            \begin{eqnarray}
      \label{DD.17}
&& \hskip-1.5cm \psi_2(\la) \le \eps_\ga d^2  e^{8  c_b \ga^{-b}}
     \end{eqnarray}
while (for $\ga$ small enough)
            \begin{eqnarray}
      \label{DD.18}
&& \hskip-1.5cm \sum_{k\ge 3}\psi_k(\la) \le  \eps_{\ga} \Big(
\sum_{k\ge 3} d^k \eps_\ga^{k/3}  e^{4k  c_b \ga^{-b}}\Big) \le c'  \eps_\ga
     \end{eqnarray}
because $k-2 \ge k/3$ for $k\ge 3$.

\vskip 3cm
\newpage
\setcounter{equation}{0}

\section{\red{Proof of Theorem \ref{thm13.1}}}
\label{app-last}
\red{In this appendix we will prove the exponential bound in Theorem \ref{thm13.1}.
The proof will use properties of the following Markov chain.}

\subsection{\red{An auxiliary Markov chain}}
\red{The auxiliary Markov chain is defined as follows. The state space is $\{0,8,9,...\}$ namely all the non negative integers
minus those from 1 to 7.  Recall that the length of a quadruple in $\mathcal R$ is $\ge 8$.  We suppose that the transition probability denoted by $p(z,z')$ is such  that:}

\begin{itemize}

\item \red{$p(0,0)=1$, $p(z,z')>0$ for all $z \ne 0$ and all $z'$ (in the state space).}

\item \red{There are $c$ and $\delta$ positive so that $p(z,z') \le c e^{-\delta z'}$ for all $z \ne 0$ and $z'$.}

    \item \red{There is $\zeta>0$ so that $p(z,0) \ge \zeta$ for all  $z \ne 0$.}

\end{itemize}

\noindent
\red{We call $P_{z_0}$ the law of the Markov chain $\{z_n\}$ starting from $z_0$ and
with transition probability  $p(z,z')$.  We denote by $n_+$ the stopping time of the chain at $0$, namely $n_+$ is such that $z_n>0$ for all $n<n_+$ and $z_{n_+}=0$.  Notice that $z_n=0$ for all $n\ge n_+$.}

\bigskip

\begin{thm}
\red{There are $C$ and $a$ positive so that for any $t>0$ and $z_0\ne 0$:}
 \begin{equation}
      \label{last.1}
\red{P_{z_0}\Big[\sum_{n =1}^{ n_+} z_n >t\Big] \le C e^{-at}}
     \end{equation}

\end{thm}

\medskip

\noindent
{\bf Proof.}  \red{The proof is a straight consequence of the following two bounds.
 \begin{equation}
      \label{last.2}
P_{z_0}\Big[ n_+>n\Big] \le  (1- \zeta)^{n}
     \end{equation}
which is proved using the third property of the transition probability.  There are $\kappa$, $C'$ and $a'$ positive so that for all $N$
\begin{equation}
      \label{last.3}
P_{z_0}\Big[ \sum_{n=1}^N z_n > \kappa N\Big] \le  C' e^{-a'N}
     \end{equation}
which follows using the Chebishev exponential inequality and the exponential bound for the transition probability.  }\qed

\bigskip

\subsection{\red{Exponential decay}}

\red{We go back to the proof of Theorem 1.  We define for any integer $u \ge 8$
 \begin{equation}
      \label{last.4}
q(u):= \sum_{\und u \in \mathcal R} \mathbf 1_{|\und u|=u} w_{\la_\ga}(\und u)
     \end{equation}
$q(u)$ is a probability on $\{u\in \mathbb N: u\ge 8\}$, $q(u)>0$ for all $u$ and
 \begin{equation}
      \label{last.5}
q(u)\le c_\om e^{-\om u},\quad \om >0
     \end{equation}
Let $x_0 \in \mathbb Z$, set $X_0=x_0$ and  for $n\ge 1$
 \begin{equation}
      \label{last.6}
X_n = x_0 +u_1+\cdots + u_n,\quad \und X = (X_n)_{n\ge 0}
     \end{equation}
where the variables $u_i$ are i.i.d.\ with law $q(u)$.  We denote by $\mathbb P_{x_0}$ the law of $\und X$.  We want to prove that
 \begin{equation}
      \label{last.7}
\lim_{x_0\to -\infty}\mathbb P_{x_0} \big[  0\in \und X\big ]  =  \alpha^{-1},\quad
\alpha = \sum_{u\ge 8} u q(u)
     \end{equation}
and that the convergence is exponentially fast.  The existence of the limit (in a much more general setup)
is well known in the literature as the
 Erd\"os, Feller and Pollard theorem. For completeness we will prove it in the remark at the end of this appendix, but for the time being we take it for proved so that we only need to show that the convergence in \eqref{last.7} is exponential.}

\red{We will first prove that there is $\zeta>0$ such that
\begin{equation}
      \label{last.8}
\mathbb P_{x_0} \big[  0\in \und X\big ]  \ge \zeta  \quad \text{for all $x_0 \le -8$}
     \end{equation}
(this is one of the assumptions used in the previous subsection).}

\noindent
{\em Proof of (G.8).} 
\red{Let $n >15$ then for any $x_0<-n$
\begin{equation*}
      \label{last.8}
\mathbb P_{x_0} \big[  \und X\cap [-n,-8] = \emptyset\big ]  \le
\sum_{y< -n} \mathbb P_{x_0} \big[ y\in \und X\big] \sum_{u>|y|-8}q(u)
 \le
 \sum_{u>n+1-8}q(u) <1 - q(8)
     \end{equation*}
     }
\red{Then, recalling that $q(8)>0$,
\begin{equation*}
      \label{last.8}
\mathbb P_{x_0} \big[  0\in \und X\big ]  \ge q(8)\;\min_{u \in[8,n]} q(u) =: \zeta
     \end{equation*}
}
 \qed
 
%

\red{We call $\und Y$
the sequence defined in \eqref{last.6} with $x_0$ replaced by $y_0$.  By  \eqref{last.7} it will be enough to prove that there are $b$ and $c$ positive so that for all $x_0 \le -8$:
  \begin{equation}
      \label{last.9}
\lim_{y_0\to -\infty}\Big |\mathbb P_{x_0} \big[  0\in \und X\big ]
-\mathbb P_{y_0} \big[  0\in \und Y\big ] \Big | \le c e^{-b|x_0|}
     \end{equation}
Given $y_0<x_0<0$ with $|x_0|$ large enough, we are going to define a coupling $Q_{x_0,y_0}$ of $\und X$ and $\und Y$ so that denoting by
$E_{x_0,y_0}$ its expectation, \eqref{last.9} becomes
  \begin{equation}
      \label{last.10}
\lim_{y_0\to -\infty}\Big |E_{x_0,y_0} \big[\mathbf 1_{ 0\in \und X} -
 \mathbf 1_{ 0\in \und Y} \big ] \Big | \le c e^{-b|x_0|}
     \end{equation}
To define  $Q_{x_0,y_0}$ we introduce ``overshooting'' variables $z_n, n\ge 0$ which are functions of $\und X$ and $\und Y$, setting $z_0 := x_0-y_0$.  Think of $x_0$ as a ``target'' for $\und Y$ which is ``shooting'' from $y_0$:
call then $n_1$ the first integer $n$ such that
$Y_{n_1} \ge x_0$ and define $z_1= Y_{n_1}-x_0$ if this is not in $\{1,..,7\}$ otherwise we set
$z_1= Y_{n_1+1}-x_0$.  $z_1$ says how much we missed the target $x_0$, which instead has been hit if $z_1=0$.}

\red{If $z_1>0$ then  $\und X$, which is shooting from $x_0$, has the new target $x_0-z_1$.  Iterating this procedure by alternating targets for $\und Y$ and $\und X$  we get
a sequence $z_n$ which is stopped as soon as the target has  been hit. Call  $n_+$ such a hitting time: i.e.\  the first $n$ such that   $z_{n}=0$.  This means that there exist integers $\bar n$ and $\bar m$ such that
 \begin{equation}
      \label{last.11}
X_{\bar n} = Y_{\bar m},\quad X_{\bar n} -x_0 = z_1+\cdots + z_{n_+}
     \end{equation}
We denote $X_n= x_0+\sum u_i$ and $Y_n= y_0+\sum v_i$ so that
the coupling  $Q_{x_0,y_0}$ is   defined by  stating how the $u_i$ and $v_i$ are coupled.  The   $u_i$  and    $v_i$ are taken  independently till $\bar n$ and $\bar m$ respectively, after that $ u_{\bar n +i}= v_{\bar m +i}$.  We also set $z_{n}=0$ for $n> n_+$.  We have
  \begin{equation}
      \label{last.12}
\Big |E_{x_0,y_0} \big[\mathbf 1_{ 0\in \und X} -
 \mathbf 1_{ 0\in \und Y} \big ] \Big | \le Q_{x_0,y_0} \big[ z_1+\cdots z_{n_+} > |x_0|\big]
     \end{equation}
The variables $z_i$ under the law $Q_{x_0,y_0}$ define a Markov chain which satisfies the properties stated in the previous subsection (the condition of exponential decay follows from \eqref{last.5}), then
\eqref{last.10}  follows from \eqref{last.1}.}

\bigskip

\noindent
\red{{\bf Remark.}  The above argument proves that the sequence $\mathbb P_{x_0} \big[  0\in \und X\big ]$ is Cauchy and it thus converges to some limit $\theta$ as $x_0\to -\infty$.  It is then easy to prove that $\theta=1/\alpha$.  Indeed, since
   \begin{equation*}
\mathbb P_{x_0} \big[ \und X \cap \mathbb N \ne \emptyset\big ] =1
     \end{equation*}
then
        \begin{equation}
      \label{last.13}
1=\mathbb P_{x_0} \big[  \text{there exists $n$: $X_n <0$ and $X_{n+1} \ge 0$ }\big ]
     \end{equation}
hence
    \begin{equation}
      \label{last.14}
1= \sum_{n>0}\mathbb P_{x_0} \big[  -n \in \und X\big ] \sum_{u\ge n} q(u)
     \end{equation}
By taking the limit $x_0\to -\infty$
    \begin{equation}
      \label{last.15}
1= \theta \sum_{n>0} \sum_{u\ge n} q(u) = \theta \alpha
     \end{equation}
}

\newpage

\bibliographystyle{9}

\end{document}